\RequirePackage{fix-cm}

\documentclass[aps,pre,twocolumn,superscriptaddress,10pt,nofootinbib,amssymb]{revtex4-2}

\usepackage[caption=false]{subfig}
\usepackage{amsfonts, amssymb, amsmath}
\usepackage{bm}
\usepackage{graphics}
\usepackage{graphicx}
\usepackage{wasysym}
\usepackage{natbib}
\usepackage{mathtools}
\usepackage{bbold}

\usepackage[utf8x]{inputenc}
\usepackage[T1]{fontenc}
\usepackage{paralist}

\usepackage[writekey=false,name={}]{notes2bib}

\usepackage[capbesideposition={right,bottom}]{floatrow}

\usepackage[usenames,dvipsnames]{xcolor}

\usepackage{color}
\definecolor{darkblue}{rgb}{0,0,0.6}
\definecolor{darkred}{rgb}{0.6,0,0}

\colorlet{lightred1}{orange!50}
\colorlet{lightred2}{orange!30}
\colorlet{lightred3}{orange!10}
\colorlet{lightred4}{orange!5}

\usepackage{tikz}
\usetikzlibrary{decorations.pathreplacing}
\usepackage{hyperref}

\hypersetup{
bookmarksopen=true,
pdftitle=Monitored Open Fermion Dynamics, 
pdfauthor= B. Ladewig et al., 
pdftoolbar=false, 
pdfstartview={FitH},		
pdfmenubar=true,			
pdfhighlight=/O,			
colorlinks=true,			
urlcolor=darkblue,
citecolor=darkblue,		
linkcolor=MidnightBlue,	
}

\usepackage[normalem]{ulem}
\usepackage{comment}

\usepackage{pifont}

\newcommand{\Fig}[2]{Fig.~\ref{#1}\textcolor{MidnightBlue}{#2}}

\newcommand{\Eq}[1]{Eq.~(\ref{#1})}
\newcommand{\eq}[1]{(\ref{#1})}

\newcommand{\rhotraj}{\hat{\rho}^{(c)}}
\newcommand{\psitraj}{\psi^{(c)}}
\newcommand{\moutcome}{m_i}
\newcommand{\moutcometime}{m_{i,t}}
\newcommand{\bosonmodes}{\Phi}
\newcommand{\trajindex}{\alpha}
\newcommand{\relativeInteractions}{\Delta S_r}
\newcommand{\subparity}{P_{|A|}}
\newcommand{\rhotworeplica}{\hat{\rho}^{(2R)}}
\newcommand{\normalorderingmass}{m}

\usepackage{abbrevs}
\usepackage{etoolbox} 
\newabbrev\RG{renormalization group (RG)}[RG]
\newabbrev\CFT{conformal field theory (CFT)}[CFT]
\newabbrev\QSD{Quantum State Diffusion (QSD)}[QSD]
\newabbrev\BKT{Berezinskii-Kosterlitz-Thouless (BKT)}[BKT]
\newabbrev\phaseone{`scale invariant' ($C$)}[($C$)]
\newabbrev\phasetwo{`scale invariant dephasing' ($C_D$)}[($C_D$)]
\newabbrev\phasethree{`measurement' ($M$)}[($M$)]

\robustify{\BKT}
\robustify{\CFT}
\robustify{\QSD}
\robustify{\RG}
\robustify{\phaseone}
\robustify{\phasetwo}
\robustify{\phasethree}

\makeatletter 
\renewcommand\maybe@space@{%
  \maybe@ictrue 
  \expandafter   \@tfor
    \expandafter \reserved@a
    \expandafter :%
    \expandafter =%
                 \nospacelist
                 \do \t@st@ic
  \ifmaybe@ic 
    \space
  \fi
}
\makeatother

\begin{document}

\title{Monitored Open Fermion Dynamics: \\ Exploring the Interplay of Measurement, Decoherence, and Free Hamiltonian Evolution}

\author{B. Ladewig}
\author{S. Diehl}
\author{M. Buchhold}
\affiliation{Institut f\"ur Theoretische Physik, Universit\"at zu K\"oln, D-50937 Cologne, Germany}

\begin{abstract}
The interplay of unitary evolution and local measurements in many-body systems gives rise to a stochastic state evolution and to measurement-induced phase transitions in the pure state entanglement. In realistic settings, however, this dynamics may be spoiled by decoherence, e.g., dephasing, due to coupling to an environment or measurement imperfections. We investigate the impact of dephasing and the inevitable evolution into a non-Gaussian, mixed state, on the dynamics of monitored fermions.
We approach it from three complementary perspectives: (i) the exact solution of the conditional master equation for small systems, (ii) quantum trajectory simulations of Gaussian states for large systems, and (iii) a renormalization group analysis of a bosonic replica field theory. For weak dephasing, constant monitoring preserves a weakly mixed state, which displays a robust measurement-induced phase transition between a critical and a pinned phase, as in the decoherence-free case. At strong dephasing, we observe the emergence of a new scale describing an effective temperature, which is accompanied with an increased mixedness of the fermion density matrix. Remarkably, observables such as density-density correlation functions or the subsystem parity still display scale invariant behavior even in this strongly mixed phase. We interpret this as a signature of gapless, classical diffusion, which is stabilized by the balanced interplay of Hamiltonian dynamics, measurements, and decoherence.
\end{abstract}

\maketitle

\section{Introduction}
Hamiltonian evolution, measurements and decoherence due to coupling to an environment (bath) are three fundamental aspects, shaping the time-evolution of quantum many-body systems. Each individual aspect, or their interplay, can give rise to collective phenomena and phase transitions in- and out of equilibrium. Recently, the interplay between Hamiltonian, or more generally, unitary evolution and measurements has gained much attention, since monitored quantum systems have been found to undergo a measurement-induced or entanglement phase transition~\cite{Skinner2019,Li2018,Li2019,Szyniszewski2019,Bao2020,Gullans2020,Choi2020,Szyniszewski2020,Fan2021,Nahum2021,Lavasani2021,Fuji2020,Lunt2020,Doggen2021arxiv,Bao2021,Turkeshi2021,Turkeshi2021arxiv,Ippoliti2021,Zabalo2021,Sierant2021arxiv,Zabalo2020,Li2021arxiv,Jian2020,Jian2020arxiv,Jian2021a,Jian2021arxiv,Alberton2021,Buchhold2021,Muller2021,Minoguchi2021,Block2021,Sharma2021arxiv,Sang2021a,Bentsen2021,Boorman2021arxiv,Biella2021,Weinstein2022arxiv}. This transition is rooted in the non-commutativity between the generators of the unitary dynamics and the measured operators, which gives rise to  macroscopically distinct stationary states. The latter is shared in common with more familiar quantum phase transitions, where the ground state is governed by a Hamiltonian $\hat{H}=\hat{H}_1 + g \hat{H}_2$ with non-commuting $[\hat{H}_1,\hat{H}_2]\neq 0$. 

However, in contrast to ground state quantum phase transitions and also to finite temperature or dissipative phase transitions \cite{Leggett1987,Tauber2014,Sieberer2016a}, measurement-induced phase transitions are not manifest at the level of the average density matrix. Rather they are detectable on the level of individual 'measurement trajectories'. For pure state trajectories, roughly three different kinds of (measurement-induced) phases can be distinguished: `area law’ entangled phases, `volume law’ phases and critical phases (`log law’). In contrast, for the linear average over trajectories, the macroscopic configurational entropy of all possible measurement outcomes eliminates all marks of the underlying quantum dynamics. 

This naturally poses the question, to what extent the fragile purity of a state is important to resolve the features of the measurement-induced evolution, and to what extent this picture is altered by sources of decoherence. This question is particularly relevant since experimental setups, such as, e.g., trapped ions or Rydberg atom arrays \cite{Schauss2012,Britton2012,Richerme2014,Jurcevic2014}, are often exposed to decoherence. At a qualitative level, unitary evolution, measurements and decoherence affect the system density matrix $\hat{\rho}$ quite differently: closed system unitary evolution, for instance, may scramble information but will never change the purity of the state $\text{tr}[\hat{\rho}^2]$. This is obvious, technically from its formulation, but also physically from the fact that it does not change the entanglement of the system with its environment. Adversely, generic decoherence,  or an imperfect measurement, roots in the coupling to an environment, i.e., in the generation of system-environment entanglement, and typically increases the mixedness of the system, apart from particularly engineered scenarios~\cite{Diehl2008,Verstraete2009}. In contrast, repeated measurements extract information from the system and monotonically reduce its entanglement with the environment, and therefore the mixedness of the state. Under the suitably combined evolution of perfect (projective or continuous) measurements and unitary gates, it has been shown that any mixed initial state purifies. The purification speed is characteristic for the underlying measurement-induced phase~\cite{Gullans2020,Gullans2020a,Bao2020,Bentsen2021},
distinguishing between, e.g., fast purification for weakly entangled phases and slow purification for strongly entangled phases. 

Adding decoherence to the picture, recent works argued that a measurement-induced area law phase, characterized by, e.g., finite long-range correlations, is robust against weak decoherence, e.g., in monitored ($\mathbb{Z}_2$-symmetric) Clifford circuits \cite{Bao2021,Li2021a}. 
However, the robustness of the transition between critical phases and area law phases, indicated by, e.g., algebraically versus exponentially decaying correlation functions (independently of the mixedness), is a priori not obvious. In fact, one might expect the critical phase to be more fragile towards perturbations. 

We address this question for a fermionic model, consisting of $U(1)$-symmetric, i.e., particle number conserving, unitary evolution stemming from a quadratic hopping Hamiltonian, which tends to delocalize the particles over the entire system. The delocalization is counteracted by continuous measurements of the local particle number $\hat{n}_i$, which tend to localize particles at individual lattice sites. Decoherence is added by coupling the system to Hermitian Lindblad operators $\hat{L}_i = \hat{n}_i$, which mimic density-dependent interactions with some Markovian environment or imperfect measurements. In the absence of decoherence, the model hosts a scale invariant, critical phase for weak measurement strengths, characterized by algebraically decaying correlations and a logarithmic entanglement growth. This phase is separated by a measurement-induced \BKT phase transition from a pinned or localized phase with exponentially decaying correlations, and an area law entanglement \cite{Alberton2021,Buchhold2021,Bao2021}. Our two main findings are: (i) we confirm the robustness of both measurement-induced phases against the environmental dephasing and detect a stable, extended critical regime and (ii) we show that the decoherence enriches the dynamics and gives rise to a novel, strongly mixed phase, characterized by an emergent decoherence-induced temperature scale.

In order to approach the dynamics analytically, we express the model as a replicated Keldysh field theory \cite{Sieberer2016a,Buchhold2021,Muller2021}, which readily allows us to include dephasing and provides access to the fermion correlation functions (see also, e.g., Refs.~\cite{Bentsen2021,Nahum2021,Bao2021,Jian2021arxiv,Jian2021a,Barratt2021arxiv} for related replica approaches). The robustness of the critical phase can be rationalized in terms of an effective bosonic, non-Hermitian variant of the sine-Gordon model, where the critical phase corresponds to the gapless theory with vanishing interactions. Using a \RG analysis, we determine the degree of relevance of perturbations, revealing the finite extent of the critical phase as well as the \emph{two} aforementioned phases (localized or decoherence-induced), dominated by different interactions. The limiting cases are related to the Gaussian \CFT studied in Ref.~\cite{Minoguchi2021}.

\section{Key Results}

\begin{figure}
    \centering
    \includegraphics[width=\textwidth]{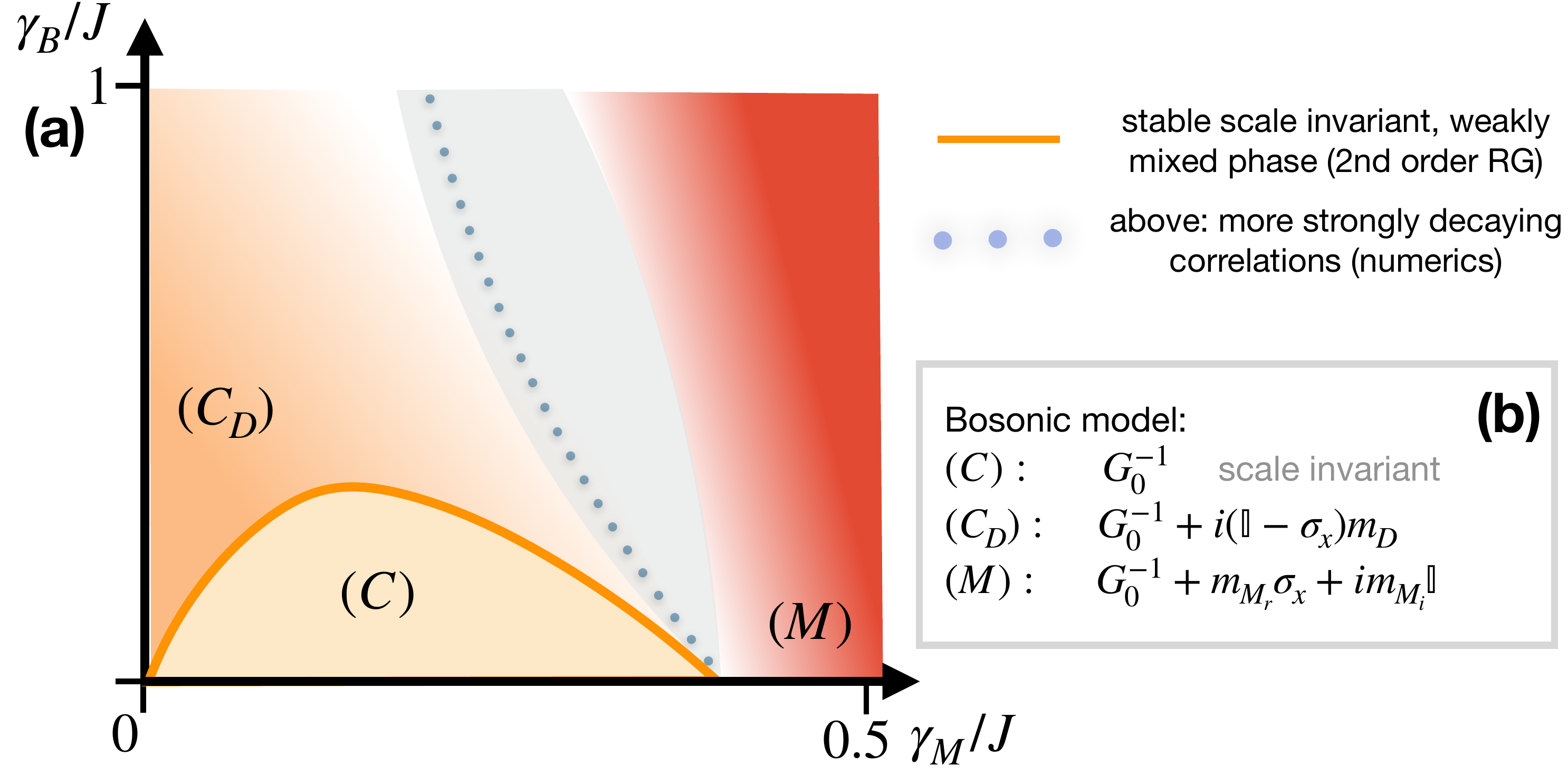}
    \caption{(a) Sketch of phase diagram, synthesized from (i) small scale numerics for the full system, (ii) larger scale simulations based on an ensemble of pure states and a (iii) \RG analysis. The three phases are a scale invariant, weakly mixed phase \phaseone, scale invariant and strongly mixed \phasetwo, and a measurement-induced phase \phasethree. The (blue) dotted line corresponds to the rough, tentative phase boundary suggested by the numerics. Including the second order \RG calculation, we indicate our synthesized estimate of the transition region from \phasetwo to \phasethree in `grey', not further resolvable in our analysis. (b) Qualitative properties of the bosonic models, see text for discussion.}
    \label{fig:SketchKeyResults}
\end{figure}

In order to treat measurements (with strength $\gamma_M$), decoherence (dephasing with strength $\gamma_B$) and unitary evolution (hopping Hamiltonian with strength $J$) of fermions on equal footing, we compare three different approaches: (i) we perform exact numerical simulations of individual measurement trajectories of the conditional (non-Gaussian) density matrix for small systems ($L=10$ sites), (ii) we numerically simulate the time evolution of the fermion correlation matrix in the framework of measured quantum trajectories, and (iii) we construct an effective bosonic replica field theory and extract the phase diagram from a perturbative \RG analysis. The synthesis of our results from (i-iii) is displayed qualitatively in Fig.~\ref{fig:SketchKeyResults}. The different phases and  transitions are extracted from the continuum field theory and the corresponding renormalization group analysis. This picture is qualitatively confirmed by the numerical analysis. However, the restricted system size for simulations does not allow us to confirm a sharp transition and to identify the precise location of the phase boundary. We will now summarize the individual aspects of our analysis.

\emph{Classification -- } We perform an analytical classification of the different phases based on the results of the replica field theory. This approach is based on bosonization, within which the fermion densities are approximated by a continuous boson field $\hat n_x\sim \partial_x\hat{\phi}_x$, see Sec.~\ref{sec:ReplicaFieldTheory} and Ref.~\cite{Buchhold2021} for details. The simultaneous presence of measurements and dephasing then makes the effective Hamiltonian for the density field generally non-Hermitian and nonlinear. However, we can extract three distinct  Gaussian fixed point theories.
Depending on the structure of each Gaussian theory, i.e., whether it is scale invariant or gapped, we associate the fixed points to different macroscopic phases. Density-dependent observables, such as density-density correlations or the subsystem parity can then be extracted readily from the corresponding Gaussian theory.

\emph{Robustness --} One key observation is that the measurement-induced dynamics, obtained previously for the decoherence-free case~\cite{Alberton2021, Buchhold2021,Bao2021}, is robust against weak decoherence. Both a scale invariant, critical phase (\phaseone in Fig.~\ref{fig:SketchKeyResults}) as well as a phase of measurement-induced pinning or localization of fermions (\phasethree in Fig.~\ref{fig:SketchKeyResults}) continue to exist for weak but nonzero dephasing rate. In the absence of decoherence, this transition is an entanglement phase transition, separating a phase with logarithmic growth of the von Neumann entanglement entropy \phaseone from an area law phase \phasethree. In the replica field theory, the former is described by a scale invariant Gaussian theory, and a propagator $(G_0^{-1})_{ab}\sim (\epsilon_{ab}^2 \partial_t^2-\eta_{ab}^2\partial_x^2) $  (for the classical and quantum fields, see Sec.~\ref{Sec:OverviewLongWavelength} for further details). The latter, on the contrary corresponds to a massive theory, for which the measurements induce a non-zero and imaginary spectral gap, $G^{-1}_0\rightarrow G^{-1}_0+m_{M_r}\sigma_x+i m_{M_i}\mathbb{1}$ (see Sec.~\ref{Sec:OverviewLongWavelength}). Increasing the dephasing strength also increases the mixedness of the system density matrix and eventually leads to a breakdown of the scale invariant phase \phaseone. Integrating the second order \RG equations provides us with a stability criterion and allows us to faithfully predict the extension of \phaseone into the regime of nonzero dephasing, see Fig.~\ref{fig:SketchKeyResults}.

\emph{Decoherence-induced temperature scale --} When increasing the strength of the decoherence, the interplay of measurements, dephasing and Hamiltonian gives rise to a new emergent scale $m_D$, which enters the action in the form of an effective temperature, $G_0^{-1}+ i m_D(\mathbb{1}-\sigma_z)$, see \Fig{fig:SketchKeyResults}{(b)}. The generation of this scale is ruled out, either by symmetry or by the structure of the \RG equations, as soon as \emph{any} of the couplings (Hamiltonian, measurement strength, dephasing strength) is set to zero, thus being a consequence of their simultaneous presence. Surprisingly, this emergent scale will \emph{not} modify the structural form of the density-dependent observables that we consider in this work. This is due to the unconventional structure of the measurement-induced propagator, which we discuss below. However, it will suppress fluctuations of the Keldysh quantum field on large distances, which is equivalent to suppressing off-diagonal entries of the density matrix and of projecting onto its diagonal. In general, as long as the system is not in a pure eigenstate of the measurement operators, this projection strongly increases the mixedness of the state. We therefore term the corresponding regime decoherent-scale invariant \phasetwo. We will discuss our phenomenological interpretation of this regime below. The second order perturbative \RG approach robustly predicts the stability of regime \phaseone against the generation of a scale for weak decoherence and weak measurements. However, estimating the exact phase boundary between the regimes \phasetwo and \phasethree from the flow equation is not always unambiguous due to the presence of runaway solutions, which are generic in \RG approaches for sine-Gordon models. The corresponding tentative phase diagram is obtained by tracing the scales with the most dominant divergent behavior. It is shown in Fig.~\ref{fig:RG_phasediagram} and sketched in \Fig{fig:SketchKeyResults}{(a)} (see Sec.~\ref{Sec:SecondOrder} for more details).

\emph{Numerical approach -- } 
The predictions from the replica field theory are complemented by numerical simulations of the conditioned density matrix $\rhotraj$. The conditioned density matrix describes the evolution of the state during one measurement trajectory in the presence of decoherence. It corresponds to the state obtained after a particular set of measurement outcomes, but in the presence of dephasing it is generally not pure. For small systems ($L=10$), we directly simulate the evolution of $\rhotraj$. For large systems we use quantum trajectories \cite{Gisin1992a,Gisin1993,Jacobs2006,Jacobs2014} to determine $\rhotraj$. In the quantum trajectories framework, decoherence is interpreted as the consequence of a series of unread (or imperfect) measurements, and $\rhotraj$ is expressed as a sum of quantum trajectories, with partly read out and partly not read out measurement outcomes \cite{Jacobs2010,Jacobs2014}. In our case, each individual quantum trajectory is described by a Gaussian state, but their \emph{sum} $\rhotraj$ itself is non-Gaussian. The quantum trajectory approach requires the simulation of a large number of auxiliary trajectories for a single \emph{measurement trajectory} $\rhotraj$, which makes it numerically costly and limits the system sizes we consider to $L \le 256$. This leaves a remaining uncertainty in the precise location of the phase transition from the numerical perspective.

In the quantum trajectory framework, entanglement properties of $\rhotraj$ are generically hard to access. Instead we use the density-density correlation function $C_{ij}$, the subsystem parity $\subparity$ (defined below in Eqs.~\eqref{Eq:Parity}, \eqref{Eq:CorrelatorFermions}) and the average purity of the density matrix $\overline{\text{tr}[\rhotraj{}^2]}$ (for small systems) as quantifiers for the different phases. In general, we use $\overline{O}$ to indicate that the observable $O$ is averaged over many different measurement trajectories $\rhotraj$. In the measurement-induced localization phase \phasethree, the state quickly evolves into a nearly pure state $\overline{\text{tr}[\rhotraj{}^2]}\approx1$, which is close to an eigenstate of the measurement operators. For large systems, we expect exponentially decaying correlations $C_{ij}\sim \exp(-|i-j|/\xi)$ and a well-defined, constant parity $\subparity$. 
In both complementary phases \phaseone and \phasetwo, the delocalizing Hamiltonian dominates over the tendency of the measurements to localize the particles, and the correlation functions decay algebraically with the distance $C_{ij}\sim |i-j|^{-2}$, and the subsystem parity decays algebraically as well \cite{Bao2021}. For small systems, we use the correlations at the largest distance $|i-j|=L/2$ to get a qualitative overview (dotted line in Fig.~\ref{fig:SketchKeyResults}). In order to distinguish the two regimes \phaseone and \phasetwo at small system sizes, we use the purity of the density matrix: it remains  large in \phaseone, and is reduced significantly in \phasetwo. 
At large system sizes, the quantum trajectory evolution does not grant efficient access to the purity. There we rely on the analytical predictions to distinguish the regimes of small and large purity via the presence or absence of the effective temperature scale in the replica field theory. This matches well with the simulations at small system sizes for a large parameter regime, see Fig.~\ref{fig:SketchKeyResults}.

\emph{Synthesis (and limitations)}: The \RG analysis and the quantum trajectory simulations confirm an extended regime with scale invariant correlation functions and strongly reduced half-system parity at weak measurement and decoherence rates, summarized in the combination of \phaseone and \phasetwo. A measurement-induced phase transition separates this regime from a localized phase \phasethree, described by almost pure states with strongly (exponentially) decaying correlations and well-defined half-system parity (consistently inferred from the analytical and numerical approaches). A transition between the two scale invariant regimes, \phaseone and \phasetwo, is indicated by the emergence of a temperature-like mass scale in the replica field theory and a significant reduction of the purity of $\rhotraj$ in the numerical simulations for small system sizes. While the replica field theory predicts a sharp phase transition between \phaseone and \phasetwo, the numerically approachable system sizes are too small to confirm a sharp transition in the purity in the thermodynamic limit. 

The precise position of the transition between phase \phasetwo and \phasethree cannot be unambiguously determined from our approaches. Numerical simulations show that strong dephasing \emph{supports} a transition into the short-ranged correlated, measurement-induced phase, and the corresponding transition region is indicated by the dotted line in Fig.~\ref{fig:SketchKeyResults}. This befits a phenomenological perspective: strong dephasing leads to a diffusive spreading of particles and at the same time suppresses the rate of diffusion, roughly as $\sim 1/\gamma_B$. This increases the tendency of the particles to become localized due to measurements and an estimate for the critical dephasing rate is $\gamma_B^c\sim 1/\gamma_M$, confirmed by the behavior of the dotted line. Integrating the second order \RG equations, however, puts the phase boundary between \phasetwo and \phasethree at larger measurement rates (right border of grey area in Fig.~\ref{fig:SketchKeyResults}). We stress that close to this particular phase boundary, the \RG equations contain a large number of flowing couplings. The determination of the most dominant ones (which in turn determine the corresponding Gaussian theories) is challenging and not entirely unambiguous. Therefore, we here rely more strongly on the prediction from the quantum trajectory simulations and our phenomenological argument for the phase boundary (dotted line). We mark the ambiguous region in parameter space as `grey' to indicate that the analytical and numerical results leave room for a discrepancy.

\emph{Phenomenological interpretation -- } The phenomenology discussed so far can be understood on an intuitive level by considering the effect of measurements, dephasing and hopping onto the conditional density matrix $\rhotraj$ in the occupation number basis. In this basis, both measurements and dephasing push the evolution of the density matrix towards the diagonal, and lead to the rapid decay of any off-diagonal elements. Measurements localize particles on individual lattice sites, and thereby evolve a diagonal density matrix into a pure state. In contrast, dephasing commutes with the diagonal and does not prefer any particular configuration, evolving any initial state that is not an eigenstate of the measurement operators into a mixed state. 

The Hamiltonian on the other hand, delocalizes particles by creating off-diagonal matrix elements. In the limit where the dephasing is weak compared to the measurement rate ($\gamma_M>\gamma_B$), the system will approach a nearly pure state, either diagonal in the occupation number when measurements dominate, or off-diagonal when the Hamiltonian dominates. In this case, measurements are the source of purity of the state, even though the Hamiltonian is dominantly delocalizing the state. If dephasing dominates over measurements, the situation becomes more subtle: the projection onto the diagonal then reduces the effect of the Hamiltonian and destroys coherent propagation of particles. In second order perturbation theory, it yields classical diffusion of particles on the diagonal of the density matrix with a rate $J^2/\gamma_B$ \footnote{For perturbative treatments of Lindblad operators, see, e.g., Refs.~\cite{Cai2013,Poletti2012,Poletti2013}. Related to our model: Refs.~\cite{Bauer2017,Bauer2019,Bernard2019} (quantum diffusive XX model, open quantum symmetric simple exclusion process, also Ref.~\cite{Eisler2011}). See Ref.~\cite{Kessler2012} for another perturbative method.}.

This classical diffusion is then counteracted by the localizing measurements. If the measurements succeed in localizing ($\gamma_M>J^2/\gamma_B$) the state again purifies and is in phase \phasethree. If the measurements do not succeed, however, the dynamics on the diagonal remains scale invariant while the state of the system is strongly mixed.

\section{Weak measurements and dephasing - from single fermions to many-body dynamics}

\subsection{Measured single fermions}

We start by giving a general introduction to the concept of continuous measurements and by deriving the time-evolution equation for fermions subject to continuous measurements and decoherence, following  Refs.~\cite{Brun2002,Turkeshi2021,Jacobs2006,Jacobs2014}. In addition, we define suitable averaged observables in the presence of measurements.

We consider an elementary model of free spinless fermions, for which both measurements and dephasing have been shown to individually lead to nontrivial modifications of the dynamics. Free fermions subject to local measurements have been studied in Refs.~\cite{Cao2019,Chen2020,Alberton2021,Minato2021,Muller2021,Turkeshi2021,Turkeshi2021arxiv,Coppola2021arxiv,Kells2021arxiv}. The impact of dephasing on fermions (or the related XX model or hard-core bosons) has been discussed in, e.g., Refs.~\cite{Znidaric2010,Znidaric2014,Medvedyeva2016,Wolff2019,Dolgirev2020,Bernier2020,Alba2021,Turkeshi2021a} (see also Ref.~\cite{Jin2021} with focus on unravellings for fermions). Here we study the situation where both measurements and dephasing are present simultaneously. 

We first illustrate the different aspects of measurements and dephasing on a simple toy model. The latter can result from either imperfect measurements or the coupling to a dephasing bath. Consider the two-dimensional Hilbert space of one fermion on a two-site lattice with basis states $\{|01\rangle,|10\rangle\}$ and creation (annihilation) operators $c_i^\dagger$ ($c_i$) for each site. Then any state has the form $|\psi\rangle = \alpha |01\rangle + \beta |10\rangle$. Now let us consider projective measurements of an operator whose eigenbasis is given by the set $\{ |\nu \rangle \}$. This measurement is described by projection operators $\hat{P}_\nu=|\nu\rangle\langle\nu|$ and Born probabilities $p_\nu$:
\begin{align}
    &\hat{P}_\nu |\psi\rangle = |\nu \rangle \langle \nu | \psi \rangle \propto |\nu \rangle, \quad \sum_\nu \hat{P}_\nu = \mathbb{1},\\
    & p_\nu = \langle \psi | \hat{P}_\nu | \psi \rangle.
\end{align}
We want to consider the more general case, where we take measurements, which only reveal very little information about the state (`weak measurements'). Then one introduces `positive operator valued measures' (POVM). Here the projectors $\hat{P}_\nu$ are replaced by operators $\hat{E}_\nu$, which fulfill \cite{Brun2002,Nielsen2010}
\begin{align}
    \sum_\nu \hat{E}_\nu = \mathbb{1},
    &&p_\nu = \text{tr}[\hat{\rho} \hat{E}_\nu],
    && \hat{E}_\nu = \hat{A}_\nu^\dagger \hat{A}_\nu.
\end{align}
The operators $\hat{A}_\nu$ are not uniquely defined by $\hat{E}_\nu$.

On the two-site lattice, the projective measurement of the particle number at site $i$, $\hat{n}_i = c_i^\dagger c_i$, is based on:
\begin{align}
 &\hat{P}_{0}^{(i)} = \mathbb{1}-\hat{n}_i, \,\ \hat{P}_1^{(i)} = \hat{n}_i,
\end{align}
where the subscript indicates whether a particle has been measured $(1)$ or not $(0)$. 
A `weak' version of this projective measurement is described by $\hat{E}_{0}^{(i)},\hat{E}_{1}^{(i)}$ (and $A_{0}^{(i)},A_{1}^{(i)}$) for generalized measurement outcomes $\moutcome \in \{0,1\}$, which correspond to measuring an ancilla instead of the system directly. It can, e.g., be written as (see in particular Ref.~\cite{Brun2002})
 \begin{align}
    &\hat{E}_{0|1}^{(i)} = \frac12\left(1 \pm p\right) \hat{P}_0^{(i)} + \frac12\left(1 \mp p \right) \hat{P}_1^{(i)}, \\
    &\hat{A}_{0|1}^{(i)} = \sqrt{\frac12(1 \pm p)}\, \hat{P}_0^{(i)} + \sqrt{\frac12(1 \mp p)}\, \hat{P}_1^{(i)}.
\end{align}
Here, $p\in [0,1]$, such that $p=0$ corresponds to performing no measurement (no information gained) and $p=1$ corresponds to a projective measurement with all information revealed. The corresponding measurement probabilities for lattice site $i$ are
\begin{align}
    &p_{0|1}^{(i)}(\hat{\rho}) = \frac12\left(1 \pm p (1-2\langle \hat{n}_i) \rangle\right), \quad \langle \hat{n}_i \rangle := \text{tr}[\hat{\rho}\, \hat{n}_i]. 
\end{align}
In this case, the different measurement outcomes are nearly equal for $p\ll 1$ and the state is only weakly altered, which opens the possibility of a continuous process in time. 

Successive weak measurements describe a stochastic dynamical process, such that for fixed value $p$ and time step between measurements $\delta t$ the `measurement rate' $\gamma$ is defined via $p = \sqrt{\gamma \delta t}$. In one time step $\delta t$, the wave function update for the state $| \psitraj \rangle$, conditioned onto the measurement outcome, is
\begin{align}
\label{eq:WeakMeasurementsOutcome}
 |\psitraj {}'\rangle = \begin{cases} \frac{\hat{A}_0^{(i)} |\psitraj\rangle}{\sqrt{p_0^{(i)}}} & \text{with prob. $p_0^{(i)}$}, \\
    \frac{\hat{A}_1^{(i)} |\psitraj \rangle}{\sqrt{p_1^{(i)}}} & \text{with prob. $p_1^{(i)}$}.
    \end{cases}
\end{align}
This turns into a continuous process for $\delta t \to 0$ (or $p\to 0$). Using the definitions of $\hat{A}_{0|1}^{(i)}$ and $p_{0|1}^{(i)}$ and expanding up to first order in $\delta t$, the evolution equation for individual measurements of site $1$ and $2$ becomes \cite{Brun2002} ($\hbar=1$)
\begin{align}
    &|\psitraj{}'\rangle - |\psitraj \rangle \approx \label{eq:QSDPureStates} \\
    &\sum_{i=1}^{2} \left[-\frac{\gamma}{2}\delta t (\hat{n}_i-\langle \hat{n}_i \rangle)^2 + \sqrt{\gamma} \Delta W_i (\hat{n}_i-\langle \hat{n}_i \rangle) \right] |\psitraj\rangle \nonumber, \\
    & \Delta W_i = \pm \sqrt{\delta t},\,\ \overline{\Delta W_i}=0,\,\ \overline{\Delta W_i \Delta W_j}=\delta t \delta_{ij}.
\end{align}
Here $\overline{(...)}$ describes the average over measurement-outcomes. This evolution is called \QSD (for the measurement of the occupation number) \cite{Gisin1992a}. Most importantly, the dynamics saturates once the state is a number eigenstate: $|\psitraj \rangle = |n_1,n_2\rangle$.

Weak/continuous measurements of this type, which alter the state only slightly per time step $\delta t$, can be realized by coupling the site $i$ to an ancilla via some Hamiltonian $H_{\text{anc}}$. We consider an ancilla qubit with basis $\{|0\rangle,|1\rangle\}$: Assume that initially the system is described by $|\psi \rangle$ and the ancilla by $|0\rangle$, such that the coupling leads to an entangled state:
\begin{align}
    |\psi \rangle \otimes |0\rangle \to \hat{A}_{0}^{(i)} |\psi \rangle \otimes |0\rangle + \hat{A}_{1}^{(i)} |\psi\rangle \otimes |1\rangle.
\end{align}
A projective measurement of the ancilla reproduces the weak measurement \emph{on the system} described before, \Eq{eq:WeakMeasurementsOutcome}, since the system is not directly measured.

A stochastic process results from repeated projective measurements of the ancillas, which after each measurement are reset in the state $|0\rangle$. An important consequence of the projective measurements is that it disentangles the ancilla from the system. In particular, any initial pure state of the system remains pure after ancilla measurements, allowing one to write \Eq{eq:QSDPureStates}.

The purity of the system state can be spoiled in several ways. This can happen, for instance due to \emph{imperfect} measurements, in which the ancillas themselves are either subject to only weak measurements or, with a certain probability, the ancilla is not measured at all \cite{Brun2002, Jacobs2006,Jacobs2014}. We consider the latter case: The measurement of the ancilla is then described by \cite{Brun2002}
\begin{align}
    &\hat{E}_0 = \eta |0 \rangle \langle 0|, \,\ \hat{E}_1 = \eta |1 \rangle \langle 1|, \,\ \hat{E}_2 = (1-\eta) \mathbb{1},
\end{align}
meaning that with probability $\eta$ (not to be confused with the measurement outcome probability) the ancilla is measured projectively and with probability $1-\eta$ it is not measured in this time step. Following the previous steps, we formulate a stochastic process for the conditional density matrix $\rhotraj$, conditioned onto the measurements outcomes (which are obtained with probability $\eta$): 
\begin{align}
    \rhotraj{}' - \rhotraj =& \sum_i -\frac{\gamma \delta t}{2} [\hat n_i,[\hat n_i,\rhotraj]] \\
     &+  \sqrt{\gamma} \Delta W_i \{ \hat{n}_i-\langle \hat{n}_i\rangle,\rhotraj \}, \label{eq:StochasticMasterEquation}\\
     \overline{\Delta W_i} = 0 &,\,\ \overline{\Delta W_i \Delta W_j} = \eta \delta t \delta_{ij}.
\end{align}
Note the occurrence of an extra factor of $\eta$ in the noise correlations here. For $\eta=1$, this equation is the density matrix formulation of \QSD. Any pure state remains pure and even an initially mixed state will purify under the evolution (for a more precise statement see, e.g., Ref.~\cite{Maassen2006}). In contrast to that, for $\eta<1$, the imperfect measurements will leave some residual entanglement between the system and the ancilla and the system state will become \emph{mixed}.

If none of the ancillas are measured, i.e., for $\eta=0$, the evolution equation reduces to the Lindblad master equation (adding a Hamiltonian for completeness) \cite{Jacobs2006}:
\begin{align}
    \partial_t \rhotraj = -i[\hat{H},\rhotraj] -\frac{\gamma}{2} \sum_i  [\hat{n}_i,[\hat{n}_i,\rhotraj]].
    \label{eq:LindbladEquation}
\end{align}
Due to the uncompensated build up of entanglement between the system and the ancillas, the state of the system evolves into a fully \emph{mixed} state, even though the joint state of system + ancillas remains pure. 

This latter situation of $\eta=0$ is equivalent to the decohering evolution of a system coupled to a dephasing bath. This yields two different physical interpretations of the time evolution including dephasing and measurements: we can either imagine that (i) dephasing is caused by an imperfect readout of the ancillas in a weak measurement setup or that (ii) the weak measurements are read out perfectly and the dephasing is caused by a separate set of ancillas, which couples to the system but is not read out at all. If not stated explicitly, we will in the following assume the latter case, and refer to the unmeasured ancillas simply as `the (dephasing) bath', for which we introduce the dephasing rate $\gamma_B$. In the measurement and bath setting, we always assume perfect continuous measurements with rate $\gamma_M$. Both scenarios (i) and (ii) bear the same theoretical description and are formally related to each other by following Tab.~\ref{tab:Conversion}.

\begin{table}
    \centering
    \begin{tabular}{c|c|c}
         & coupling to a bath &  imp. measurement\\ \colrule
    Lindblad prefactor &    $\gamma_M+\gamma_B$ & $\gamma$ \\ \colrule
    Noise strength & $\gamma_M$ & $\eta \gamma$ \\ \colrule
    Conversion & $\eta = \frac{\gamma_M}{\gamma_M+\gamma_B}$  
    \end{tabular}
    \caption{Relation of a dephasing bath and imperfect measurements. Any set $(\gamma_M,\gamma_B)$ for measurements in the presence of a bath corresponds to a set $(\eta,\gamma)$ for imperfect measurements.}
    \label{tab:Conversion}
\end{table}

To conclude this introduction, we briefly discuss suitable  `observables' and their averages in the presence of measurements. Each individual trajectory with a particular set of measurement outcomes $\{\moutcometime\}$ at sites $i$ at time $t$ yields a random state $\rhotraj_t$ after time $t$, which will, just like the measurement outcomes, differ from realization to realization. In order to make meaningful statements about the dynamics, we have to define a reproducible average over observables, which is independent of the set of measurement outcomes. This is similar to disordered systems, where one has to consider suitable averages over disorder realizations. When averaging over all possible measurement outcomes, only objects nonlinear in the state $\rhotraj$ reveal non-trivial information \cite{Skinner2019}. To see this, consider the expectation value of some operator $\hat{\mathcal{O}}$:
\begin{align}
    \overline{\langle \hat{\mathcal{O}} \rangle} = \text{tr}\left[ \overline{\rhotraj} \hat{\mathcal{O}}\right],
\end{align}
where the overbar in $\overline{\rhotraj}$ denotes that one has taken the average over all possible measurement outcomes. This is equivalent of taking the average over all possible measurement outcomes in each time step $\delta t$, which eliminates the randomness in \Eq{eq:StochasticMasterEquation} and yields  \Eq{eq:LindbladEquation} for the averaged evolution (due to $\overline{\Delta W}=0$). The stationary solution for  $\overline{\rhotraj_t}$ is then the totally mixed state $\overline{\rhotraj_{t \to \infty}} \propto \mathbb{1}$  under certain conditions \footnote{For conditions, see, e.g., Refs.~\cite{Spohn1976,Spohn1977,Nigro2019}.}. Obviously, this state reveals no information about the interplay of measurements and the Hamiltonian $\hat{H}$. 

Instead, one has to focus on higher moments of $\rhotraj$, which coincide with the nonlinear observables mentioned above. An example for a higher moment is
\begin{align}
\overline{\langle \hat{\mathcal{O}}\rangle \langle \hat{\mathcal{O}}\rangle} = \overline{\text{tr}[\hat{\mathcal{O}} \rhotraj]^2} \neq \text{tr}[\hat{\mathcal{O}} \overline{\rhotraj}]^2,
\end{align}
which has a similar appearance as the Edwards-Anderson order parameter \cite{Edwards1975,Castellani2005,Mezard1986,Bao2021,Li2021arxiv,Sang2021a} in the theory of spin glasses.

\subsection{Continuously measured fermions with decoherence - many-body formulation \label{sec:ManyBodyFormulation}}

Here we introduce the concrete fermionic model that we study in this work and discuss the qualitative picture of the particular limiting cases in the dynamics. We focus on (free) spinless fermion models in $(1+1)$ dimensions \cite{Cao2019,Chen2020,Jian2020arxiv,Alberton2021,Turkeshi2021,Turkeshi2021arxiv,Buchhold2021}, and consider a particle number conserving evolution on a lattice. The unitary part of the evolution is described by the nearest-neighbor hopping Hamiltonian:
\begin{align}
\hat{H} = J \sum_{i=1}^L c_i^\dagger c_{i+1} +c_{i+1}^\dagger c_i,
\end{align}
for $L$ lattice sites with $N=L/2$ fermions (half-filling) and periodic boundary conditions. The unitary dynamics competes with weak measurements of the local density $\hat{n}_i$ (with a strength $\gamma_M$), and a dephasing (Markovian) bath coupled to each site. The bath is described by Lindblad-operators $\hat{L}_i=\hat{n}_i$ (with strength $\gamma_B$) as in \Eq{eq:LindbladEquation}. Both the measurements and the dephasing result from coupling each lattice site to an ancilla system as discussed above. This yields the continuous, stochastic master equation
\begin{align}
d\rhotraj =& -i[ \hat{H},\rhotraj] dt - \frac{\gamma_B+\gamma_M}{2} \sum_{i=1}^L  [\hat{n}_i,[\hat{n}_i,\rhotraj]] dt \nonumber \\
&+ \sqrt{\gamma_M} \sum_{i=1}^L dW_i(t) \{ \hat{n}_i -\langle \hat{n}_i \rangle,\rhotraj \}.\label{eq:FullFermionicModel}
\end{align}
Here, $\overline{dW_i(t) dW_j(t')} = dt \delta_{ij} \delta(t-t')$ is the Gaussian measurement noise and $\overline{(...)}$ denotes the average over all possible measurement outcomes, which is equivalent to all possible noise realizations.

\begin{figure}
    \centering
    \includegraphics[width=\textwidth]{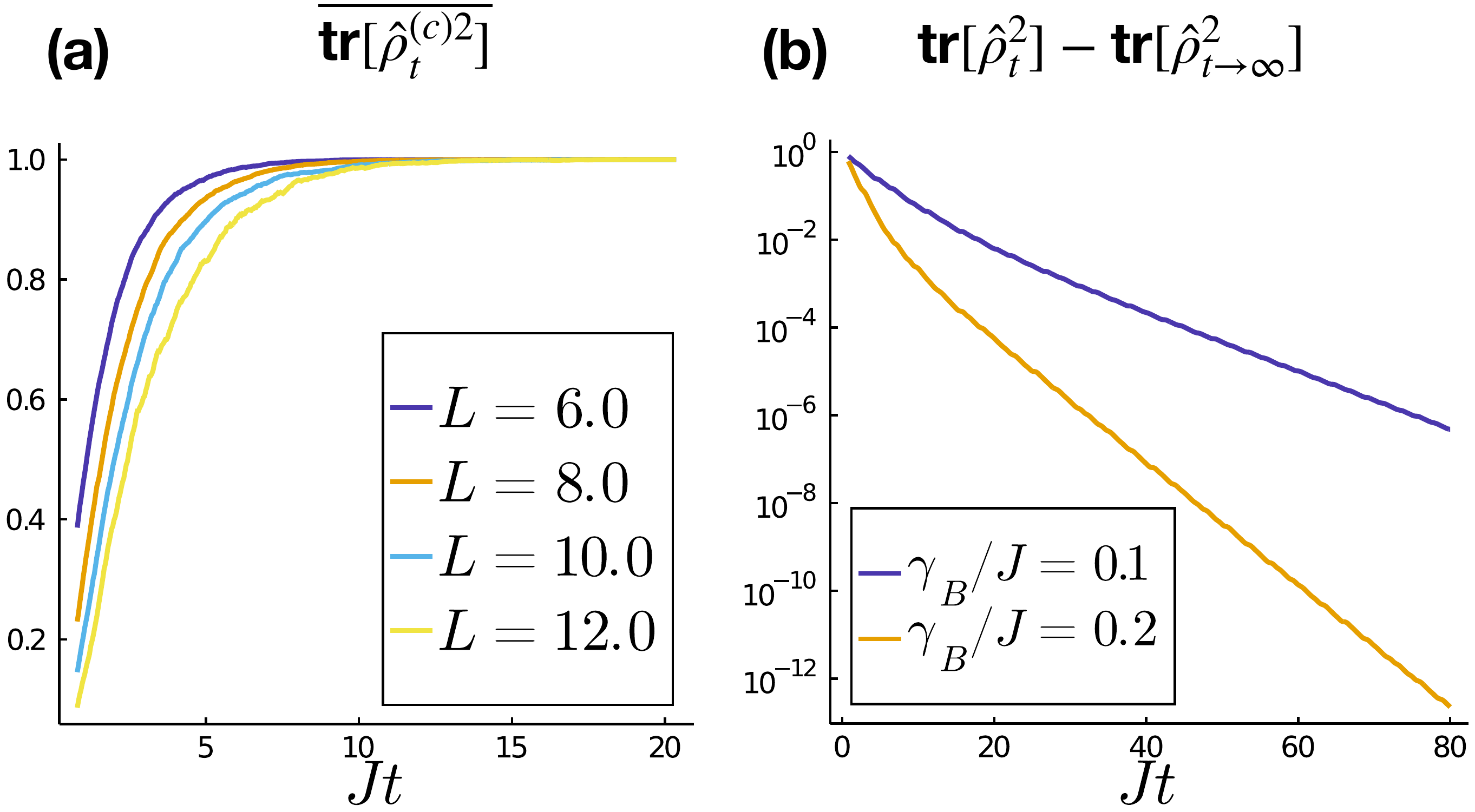}
    \caption{(a) Purification in the presence of measurements and a Hamiltonian ($\gamma_M/J=1$), starting from $\rhotraj_{t=0}\sim \mathbb{1}$; (b) mixing without measurements in the presence of a bath for $L=10$ (decay towards the (maximally low) purity of the infinite temperature state  $\text{tr}[\hat{\rho}_{t\to \infty}^2]$ with $\hat{\rho}_{t\to \infty} \sim \mathbb{1}$).}
    \label{fig:extreme_cases}
\end{figure}

\subsection{Qualitative picture of the dynamics}\label{sec:qualitative_Pic}
To set the stage for the later discussion, we will now individually discuss the three limiting cases: (1) $\gamma_B=0$; (2) $\gamma_M=0$ and (3) $J=0$ at a qualitative level.

\subsubsection{Measurements and Hamiltonian}
\begin{figure}
    \centering
    \includegraphics[width=\textwidth]{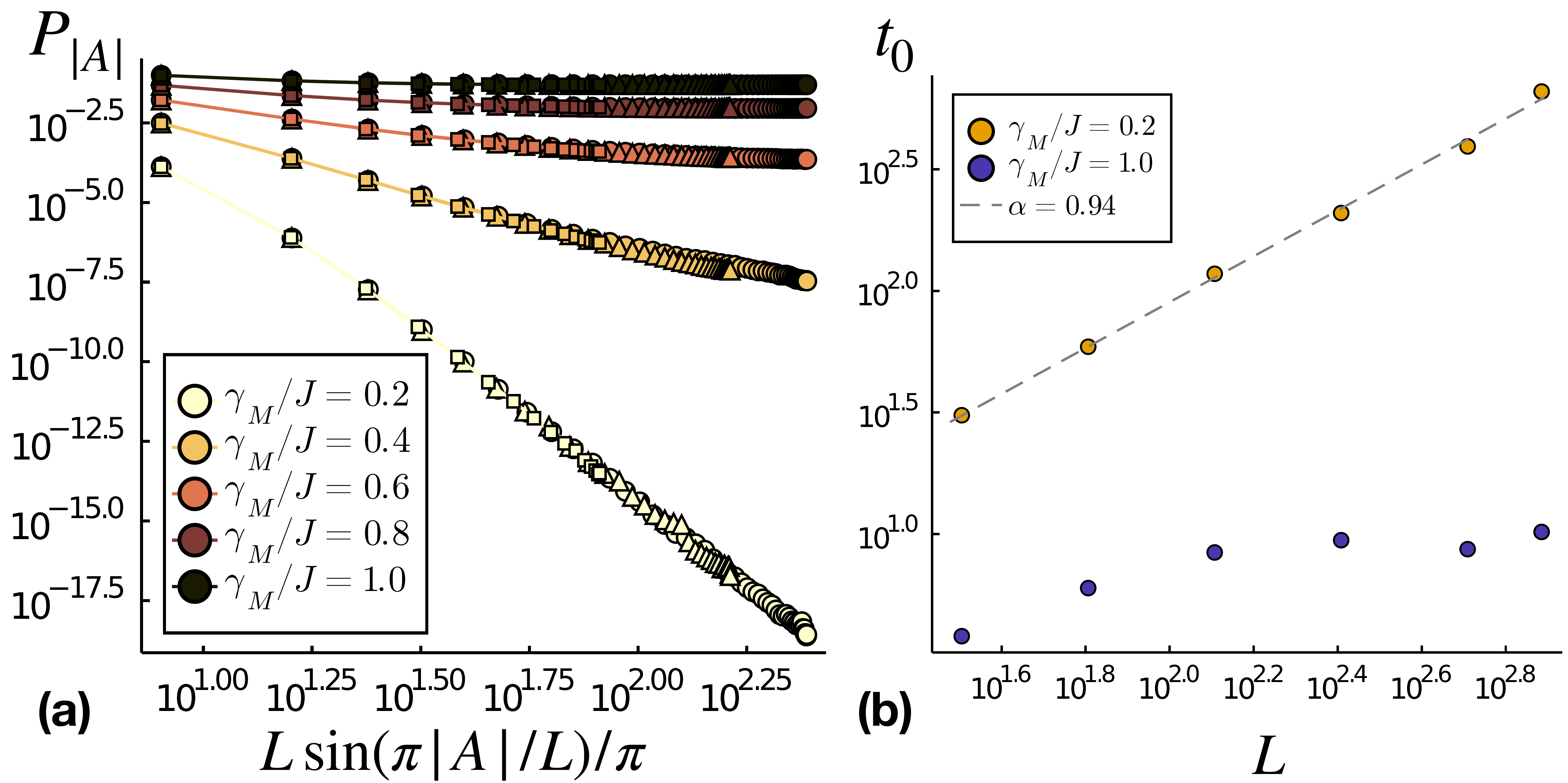}
    \caption{Overview of the system subject to measurements: (a) Subsystem size ($|A|$) resolved parity $\subparity$ for $L=256,512,768$ (square, triangle, circle) on a log-log scale, plotted for the rescaled length $L/\pi \sin(\pi |A|/L)$ (anticipating a conformal field theory). For $\gamma_M/J \ll 1$, the parity decays roughly algebraically and for $\gamma_M/J \sim 1$, the parity saturates, indicating a phase transition in between. (b) Scaling of the purification time scale $t_0$ with $L$ (log-log scale) of an ancilla coupled to two sites ($L=32,64,128,256,512,768$); $\alpha$ denotes the exponent of an algebraic best fit. For $\gamma_M/J=0.2$, $t_0(L)$ grows with $L$, whereas for $\gamma_M/J=1.0$ it saturates, giving a dynamical indicator of the two different phases.}
    \label{fig:OverviewMeasurements}
\end{figure}

The Hamiltonian tends to delocalize the fermions over the lattice, whereas the local measurements tend to `localize' the state into an occupation number eigenstate. Intuitively, this effect is shown in the distribution of the expectation values $\langle \hat n_i \rangle$: delocalization would correspond to $\langle \hat n_i \rangle \approx 1/2$ for each individual trajectory most of the time, and localization to $\langle \hat n_i \rangle \approx 0,1$. A quantitative measure of how close the state is to a number eigenstate is the local parity: $2\hat{n}_j -1$ or globally, the subsystem parity \cite{Bao2021,Li2021arxiv}:

\begin{align}
&\subparity(t) =\overline{\langle \prod_{j\in A} (2\hat{n}_j-1) \rangle^2} = \overline{\langle \prod_{j\in A} \exp(i\pi \hat{n}_j) \rangle^2},
\label{Eq:Parity}
\end{align}
where $A$ is a contiguous subset of the lattice of length $|A|$.
Similarly, we compute local density correlations \cite{Alberton2021}
\begin{align}
    &C_{ij}(t) = \overline{\langle \hat{n}_i \rangle \langle \hat{n}_j \rangle -  \langle \hat{n}_i \hat{n}_j \rangle }.
\label{Eq:CorrelatorFermions}
\end{align}
At large scales, the behaviour of $C_{ij}$ becomes qualitatively different depending on the phase: it either decays algebraically or exponentially, hallmark of the \BKT phase transition \cite{Alberton2021,Bao2021}, for an overview of \BKT-physics see, e.g., Ref.~\cite{Kogut1979}. Similarly, $\subparity$ decays algebraically with the subsystem size $|A|$ or becomes independent of $|A|$, see Fig.~\ref{fig:OverviewMeasurements}(a). Anticipating conformal invariance for weak measurements, we plot $\subparity$ as a function of $L/\pi \sin(\pi |A|/L)$ \cite{Calabrese2004,Chen2020,Alberton2021}.

Besides this stationary picture, also the dynamics can be used to distinguish the different phases. Here we focus on the bridging topic of \emph{purification} \cite{Gullans2020,Gullans2020a}. In the presence of any finite strength of measurements, an initially mixed state will purify over time \footnote{For a more precise statement and necessary conditions see, e.g., Ref.~\cite{Maassen2006}.}, see Fig.~\ref{fig:extreme_cases}(a). The system size dependence of the time it takes to approach the state can differ qualitatively in parameter space, indicating the presence of phase transitions. As was introduced in Refs.~\cite{Gullans2020,Gullans2020a} (see also Refs.~\cite{Li2021b,Block2021}), this time scale $t_0(L)$ will grow with the system-size if we are in the weakly measured phase. For large measurement rates, $t_0(L)$ will only weakly depend on the system size (see Sec.~\ref{Sec:DetailsPurificationTimeScale} for further details of the calculation and definition of $t_0(L)$).

A convenient approach to detect the purification dynamics is to couple a few sites of the system $(S)$ to a reference ancilla qubit $(R)$ with basis $\{ |0_R\rangle, |1_R\rangle \}$ and to initially prepare the system and ancilla in a fully entangled state \cite{Gullans2020a,Gullans2020}:
\begin{align}
\begin{aligned}
    & |\psi_t\rangle_{SR} = \sum_{\trajindex=0,1} \sqrt{p_\trajindex} |\psi^{(\trajindex)}_t\rangle |\trajindex_R\rangle, \,\ \hat{\rho}_{SR} = |\psi_t \rangle \langle \psi_t|,\\
    & |\psi_{t=0} \rangle_{SR} = \sqrt{\frac12} |\psi^{(0)} \rangle |0_R\rangle + \sqrt{\frac12} |\psi^{(1)} \rangle |1_R\rangle,
    \label{eq:RefAncillaConstruction}
    \end{aligned}
\end{align}
such that $\hat{\rho}_{R,t=0}=\text{tr}_S[\hat{\rho}_{SR}] = \frac12 \mathbb{1}_{2\times2}$ for $\langle \psi^{(0)} | \psi^{(1)} \rangle =0$ initially. The reduced density matrix of the system evolves according to \Eq{eq:FullFermionicModel} for $\gamma_B=0$. The purification of the system is then related to the entanglement between the system and the ancilla qubit. Tracking the entanglement for long times, the purification time scale $t_0$ can be extracted, see Fig.~\ref{fig:OverviewMeasurements}(b). Time scales for weak measurements ($\gamma_M/J=0.2$) and strong measurements ($\gamma_M/J=1.0$) are shown, exhibiting either a linear growth in $L$ or a saturation. The entanglement vanishes over time, because initially orthogonal states $|\psi^{(0)}\rangle$ and $|\psi^{(1)}\rangle$ start to overlap more and more and finally collapse onto each other, describing the very purification process. The growth of $t_0(L)$ with the system size for weak measurements supports the earlier findings of an extended critical phase, featuring logarithmic entanglement growth and algebraically decaying correlations \cite{Chen2020,Alberton2021,Bao2021,Turkeshi2021} (see also Refs.~\cite{Jian2020arxiv,Jian2021a}). The growth being nearly linear fits as well to a scale invariant (`conformal') phase \cite{Li2021b} with a dynamical critical exponent $z=1$.

\subsubsection{Bath and Hamiltonian}
In stark contrast, in the absence of any measurements but in the presence of a dephasing bath, any initial state will become fully mixed under the combined dynamics of unitary evolution and dephasing. This is a consequence of the Hamiltonian not commuting with the dephasing (local particle density) operators, which here gives rise to a unique stationary state.

As discussed above, from the perspective of measurements, `dephasing' is equivalent to not reading out any measurement outcomes, and thus to averaging over all possible outcomes in each step. Over time, this leads to a maximum uncertainty (or to a minimum of knowledge) on the underlying state, reflected in a maximum entropy, i.e., fully mixed, state. Even though each single measurement trajectory $|\psi^{(\trajindex)}\rangle$ may be in a pure state, the maximum uncertainty of the measurement outcomes 
\begin{align}
    \overline{\rhotraj} = \sum_\trajindex p_\trajindex |\psi^{(\trajindex)} \rangle \langle \psi^{(\trajindex)} | \stackrel{t \to \infty}{\rightarrow}  \sim\mathbb{1},
\end{align}
describes a maximally mixed state, see Fig.~\ref{fig:extreme_cases}(b). Here, $\mathbb{1}$ is the identity in the particle number conserving Hilbert space of $L/2$ fermions for $L$ lattice sites. (For this particular model exact results regarding the dynamics, e.g., its integrability, have been obtained, see, e.g., Ref.~\cite{Medvedyeva2016}.)

\subsubsection{Measurements and bath} Without a scrambling Hamiltonian, the dephasing bath will increase the mixedness of any initial state by projecting it onto the diagonal elements in the particle number basis. At the same time, the measurements, which do commute with the dephasing operators, increase the knowledge on the system by projecting the diagonal elements into states with well-defined local particle number. The asymptotic dynamics without Hamiltonian is therefore always \emph{purifying} \footnote{Given that measurement and bath operators commute and measurements are applied on each lattice site.}. If we consider a generic initial state in the occupation number basis
\begin{align}
    \rhotraj_{t=0} = \sum_{\trajindex,\trajindex'} \rho_{\trajindex \trajindex'} |\{n\}_\trajindex\rangle \langle \{n\}_{\trajindex'}|,
\end{align}
the off-diagonal elements will decay exponentially fast and $\rhotraj_t$ will evolve into an eigenstate of all the local particle number operators. The probabilities $p_\trajindex \equiv\rho_{\trajindex \trajindex}$ for each eigenstate are dictated by the diagonal elements of the initial state:
\begin{align}
    & \rhotraj_{t\to \infty} =  |\{n\}_\trajindex \rangle \langle \{n\}_\trajindex |\quad  \text{with prob. $p_\trajindex$}. \label{eq:MeasurementBathPurifiedState}
\end{align}

\subsubsection{Measurements, Bath and Hamiltonian}
This yields an overall picture of the dynamics, where the bath tends to eliminate the off-diagonal elements and to project the system onto its diagonal in the measurement/dephasing basis. The measurements then purify the diagonal by projecting it onto eigenstates with well-defined particle number. The Hamiltonian scrambles this information and encodes it in the off-diagonal elements of $\rhotraj$, where it is  either destroyed by the dephasing or recovered by the measurements. The interplay of all three mechanisms gives rise to a diverse phase diagram, discussed below.

\begin{figure*}
    \centering
    \includegraphics[width=\textwidth]{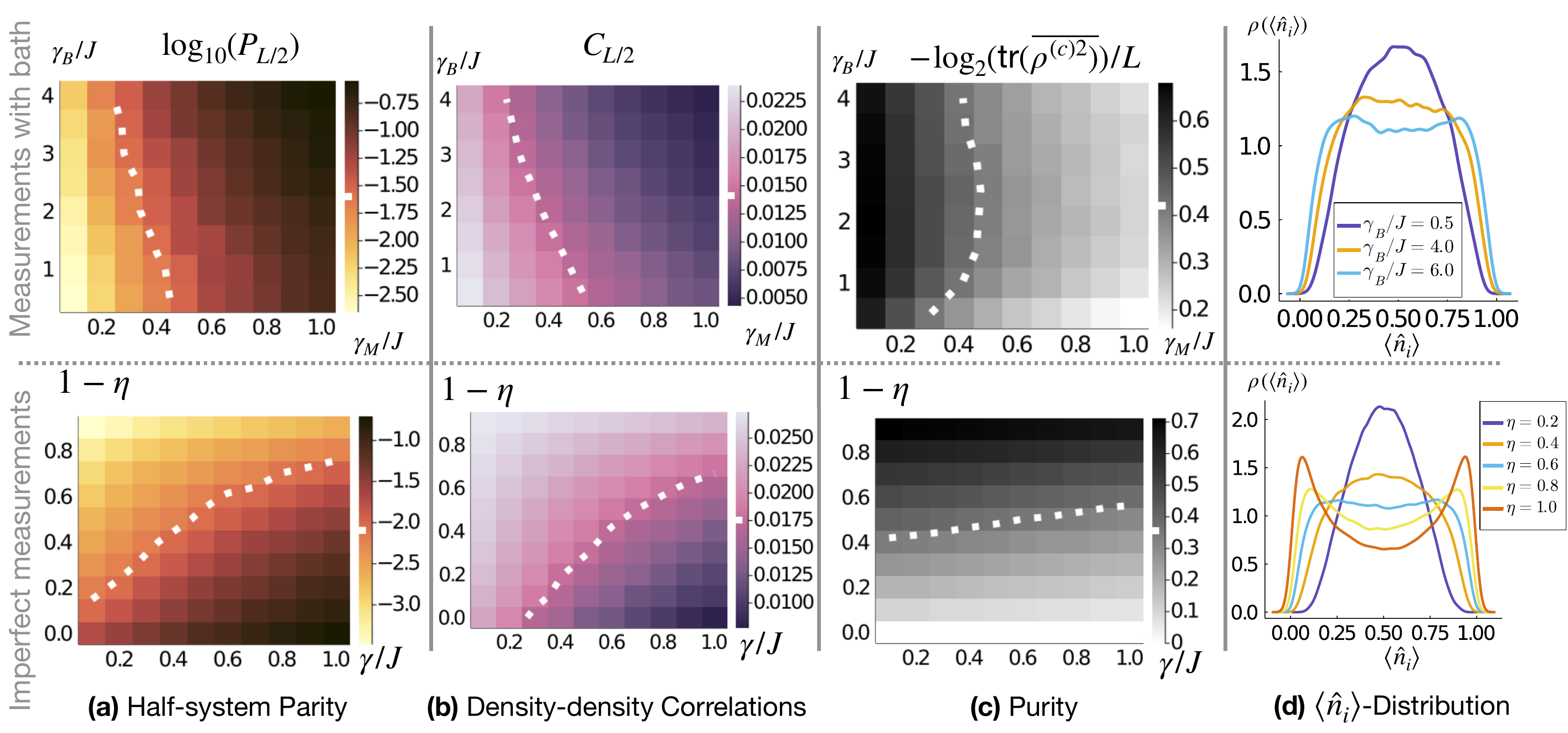}
    \caption{Overview of measured systems coupled to a bath (top) and the alternative view point of imperfect measurements (bottom) for $L=10$ (and $L/2$ fermions), the formal relation is given in Tab.~\ref{tab:Conversion}; dashed lines indicate contours as a visual add. (a) Log of the half-system parity $P_{|A|=L/2}$ in the $(\gamma_M/J,\gamma_B/J)$ and $(\gamma/J,1-\eta)$ plane respectively, indicating roughly two regimes of large and small parity. (b) Density-density correlations at distance $l=L/2$, giving rise to a similar bipartition. (c) Log of the average purity of the conditional density matrices, indicating a third regime with low purity but still larger correlations to the left(above) of the dashed line in the top(bottom) plot. (top) For large $\gamma_B/J$ the purity is again increasing (in accord with a measurement-induced phase, supported by strong dephasing). (d) Probability density $\rho(\langle \hat{n}_i\rangle)$ of the local expectation values $\langle \hat{n}_i \rangle$ (extracted from histograms), indicating a qualitative difference between the regimes ($\gamma_M/J=0.3$ (top) and $\gamma/J=1.0$ (bottom)).}
    \label{fig:OverviewSmallSystemsNew}
\end{figure*}

\subsection{Phase structure for small system sizes}
In the presence of dephasing or imperfect measurements, the state $\rhotraj$ of the system is mixed and no longer Gaussian (e.g., Wick's theorem is not applicable for $\gamma_B \neq 0$). Therefore it cannot be efficiently simulated by the numerical methods developed for free fermions in previous works~\cite{Cao2019, Alberton2021, Muller2021, Turkeshi2021}. Larger system sizes ($L=128-256$) are then only numerically accessible via quantum trajectory simulations of the fermion covariance matrix (see below). However, due to the non-Gaussianity of the state, the covariance matrix cannot be used to extract typical entanglement measures, such as, e.g., the entanglement negativity, the mutual information or the purity of the state. 

In order to gain some insight into the dynamics of the purity, we perform exact numerical simulations of \Eq{eq:FullFermionicModel} for small system sizes ($L\le10$ sites). We identify the three qualitatively different regimes (\phaseone, \phasetwo, \phasethree) discussed in the summary for $\gamma_B\ge0$ and $\gamma_M\ge0$. These regimes are anticipated in Fig.~\ref{fig:OverviewSmallSystemsNew} and characterized by:
\begin{enumerate}
\item[\phaseone:  ]{scale invariant, weakly mixed,}
\item[\phasetwo:]{scale invariant, strongly mixed,}
\item[\phasethree: ]{short correlation length and weakly mixed.}
\end{enumerate}
For such small systems, we can only distinguish strongly and weakly decaying correlations, but will not be able to identify, e.g., scale invariance. Anticipating the field-theoretic discussion, we nevertheless already label the regimes as 'scale invariant' or having a short correlation length.

\subsubsection{Competition (small systems) - Hamiltonian vs. measurement vs. bath}
From the limiting cases discussed in Sec.~\ref{sec:qualitative_Pic}, we can deduce the behaviour for competing measurements $\gamma_M$ and dephasing bath $\gamma_B$. \\
\underline{$\gamma_M>0,\gamma_B=0$ } --  This is the well-studied limit of continuous measurements, discussed in Refs.~\cite{Alberton2021, Buchhold2021,Minato2021,Muller2021, Turkeshi2021}. The measurement strength induces a transition (or crossover for small system sizes) from a scale invariant phase to a gapped and short-ranged correlated phase, which is indicated by, e.g., the subsystem parity $\subparity$ or the correlator $C_{ij}$. \\
\underline{$\gamma_M=0,\gamma_B>0$ } -- The state evolves towards the fully mixed state with exponentially small purity. 
\underline{$\gamma_M >0, \gamma_B >0$ } --  When all scales in the problem start to compete, the dynamics becomes more diverse and we show the results from the exact numerical simulations ($L=10$) in Fig.~\ref{fig:OverviewSmallSystemsNew} for measurements+bath (top) and imperfect measurements (bottom). Starting from $\gamma_B=0$, we observe a qualitative change in the parity and the correlations along the $\gamma_M$-axis. Notably, the regime of large half-system parity is increasing, once we increase the bath strength, see \Fig{fig:OverviewSmallSystemsNew}{(a)}. In a similar manner, the regime of longer-ranged (`scale invariant') correlations is decreasing, \Fig{fig:OverviewSmallSystemsNew}{(b)}. For $\gamma_B > \gamma_M$, we observe a significantly decreased purity due to the faster decay of the off-diagonal elements of $\rhotraj$, see \Fig{fig:OverviewSmallSystemsNew}{(c)}. 

When approaching the limit of strong dephasing, $\gamma_B \gg \gamma_M$, the situation is reversed again and we observe states with higher purity, enforced by strong dephasing and moderate measurements, see \Fig{fig:OverviewSmallSystemsNew}{(c)}. The behavior of the observables in this regime corresponds to the strong measurement phase \phasethree: In this limit, strong dephasing prevents the Hamiltonian from scrambling the information on the local particle numbers gained by previous measurements and traps the system into a quantum-Zeno regime, where even moderate measurements can lead to a purification of the diagonal of $\rhotraj$.

The simulations confirm a phase diagram consisting of three regimes: \phaseone longer-ranged correlations, weakly mixed; \phasetwo longer-ranged correlations, strongly mixed; \phasethree short-ranged correlated and weakly mixed. For small systems, there cannot be a sharp phase transition separating the different regimes, but we can already extract qualitative differences on short distances. To understand how this translates into potentially different phases in the thermodynamic limit, we complement this numerical analysis below with an analytical field theory approach and numerical simulations of the correlation matrix for larger systems.

\subsubsection{Competition (small systems) - Imperfect measurements}

The coupling to a dephasing bath and imperfect weak measurements are different sides of the same coin, where instead of $\gamma_M$ and $\gamma_B$ we use $\gamma$ and the imperfection rate $\eta \le 1$ (see Tab.~\ref{tab:Conversion}). We therefore extract a similar qualitative picture as above. \\
\underline{$\eta<1$ -- } Now regime \phasetwo roughly covers the upper part in the $(\gamma/J,1-\eta)$-plane, see \Fig{fig:OverviewSmallSystemsNew}{(a-c)} (bottom). Picking up the qualitative discussion from the beginning, the different regimes can be distinguished from the statistics of the local expectation values $\langle \hat n_i \rangle$. Starting from the strong measurement phase, increasing the imperfection rate, the distribution starts from a bimodal distribution and turns into a distribution closely centered around $\langle \hat n_i \rangle = 1/2$, see \Fig{fig:OverviewSmallSystemsNew}{(d)}.

\subsection{Long-wavelength theory: Replicas, Dirac fermions and Bosonization \label{Sec:OverviewLongWavelength}}
In order to confirm the picture developed for the dynamics of small systems, we develop and analyze an analytical approach, which is based on (i) a replica theory \cite{Mezard1986,Holzhey1994,Calabrese2004,Calabrese2009,Castellani2005}, and (ii) the mapping to a bosonic description in the continuum limit (see Refs.~\cite{Buchhold2021,Muller2021, Fradkin2013}). This allows us to study higher moments of the density matrix and to perform a \RG analysis of the different regimes. We show below that the three regimes discussed above correspond to three different quadratic models, which in turn can be viewed as different fixed points of the bosonized replica field theory action.

The analytical approach needs to account for the computation of higher moments in the stochastic variable $\rhotraj$. This can be done efficiently  by analyzing the replicated object $\rhotraj \otimes \rhotraj$, i.e., two \emph{identical} copies of the state in the replicated Hilbert space $\mathcal{H}^{(2R)} = \mathcal{H} \otimes \mathcal{H}$. In terms of the replicated state, the quadratic moments are (defining $\hat{P}_A = e^{i\pi \sum_{j\in A} \hat{n}_j}$):
\begin{align}
C_{ij} &= -\frac12 \text{tr}\left[ \left(\hat{n}_i^{(1)}-\hat{n}_i^{(2)} \right)\left(\hat{n}_j^{(1)}-\hat{n}_j^{(2)}\right) \overline{\rhotraj \otimes \rhotraj}\right], \\
 \subparity &= 1- \frac12 \text{tr}\left[(\hat{P}_A^{(1)}-\hat{P}_A^{(2)})^2 \,\ \overline{\rhotraj\otimes \rhotraj}\right] \nonumber \\
 &= \text{tr}\left[ e^{i\pi \sum_{j\in A} (\hat{n}_j^{(1)}-\hat{n}_j^{(2)})}\overline{\rhotraj\otimes \rhotraj }\right].
\label{eq:OperatorsReplicaSpace}
\end{align}
Here $\hat{\mathcal{O}}^{(1)} = \hat{\mathcal{O}} \otimes \mathbb{1}, \hat{\mathcal{O}}^{(2)} = \mathbb{1} \otimes \hat{\mathcal{O}} $ for any operator $\hat{\mathcal{O}}$ defined on the single replica Hilbert space. Both observables are determined from the \emph{deterministic} evolution of the average $\rhotworeplica = \overline{\rhotraj \otimes \rhotraj}$.

We follow the approach outlined in Refs.~\cite{Buchhold2021,Muller2021}, where the universal long-wavelength dynamics of monitored fermions was analyzed in terms of a continuum field theory. This directly enables the powerful tool of bosonization, which in turn has the major advantage that the properties of the different (thermodynamic) phases are encoded in a \emph{quadratic} theory, serving as an effective description for \RG fixed points. In terms of Dirac fermions, the free Hamiltonian is
\begin{align}
\hat{H} = i \nu \int_x \hat{\Psi}_x^\dagger \sigma_z \partial_x \hat{\Psi}_x,
\end{align}
with $\hat\Psi = (\hat\Psi_R,\hat\Psi_L)^T$ being the common left- and right-moving fermion operators \cite{Fradkin2013}. The measured density operators (both measured with the same strength $\gamma_M$) are
\begin{align}
&\hat{\mathcal{O}}_{1,x} = \hat{\Psi}^\dagger_x \hat{\Psi}_x, \\
& \hat{\mathcal{O}}_{2,x} = \hat{\Psi}_x^\dagger \sigma_x \hat\Psi_x.
\end{align}
This formulation allows for an equivalent description in terms of bosons, described by the operators \cite{Buchhold2021}
\begin{align}
&\hat{H} = \frac{\nu}{2\pi} \int_x \left[ \left(\partial_x \hat{\theta}_x\right)^2 + \left( \partial_x \hat{\phi}_x\right)^2 \right] ,\\
& \hat{\mathcal{O}}_{1,x} = -\frac{1}{\pi} \partial_x \hat{\phi}_x, \\
&\hat{\mathcal{O}}_{2,x} = m \cos \left(2\hat{\phi}_x \right),\label{eq:36}
\end{align}
where $m$ is a regularization dependent constant and the operators fulfill: $[\partial_x \hat{\theta}_x,\hat{\phi}_{y}]=-i \pi \delta(x-y)$.

In this setting, $\hat{H}$ and $\hat{\mathcal{O}}_{1,x}$ constitute a quadratic theory, which is exactly solvable and will be the ground for further treatments of the nonlinearities, Eq. \eqref{eq:36}.

At this level, $\rhotworeplica$ can be decomposed into a product state in the `absolute/relative'-basis: 
\begin{align}
\hat{\phi}_x^{(a)} := \frac{\hat{\phi}_x^{(1)}+\hat{\phi}_x^{(2)}}{\sqrt{2}}, && \hat{\phi}_x^{(r)} := \frac{\hat{\phi}_x^{(1)} - \hat{\phi}_x^{(2)}}{\sqrt{2}}
\label{Eq:DefRelativeAbsoluteModes}
\end{align}
(discussed in detail in the Appendix~\ref{App:DetailsRGAnalysis}), such that $\rhotworeplica = \hat{\rho}^{(a)} \otimes \hat{\rho}^{(r)}$, where $\hat{\rho}^{(a)}$ heats up indefinitely, while $\hat{\rho}^{(r)}$ encodes the non-trivial correlations. To see this, consider $C_{ij}$ and $\subparity$: in both expressions only the combination $\hat{n}_i^{(1)}-\hat{n}_i^{(2)}$ appears. Slightly simplified, we can identify 
\begin{align}
\hat{n}_i^{(1)}-\hat{n}_i^{(2)} \stackrel{\sim}{\to} -\frac{1}{\pi} \partial_x \left( \hat{\phi}_x^{(1)}-\hat{\phi}_x^{(2)}\right) = - \frac{\sqrt{2}}{\pi}\partial_x \hat{\phi}^{(r)}_x ,
\end{align}
where in the last part, we have used \Eq{Eq:DefRelativeAbsoluteModes}. Most importantly, these observables only depend on the relative operators. Furthermore, we are able to write down a path integral and effective action for the relative modes only \cite{Buchhold2021}. The replica field theory for the relative coordinate is derived in two major steps: (1) performing a coordinate transformation to absolute and relative replica coordinates, \Eq{Eq:DefRelativeAbsoluteModes}; (2) introducing  Keldysh coordinates and integrating out the absolute modes (as detailed in the Appendix~\ref{App:DetailsRGAnalysis}). Observables, which only depend on the relative coordinate, $\hat{\mathcal{O}}^{(r)}$, can therefore be calculated from this path integral ($X:=(t,x)$):
\begin{align}
\begin{aligned}
    &\text{tr}[\hat{\mathcal{O}}^{(r)} \, \overline{\rhotraj \otimes \rhotraj}] \approx  \\
    &\int \mathcal{D}[\phi_c^{(r)},\phi_q^{(r)}] \, \mathcal{O}^{(r)} \, e^{i \int_X \frac{1}{2}\bosonmodes_X^T G_0^{-1}\bosonmodes_X + i\relativeInteractions},
    \end{aligned}
\end{align}
with $\bosonmodes_X^T = (\phi_{c,X}^{(r)},\phi_{q,X}^{(r)})$. After rescaling $t \to \nu t$, the inverse propagator $G_0^{-1}$ reads:

\begin{align}
G_0^{-1}= -\frac{1}{\pi}\begin{pmatrix} i \frac{1}{\pi} 2 \frac{\gamma_M}{\nu} \partial_x^2 & \partial_t^2 - \partial_x^2 \\ \partial_t^2 - \partial_x^2 & i\frac{1}{\pi}2 \frac{(\gamma_M+\gamma_B)}{\nu} \partial_x^2 \end{pmatrix}.
\label{eq:RelativeModeGreensFunction}
\end{align}

Most importantly, the quadratic part of the action is scale invariant, the property which determines the phase of weak measurements, e.g., the linear growth of the purification time (with system size). At second order in the nonlinearities, also additional derivative terms are generated under \RG transformations in $G_0^{-1}$ (see Appendix \Eq{eq:GeneralDefinitionInversePropagator} for the more general form). The interaction part takes the sine-Gordon form (where we suppress the index $(r)$ in the fields for now):
\begin{align}
&\relativeInteractions  = \int d^2 X \left[i \lambda_c \cos(4\phi_{c,X}) + i \lambda_q \cos(4\phi_{q,X}) \right. \label{eq:RelativeInteractions}\\
&\left. + i \lambda_{cq}^{(c)} \cos(2\phi_{c,X})\cos(2\phi_{q,X})+  \lambda_{cq}^{(s)} \sin(2\phi_{c,X})\sin(2\phi_{q,X})\right], \nonumber
\end{align}
featuring four different, real valued, interactions described by $\lambda_{cq}^{(c)},\lambda_{cq}^{(s)},\lambda_c,\lambda_q$.
In the next section, we will analyze in more detail, which of these interactions can turn relevant, at large scales, dominating the physics. For now it is sufficient to distinguish three possible, different scenarios: 
\begin{enumerate}
\item[\phaseone:  ]
 no interaction is relevant; \\
\item[\phasetwo:  ] $\lambda_q$ is relevant \footnote{At first order in the \RG, $\lambda_c$ will never turn relevant before any of the other couplings as we will see later.}, or\\ \item[\phasethree:  ] $\lambda_{cq}$'s are relevant. 
\end{enumerate}
These three cases lead to an effective, \emph{quadratic} model with a modified inverse propagator $G^{-1}$:
\begin{align}
    &\text{\phaseone :} && G^{-1} = G_0^{-1}, \\
    &\text{\phasetwo :} && G^{-1} = G_0^{-1}+ i m_D(\mathbb{1}-\sigma_z),  \label{eq:EffectiveT}\\
    &\text{\phasethree :} && G^{-1} = G_0^{-1} +m_{M_r}\sigma_x + i m_{M_i} \mathbb{1}.
\end{align}
Here, $m_{M_{r,i}}$ are measurement-induced parts of a complex mass term $m_M= m_{M_{r}} + i m_{M_{i}}$,  and $m_D$ is a dephasing-induced real valued mass.

The observables in this effective description are determined by quadratic correlators \footnote{We assume that the equal time correlation function $\langle \hat{\phi}_X \hat{\phi}_Y \rangle$ is determined by the dominant contribution of $\langle \phi_{c,X}\phi_{c,X}\rangle$ or $\langle \phi_{q,X} \phi_{q,Y}\rangle$ .},
\begin{align}
    &C_y \approx -\frac{1}{\pi^2} \langle \partial_x\phi_x^{(r)} \partial_x\phi_{x+y}^{(r)} \rangle, \\
    &\subparity\approx \langle e^{i\sqrt{2} (\phi_0^{(r)}-\phi_{|A|}^{(r)})}\rangle \approx e^{- \langle (\phi_0^{(r)}-\phi_{|A|}^{(r)})^2\rangle},
\end{align}
which can be directly evaluated using $iG(x,y)=\langle\phi^{(r)}_x\phi^{(r)}_y\rangle$. The equal-time correlators in real space are

\begin{align}
    \langle\phi^{(r)}_0\phi^{(r)}_y\rangle \sim \begin{cases} &\log(|y|) \qquad\qquad \text{in \phaseone}, \\
    & \log(|y|)\qquad \qquad \text{in \phasetwo}, \\
    &\exp(-|y|/\xi)\quad \quad \text{in \phasethree},
    \end{cases}
\end{align}
and the corresponding observables are shown in Tab.~\ref{tab:ReplicaObservables}.

\begin{table}[t]
    \centering
    \begin{tabular}{l|c|c}
    \toprule
         & $C_y$ & $\subparity$ \\ \colrule
       case \phaseone,\phasetwo  & $\sim |y|^{-2}$ & $|A|^{-K}$\\ \colrule
       case \phasethree & $\sim \exp(-|y|/\xi)$ & const. \\ \botrule
    \end{tabular}
    \caption{Overview of the subsystem parity and density-density correlations in the different regimes. $K$ is a real valued exponent, depending on the details of the propagator $G$.}
    \label{tab:ReplicaObservables}
\end{table}

In the absence of interactions, correlations and subsystem parity are algebraic functions of the distance, reflecting the scale invariant nature of the theory \phaseone. As expected, the measurement-induced mass(es) $m_M$ leads to short-ranged density-correlations for \phasethree. Surprisingly though, these `observables' are insensitive to the presence of the dephasing-induced scale $m_D$. There is an important difference between \phaseone and \phasetwo: the dephasing-induced mass $m_D$ enters the propagator, \Eq{eq:EffectiveT}, in the same way as an effective temperature scale ($qq$-sector) in a single-replica Keldysh theory~\cite{Sieberer2016a, Kamenev2011}. In a single-replica, equilibrium framework, this scale would impose all correlations to be exponentially suppressed in time and in space. However, in the replica and measurement framework here, nonzero entries in the $cc$-sector of Eqs.~\eq{eq:RelativeModeGreensFunction},\eq{eq:GeneralDefinitionInversePropagator} prevent correlations from becoming exponentially localized (in equilbrium, a non-zero $cc$-entry is ruled out by causality). The effect of a non-zero $m_D$ in the measurement setup is an exponential suppression of fluctuations of the 'quantum field' $\phi_q$. The quantum field has a direct correspondence to the off-diagonal entries of the density matrix~\cite{Sieberer2016a, Kamenev2011} and a non-zero $m_D$ therefore also implies an exponential decay of the off-diagonal elements of $\rhotworeplica$. This allows for the following interpretation: the emergence of a non-zero scale $m_D$ indicates a strong confinement of the dynamics onto the diagonal of the density matrix, due to a dominant impact of dephasing over the coherent scrambling. This behavior is very similar to the \emph{thermalization} in generic Hamiltonian systems \cite{Deutsch1991,Srednicki1994}, where the density matrix is effectively projected onto its diagonal in the energy eigenbasis. In contrast to thermalization, the structure on the diagonal depends on the strength of the measurements. If the measurements dominate, the density matrix is close to a pure state, corresponding to $m_{M_r}, m_{M_i}>0$. If the measurements are not dominant, diffusion of particles leads to a homogenization of the diagonal and to a diagonal ensemble with very high mixedness. Thus $m_D>0$ and $m_{M_r}, m_{M_i}=0$ corresponds to the mixed, but scale invariant phase $\phasetwo$.

\section{Results: RG analysis and quantum trajectory simulations}
In the following, we analyze the phase structure of the fermion model at large distances. To do so, (i) we use a \RG calculation of the boson model to track the competition of the different nonlinearities with the free, scale invariant part, and (ii) we simulate the dynamics of the fermionic density matrix numerically by using quantum trajectories for an ensemble of (Gaussian) states. Our main findings are:
\begin{itemize}
    \item In a first order perturbative \RG approach, we identify four different scenarios: (1) all interaction couplings are irrelevant (corresponding to phase \phaseone), (2) only $\lambda_q$ is relevant (phase \phasetwo), (3) only $\lambda_{cq}$'s are relevant (phase \phasethree), and (4) multiple interaction couplings are relevant (phase \phasetwo or \phasethree, depending on the dominant divergence). Including second order \RG equations, i.e., the renormalization of the kinetic coefficients, a robust and extended, scale invariant phase \phaseone with vanishing interaction strengths can still be identified and fits well to the numerical results for phase \phaseone. Outside of this phase, several coefficients of the nonlinear contributions diverge, which is generic for sine-Gordon models. The different phases can then be estimated by identifying the most strongly diverging interaction coupling. It corresponds to the shortest and therefore most dominant length- or time scale.
    \item{The numerical quantum trajectory simulations reveal a regime of algebraically ($\sim 1/l^2$ where $l$ is the distance) decaying correlations, corresponding to either phase \phaseone or phase \phasetwo. In addition, we identify a regime of strongly suppressed correlations, qualitatively matching exponential decay in space as expected from the field theory for phase \phasethree. The distinction between the different regimes is further confirmed by a qualitative change in the distribution function of the local averages $\langle \hat n_i \rangle$, though the numerically accessible system sizes are too small to observe a sharp transition.}
    \item{These findings provide strong evidence for the robustness of the critical phase at weak measurement, including in the presence of dephasing or imperfect measurements.}
\end{itemize}

\subsection{Construction replica-field theory and RG analysis \label{sec:ReplicaFieldTheory}}
To set the stage for the detailed \RG analysis, we briefly discuss a set of symmetries, which distinguish between the presence or absence of a bath.

The path integral is constructed from the two-replica master equation according to the conventional Keldysh path integral technique~\cite{Sieberer2016a, Kamenev2011}. It contains four different fields $\{\phi_\pm^{(1)},\phi_\pm^{(2)}\}$: one field per each replica and per each Keldysh contour. In the presence of both measurements and dephasing, the action is invariant under exchanging labels on \emph{all} contours simultaneously: 
\begin{align}
&\phi_+^{(1)} \leftrightarrow \phi_+^{(2)} \text{ and }\phi_-^{(1)} \leftrightarrow \phi_-^{(2)}.
\end{align}
In the \emph{absence} of dephasing, there is an additional symmetry: the action is invariant even under exchanging labels only on a \emph{single} contour:
\begin{align}
&\phi_+^{(1)} \leftrightarrow \phi_+^{(2)} \text{ or }\phi_-^{(1)} \leftrightarrow \phi_-^{(2)}.\label{eq:sym2}
\end{align}
The second symmetry, Eq.~\eqref{eq:sym2}, forbids the generation of a temperature scale, i.e., it pins $m_D=0$. This implies that phase \phasetwo is excluded by symmetry in the absence of dephasing, a fact that is reflected in the \RG equations. In terms of the relative and absolute modes (and in Keldysh coordinates) with the bare propagator \Eq{eq:RelativeModeGreensFunction}, the additional coupling to a bath, $\gamma_B \neq 0$ only enters the last entry in \Eq{eq:RelativeModeGreensFunction}. At this level, the additional symmetry for $\gamma_B=0$ is given by $\phi_q^{(r)} \leftrightarrow \phi_c^{(r)}$ (the still remaining symmetry for $\gamma_B\neq 0$ is $\{\phi_c,\phi_q\} \to \{-\phi_c,-\phi_q\}$).

\subsubsection{First-order RG analysis \label{Sec:FirstOrderRG}}

To study the fate of the different interaction terms in $\Delta S_r$ at large length scales, we use a perturbative momentum shell \RG scheme, see, e.g., Refs.~\cite{Kogut1979,Buchhold2021} and further details in Appendix~\ref{App:FirstOrderRG}. We show that for $\gamma_B=0$, only one kind of interaction can turn relevant, but for $\gamma_B \neq 0$ a competition between different interactions becomes possible. A summary of the relevant terms, depending on the parameters in the $(\gamma_M/\nu,\gamma_B/\nu)$-plane, is shown in \Fig{fig:RG_phasediagram}{(c)}.

For the \RG scheme, momentum modes in a shell below the ultraviolet cutoff $\Lambda$, $k \in [\Lambda/b,\Lambda]$ for $b>1$, are integrated out. To this end, we separate the fields $\phi(X) = \phi^>(X) + \phi^<(X)$ into long-distance ($>$) and short-distance modes ($<$), and integrate out $\phi^<(X)$ (where $X=(t,x)$) \cite{Kogut1979,Buchhold2021}. Since we are interested in the stability of the 'conformal' phase, we use an appropriate symmetric rescaling of space and time (dynamical critical exponent $z=1$):
\begin{align}
\begin{aligned}
x \to \tilde{x}= x/b, && k \to \tilde{k}=  kb, \\
t \to \tilde{t}=t/b, && \omega \to \tilde{\omega} = \omega b.
\end{aligned}
\end{align}
The fields $\phi_c,\phi_q$ are dimensionless (at the Gaussian fixed point) and will not be rescaled. Combining the renormalization and the rescaling, we get the flow equations (using $b=e^s$) \footnote{Note that $\alpha_{cc},\alpha_{qq}$ and $\alpha_{cq}$ are real, therefore $\lambda_{cq}^{(s)}$ is generated as a real coupling and $\lambda_{cq}^{(c)}$ and $\lambda_{cq}^{(s)}$ stay purely real during the flow.} 
\begin{align}
&\partial_s \lambda_c \approx \left(2-\frac{8}{\pi}\alpha_{cc} \right) \lambda_c, \\
&\partial_s \lambda_q \approx \left(2-\frac{8}{\pi}\alpha_{qq} \right) \lambda_q, \\
& \partial_s \lambda_{cq}^{(c)} \approx \left(2-\frac{2}{\pi}(\alpha_{cc}+\alpha_{qq})\right) \lambda_{cq}^{(c)} - \frac{4}{\pi}\alpha_{cq} \lambda_{cq}^{(s)}, \\
& \partial_s \lambda_{cq}^{(s)} \approx \left(2-\frac{2}{\pi}(\alpha_{cc}+\alpha_{cq})\right) \lambda_{cq}^{(s)} +  \frac{4}{\pi} \alpha_{cq} \lambda_{cq}^{(c)}.
\end{align}
The last two flow equations can be decoupled, introducing complex parameters $\lambda_\pm = (\lambda_{cq}^{(c)} \pm i \lambda_{cq}^{(s)})/2$:
\begin{align}
&\partial_s \lambda_+ \approx \left(2-\frac{2}{\pi}(\alpha_{cc}+\alpha_{qq}) +i \frac{4}{\pi}\alpha_{cq}\right)\lambda_+, \\
&\partial_s \lambda_- \approx \left(2-\frac{2}{\pi}(\alpha_{cc}+\alpha_{qq}) -i \frac{4}{\pi}\alpha_{cq}\right)\lambda_- .
\end{align}
They are then seen to follow the typical form of \BKT flow equations at first order \cite{Kogut1979,Fradkin2013}, describing a threshold phenomenon: only once the prefactor is positive, the operators turn relevant. The coefficients $\alpha_{ab}$ in the prefactors are directly related to the equal-time correlators in momentum-space (see again \Eq{eq:RelativeModeGreensFunction}):
\begin{align}
\langle \phi_a(0,k),\phi_b(0,-k)\rangle = \chi_{ab}\frac{\alpha_{ab}}{|k|}, && \chi_{ab} = \begin{cases} 1 & a=b \\ -i & a\neq b \end{cases}.
\end{align}
In particular, we find
\begin{align}
\left(\begin{array}{c}\alpha_{cc}\\ \alpha_{qq}\\ \alpha_{cq}\end{array}\right) = \left(\begin{array}{c}\frac{\gamma_M + \gamma_B}{\nu}\\ \frac{\gamma_M}{\nu}\\  \frac{\pi}{2} (1 -|z|^2) \end{array}\right) \cdot f(z),
\end{align}
where we have defined:
\begin{align}
\begin{aligned}
&z^2 = 1 + i \frac{2}{\pi} \sqrt{\frac{\gamma_M(\gamma_M+\gamma_B)}{\nu^2}},\\
&f(z) = i\, \frac{1}{|z|^2( z-z^*)}.
\end{aligned}
\end{align}
In the absence of a bath ($\gamma_B=0$), we have the additional symmetry $\phi_c \leftrightarrow \phi_q$, which implies $\alpha_{cc}=\alpha_{qq}$ and therefore the $\lambda_{cq}$'s are always more relevant than $\lambda_c,\lambda_q$. This gives rise to either phase \phaseone, where no interaction is relevant, or phase \phasethree, where measurements induce an effective mass $m_M$.

For nonzero $\gamma_B$, the symmetry Eq.~\eqref{eq:sym2} is no longer present and $\lambda_q$ can become more relevant than $\lambda_{cq}$ as soon as $\gamma_B \ge 2\gamma_M$. This gives rise to the phase \phasetwo. The corresponding regimes in the $(\gamma_M/\nu,\gamma_B/\nu)$-plane, where the individual couplings become relevant, are shown in \Fig{fig:RG_phasediagram}{(c)}. In the limit where both $\gamma_B=0$ and $\gamma_M/\nu \to 0$, the couplings $\lambda_{cq}$ become marginal (in agreement with the limit of free fermions at half filling \cite{Fradkin2013}). Nevertheless, the interaction couplings will become less relevant at second order (for small $\gamma_M/\nu$), similar to a sine-Gordon model with \emph{imaginary} couplings \cite{Fendley1993} (see also Ref.~\cite{Buchhold2021} for more details).

\subsubsection{Second-order RG analysis \label{Sec:SecondOrder}}

The first order \RG equations offer three different regimes, where either \phaseone no interaction is relevant, \phasethree (amongst others) a $\cos(2\phi_c)\cos(2\phi_q)$-term is relevant or \phasetwo only $\cos(4\phi_q)$ is relevant. 
While this explains the origin of the three different phases, in terms of different couplings, it yields a premature estimate of the actual phase boundaries, especially in the presence of real and imaginary couplings \cite{Fendley1993,Buchhold2021}. For instance, without a bath, the Gaussian fixed point is always unstable according to the first order equations. In order to obtain an improved estimate for the phase boundaries, we consider a second order \RG approach, for which the derivative terms are renormalized. Then the scale invariant Gaussian phase is stable in an extended parameter regime of small but non-zero measurement strengths. At second order, one needs to track the \RG flow of $10$ couplings: $4$ interaction couplings and $6$ derivative terms in the quadratic sector, see \Eq{eq:GeneralDefinitionInversePropagator} \footnote{Compared to (essentially) $2$ complex ones without a bath, see Appendix~\ref{App:RecoveringSymmetricCase} and Ref.~\cite{Buchhold2021}.}. The full set of flow equations is rather involved and is not discussed here. We refer to Appendix~\ref{App:DiscussionFlowEquations}, and \Eq{eq:FullSetFlowEquations}.

\begin{figure}
    \centering
    \includegraphics[width=0.95\textwidth]{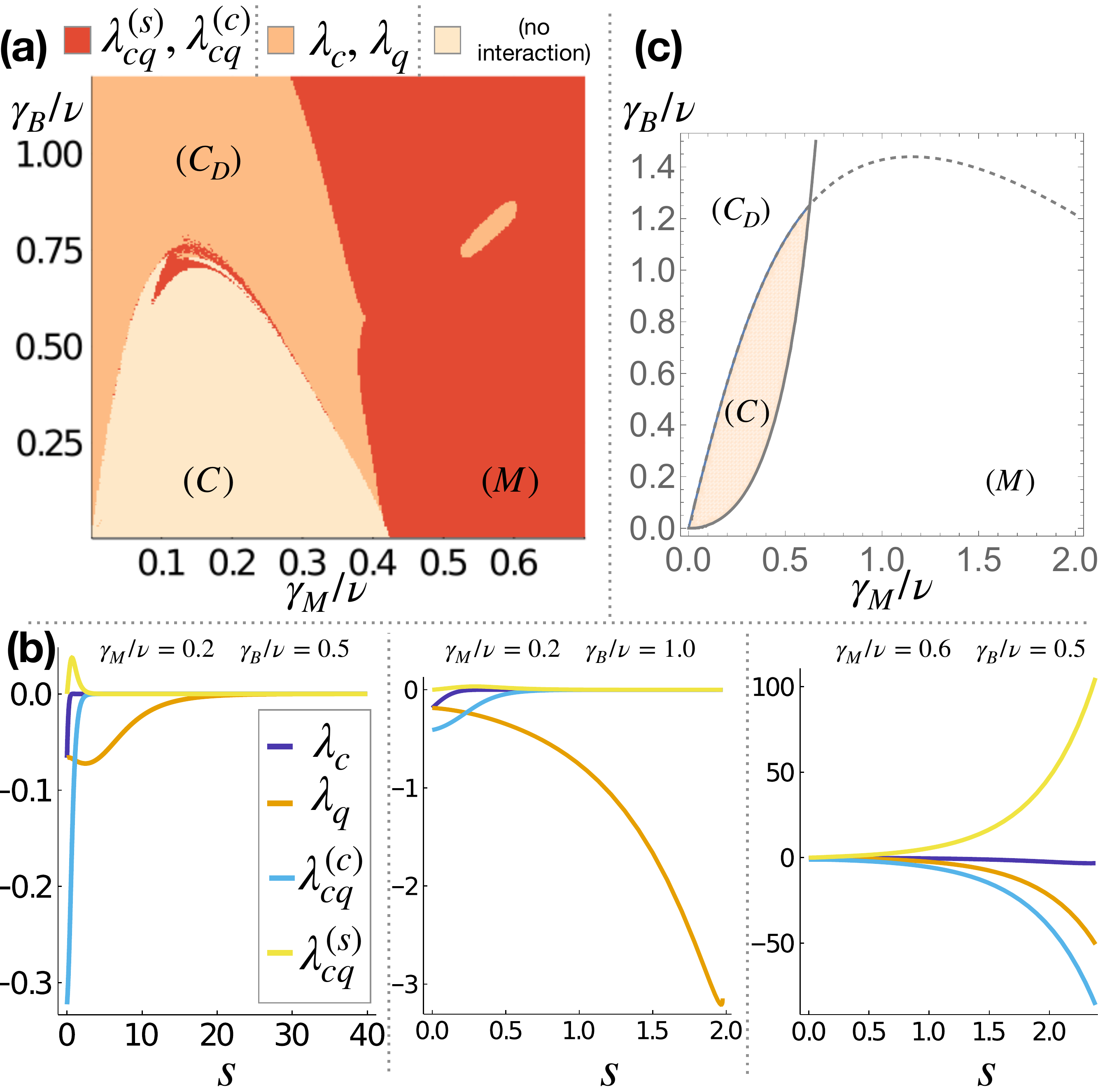}
    \caption{(a) Tentative phase diagram from the bosonic theory at second order (including derivative corrections). We identify a stable regime \phaseone (light orange), where all interactions vanish. Beyond this regime, interactions diverge and/or the flow breaks down at $s=s_f$ (see text). To get a rough estimate of the phase structure beyond \phaseone, we indicate the most strongly growing interaction strength at $s_f$, corresponding to $\phasetwo$, if either $\lambda_q$ (or $\lambda_c$) is growing, or $\phasethree$, if $\lambda_{cq}^{(c)}$ or $\lambda_{cq}^{(s)}$ grows the strongest. In the region between \phaseone and \phasetwo, $\lambda_{cq}$'s display in parts an oscillatory behaviour (red island). (b) Resolved flows of the interactions $\lambda$ (in units of $\normalorderingmass^2$) are shown for the pairs $(\gamma_M/\nu,\gamma_B/\nu)$: $(0.2,0.5)$ (left, \phaseone), $(0.2,1.0)$ (middle, \phasetwo) and $(0.6,0.5)$ (right, \phasethree). (c) First order phase diagram: in the orange area no interaction is relevant. In \phasetwo, only $\lambda_q$ is relevant and in \phasethree only $\lambda_{cq}$'s. In the top right corner, multiple interactions are relevant.}
    \label{fig:RG_phasediagram}
\end{figure}

Integrating the \RG equations numerically, we identify three main features: (i) the scale invariant phase \phaseone is robust in an extended parameter regime in the parameter plane, shown in light orange in \Fig{fig:RG_phasediagram}{(a)}. In particular, at $\gamma_B=0$, a measurement-induced transition takes place at a \emph{non-zero} measurement-strength $\gamma_M/\nu$ \footnote{The quantitative extension of the regime depends on the initial conditions and two constants in the \RG equations, whose values are a priori not known from the microscopic model.}. (ii) for non-zero $\gamma_B$, but $\gamma_M \to 0$ a dephasing-induced transition into the phase \phasetwo, as predicted by the first order equations, is confirmed \Fig{fig:RG_phasediagram}{(a),(b)}. 
(iii) Increasing the dephasing strength, the regime in which $\lambda_{cq}$'s become relevant and which is associated with the measurement-induced phase \phasethree, extends to smaller values of the measurement strength $\gamma_M/\nu$. This is qualitatively in line with the quantum trajectory simulations and the exact simulations for small systems. However, the predicted value of the phase boundary between phase \phasethree and \phasetwo fluctuates between the \RG approach and the quantum trajectory simulations (see below), as expected for non-universal quantities determined in a long wavelength effective theory.

A word of caution is at order at this point. In the parameter regime associated with phases $\phasetwo$ and \phasethree, it may happen that the poles of the propagator fully move onto the real axis and their imaginary part vanishes. In this case, the \RG flow breaks down and all correlation functions become formally infinite. This behavior is familiar from the situation of pure dephasing ($\gamma_M=0, \gamma_B>0$) and, e.g., discussed in Ref.~\cite{Buchhold2015}. It reflects the unbounded fluctuations of the fields $\phi$ if the system reaches an infinite temperature (maximally mixed) state. To which extent this behavior reflects a truly physical approach towards the infinite temperature state, or may be cured by higher order contributions to the \RG equations, is beyond the scope of this paper. For the tentative phase diagram in \Fig{fig:RG_phasediagram}{(a)}, we take the dominant coupling to be the one with the largest derivative when the integration of the flow equations breaks down.

\begin{figure*}
    \centering
    \includegraphics[width=\textwidth]{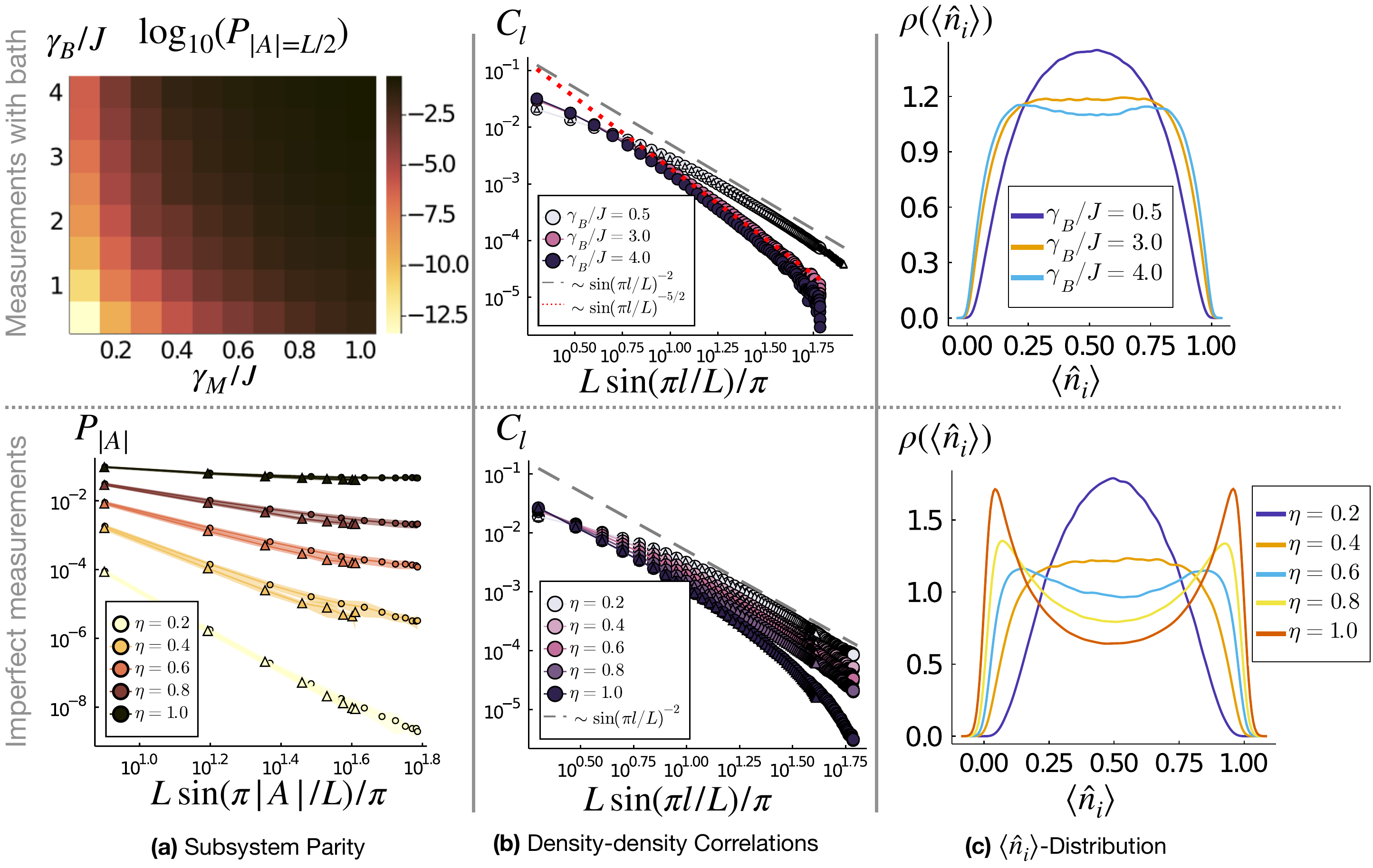}
    \caption{Overview of measured systems coupled to a bath (top) and imperfect measurements (bottom), the formal relation is given in Tab.~\ref{tab:Conversion}. (a) Overview of the half-system ($|A|=L/2$) parity for $L=128$ (top) and subsystem resolved parity for $L=128,192$ (triangles, circles) on a log-log scale for $\gamma/J=1.0$ (bottom). A regime of low half-system parity can be identified (top), which results from a strongly ($\sim$ algebraically) decaying $\subparity$ as a function of $|A|$. For imperfect measurements (bottom), examples are shown, ranging from saturation ($\eta=1.0$) to strong decay ($\eta=0.2$). (b) Correlations $C_{l}$ (log-log) for different bath strengths for $L=192$ and $\gamma_M/J=0.3$ (top) (also $L=256$ (triangles) for $\gamma_B/J=0.5$) and for different imperfection rates $\eta$ and $\gamma/J=1.0$ (bottom); (c) Probability density $\rho(\langle \hat{n}_i\rangle)$ of the local expectation values $\langle \hat{n}_i \rangle$ (extracted from histograms), turning from unimodal to bimodal for $\gamma_M/J=0.3$ (top; $L=192$) and $\gamma/J=1.0$ (bottom; $L=192$).}
    \label{fig:OverviewLargeSystems}
\end{figure*}

\subsection{Numerical investigation: Ensemble of Gaussian states}

For $\gamma_B=0$, any initial Gaussian state (product state), remains Gaussian due to the quadratic nature of the Hamiltonian and the measurement operators $\hat{n}_i^2 = \hat{n}_i$ for fermions. This allows for an efficient numerical simulation \cite{Cao2019,Chen2020,Alberton2021,Muller2021,Turkeshi2021,Kells2021arxiv,Coppola2021arxiv} as well as evaluation of correlators (using Wick theorem) and, e.g., entanglement \cite{Cao2019}.
For $\gamma_B>0$, the system is  described by a statistically weighted sum, i.e., an \emph{ensemble} of $n_\text{ens}$ (Gaussian) states, labeled by the index $\alpha$
\begin{align}
    \rhotraj_t = \sum_{\trajindex=1}^{n_\text{ens}} p_\trajindex(t) |\psi^{(\trajindex)}_t \rangle \langle \psi^{(\trajindex)}_t |.
    \label{eq:TrajectorySumWithWeights}
\end{align}
Such an ensemble approach was introduced in Refs.~\cite{Jacobs2010,Jacobs2014} and we use a similar scheme, explained below, combined with the efficient simulation of pure, Gaussian states presented in Ref.~\cite{Cao2019}. The necessary numerical approximation is to limit the ensemble size in \Eq{eq:TrajectorySumWithWeights} to a fixed number of states $n_{\text{ens}}$.

The \emph{advantage} of this approach is that the each ensemble member is still a Gaussian and therefore observables are still easy to evaluate, e.g., 
\begin{align}
    \Big(\text{tr}\left[ \hat{\mathcal{O}} \rhotraj \right]\Big)^2 = \Big(\sum_\trajindex p_\trajindex \langle \hat{\mathcal{O}}\rangle_\trajindex \Big)^2, \,\ \langle \hat{\mathcal{O}}\rangle_\trajindex:= \langle \psi^{(\trajindex)} | \hat{\mathcal{O}} | \psi^{(\trajindex)} \rangle.
\end{align} 
The \emph{drawback} of this representation is that we do not have direct access to quantities like entanglement entropies anymore, which for Gaussian states can be inferred directly from the correlation matrix. In addition, this approach is challenged by strongly mixed states, which, due to the limited ensemble size (i.e., much smaller than the exponentially large dimension of the Hilbert space of $L/2$ fermions on a lattice of $L$ sites), cannot be efficiently represented.
In the following, we sketch the (approximate) time evolution of the states and weights $p_\trajindex$ in this framework. A major simplification for the numerical simulation of the ensemble in Eq.~\eqref{eq:TrajectorySumWithWeights} is the above outlined equivalence between dephasing and unread measurements. Dephasing corresponds to an evolution where each individual state $|\psi^{(\trajindex)}_t\rangle\langle\psi^{(\trajindex)}_t|$ is measured separately, and the measurement outcome is not recorded. In contrast, in a true (read out) weak measurement, the whole ensemble is measured collectively and the result is recorded. 

In order to realize this evolution protocol, we define two types of Trotterized evolution operators \footnote{Formally, $\hat{U}_\trajindex$ can be any unravelling of the Lindblad dynamics. In practice, we use a unitary unravelling of the Lindblad dynamics, see Appendix~\ref{App:NumericalMethods},\ref{App:EnsembleSimulations}.} 
\begin{align}
    \hat U_\trajindex&\equiv \exp\Big[\sum_j \sqrt{\gamma_B}\Delta W_j^{(\trajindex)}(\hat n_j-\langle \hat n_j\rangle_\trajindex)-\gamma_B\delta t (\hat n_j-\langle \hat n_j\rangle_\trajindex)^2\Big],\nonumber\\
    \hat V&\equiv \exp\Big[\sum_j \sqrt{\gamma_M}\Delta W_j(\hat n_j-\langle\langle \hat n_j\rangle\rangle)-\gamma_M\delta t (\hat n_j-\langle\langle \hat n_j\rangle\rangle)^2\Big],\nonumber
\end{align}
where $\langle \hat n_j\rangle_\trajindex=\langle\psi^{(\trajindex)}_t|\hat n_j|\psi^{(\trajindex)}_t\rangle$ is the individual state average and $\langle\langle \hat n_j\rangle\rangle:=\sum_\trajindex p_\trajindex \langle\hat n_j\rangle_\trajindex$ is the ensemble average of $\hat n_j$. The dephasing noise $\Delta W_j^{(\trajindex)}$ is state specific $\overline{\Delta W_j^{(\trajindex)}\Delta W_{j'}^{(\trajindex  ')}}=\delta_{\trajindex,\trajindex '}\delta_{j,j'}\delta t$ and the measurement noise $\Delta W_j$ is the same for the whole ensemble $\overline{\Delta W_j\Delta W_{j'}}=\delta_{j,j'}\delta t$. The update reads \cite{Jacobs2010}:

\begin{align}
\begin{aligned}
     &\rhotraj_{t+\delta t}& = \sum_\trajindex p_\trajindex(t+\delta t) |\psi_{t+\delta t}^{(\trajindex)}\rangle \langle \psi_{t+\delta t}^{(\trajindex)}|,\\
    &|\psi^{(\trajindex)}_{t+\delta t}\rangle&=\frac{e^{-i\hat H\delta t}\hat V\hat U_\trajindex |\psi^{(\trajindex)}_{t}\rangle}{\sqrt{\langle\psi^{(\trajindex)}_t|\hat U_\trajindex\hat V^2\hat U_\trajindex |\psi^{(\trajindex)}_{t}\rangle}},\\
    &p_\trajindex(t+\delta t)&=p_\trajindex(t)\langle\psi^{(\trajindex)}_t|\hat U_\trajindex\hat V^2\hat U_\trajindex |\psi^{(\trajindex)}_{t}\rangle.
    \label{eq:EnsembleUpdateRules}
    \end{aligned}
\end{align} 
Therefore, $\rhotraj$ can still be simulated using Gaussian states, though with the additional overhead of evolving $n_\text{ens}$ states with individual weights $p_\trajindex$ at the same time (significantly reducing the accessible system sizes).

\subsubsection{Application: Purification for $\gamma_B=0$ \label{Sec:DetailsPurificationTimeScale}}

As we already discussed in Sec.~\ref{sec:ManyBodyFormulation} (see also again Refs.~\cite{Gullans2020,Gullans2020a}), the time scale $t_0$ of disentangling or decorrelation of the system with an ancillary system is another indicator of a (critically) entangled phase ($t_0$ grows with system size) or weakly entangled phases ($t_0$ saturates with system size). Here we briefly review the numerical side of the analysis, as it connects the physical intuition of purification and the numerical implementation (for mixed states) discussed before. 

The general idea is to couple the system to a reference ancilla $\{|k_R\rangle \} = \{|0_R\rangle,|1_R\rangle\}$, such that system+ancilla are described by a pure state \cite{Gullans2020,Gullans2020a}, see again \Eq{eq:RefAncillaConstruction}, where we choose
\begin{align}
    & | \psi^{(0)} \rangle = | 0 1 0 1 ...\rangle, \,\ |\psi^{(1)} \rangle = |1 0 0 1 01 ..\rangle, \,\ \langle \psi^{(0)} | \psi^{(1)} \rangle = 0, \nonumber
\end{align}
and initially the ancilla is fully entangled with the system. The overall pure state is not necessarily a Gaussian state, but the state of the system itself, $\text{tr}_R[|\psi_t\rangle \langle \psi_t|_{SR}] = \rhotraj_{S,t}$ is still a sum of two Gaussian states: 
\begin{align}
    &\rhotraj_{S,t} = p_0(t)|\psi^{(0)}_t \rangle \langle \psi^{(0)}_t| +p_1(t)|\psi^{(1)}_t \rangle \langle \psi^{(1)}_t|,
\end{align}
which can be directly simulated, using the ensemble-approach discussed above, \Eq{eq:EnsembleUpdateRules}. Purification means that the overlap $\langle \psi^{(0)}_t| \psi^{(1)}_t\rangle \to 1$ with the time-scale $t_0(L)$ shown in Fig.~\ref{fig:OverviewMeasurements}(b), as already discussed. Since the overall state $|\psi\rangle_{SR}$ is pure, this time scale can be extracted from the decay of the averaged entanglement $S_E$ between the reference ancilla and the system $\overline{S_E}(t) \sim e^{- t/t_0(L)}$, expected for long times \footnote{In the scale invariant (or critical) phase, initially an algebraic decay in time is expected \cite{Gullans2020,Gullans2020a,Li2021b,Buchhold2021}, turning into an exponential decay for long times.} with
\begin{align}
    \overline{S_E}(t) = -\overline{\text{tr}[ \rhotraj_{R,t} \log \rhotraj_{R,t}]} = -\overline{\text{tr}[ \rhotraj_{S,t} \log \rhotraj_{S,t}]}
\end{align}
and $\rhotraj_R$ and $\rhotraj_S$ the corresponding reduced density matrices (the entanglement entropy as a function of time is plotted in the Appendix, Fig.~\ref{fig:TimeResolvedPurification}).

\subsubsection{Application: Finite coupling to a bath}
In the following, we discuss the results from simulating ensembles for a finite bath strength or finite imperfection rate \footnote{A technical remark: Due to the necessity to use an ensemble, the accessible system sizes are smaller than for single Gaussian states, due to a massive increase in the number of trajectories and the need to keep track of $n_{\text{ens}}$ of them at once. Therefore, we investigate system sizes from $L=128$ to at most $L=256$. We use $n_\text{ens}=500$, $50$ realizations and $\delta t=0.05$ if not stated differently.}. From simulations of small systems and the analytical results, we expect a regime with $C_l \sim |l|^{-2}, \subparity \sim |A|^{-K}$ and a regime with $C_l < |l|^{-2}, \subparity \sim \text{const.}$ (we will not investigate the purity here and will not distinguish between phase \phaseone and \phasetwo). To get an overview, we plot the half-system parity in \Fig{fig:OverviewLargeSystems}{(a)} for $L=128$, suggesting a similar qualitative bipartition of the phase diagram compared to the small-scale simulations, \Fig{fig:OverviewSmallSystemsNew}{(a)}. More quantitatively, the decay of correlations is shown in \Fig{fig:OverviewLargeSystems}{(b)}. For small $\gamma_B/J$, the decay is still roughly $|l|^{-2}$, but for $\gamma_B/J \gg 1$ the decay is stronger ($\sim |l|^{-5/2}$ as a guide to the eye) \footnote{We use the rescaled coordinates $l \to L/\pi \sin(\pi l/L)$, anticipating that \phaseone is described by a conformal theory.}. At these intermediate system sizes, the behaviour could still be associated with an algebraic decay, nevertheless it could also be a transient towards an exponential decay, such that for the accessible system sizes a prediction of a sharp transition is not possible. A qualitative change though is again supported by the change in the $\langle \hat n_i \rangle$-histograms towards a bimodal distribution, see \Fig{fig:OverviewLargeSystems}{(c)}. As expected for larger noise strengths, the observables become more noisy towards larger $\gamma_B/J$, partly due to insufficient number of runs and partly due to limitations of the method itself. \\

\emph{Imperfect measurements}: Starting in the measurement dominated phase ($\gamma_M/J \sim 1$), reducing $\eta$ leads to less strongly decaying correlation functions, \Fig{fig:OverviewLargeSystems}{(b)} and a change in the $\langle \hat n_i \rangle$-distribution (comparable to \Fig{fig:OverviewSmallSystemsNew}{(d)}), \Fig{fig:OverviewLargeSystems}{(c)}. Complementing the correlation picture, we plot the subsystem-resolved parity for a fixed $L=128,192$, which turns from saturation into an ($\sim$ algebraically) decaying function for small $\eta$, see \Fig{fig:OverviewLargeSystems}{(a) (bottom)}. \\

In summary, the numerical findings qualitatively support the existence of a finitely extended scale invariant phase on the one hand, and a measurement-induced phase with more strongly decaying correlations as well as $\langle \hat n_i \rangle$ being closer to $0$ or $1$ on the other hand. Nevertheless, the accessible system sizes do not allow us to claim a sharp phase transition. Furthermore, the used method should become less trustworthy the larger the bath strength is (or the lower $\eta$), since the fixed-size ensemble we use will turn too small.

\section{Conclusion and Outlook}

The interplay of local measurements and unitary evolution can give rise to phase transitions, manifesting in, e.g., either delocalized, strongly entangled or localized, weakly entangled conditional states $\rhotraj$. 

A tractable example, numerically as well as analytically, are spinless fermions subject to a hopping Hamiltonian and local measurements of the particle number, featuring a \BKT-transition \footnote{The universal behavior at the transition was analyzed numerically by performing finite size scaling in Refs.~\cite{Alberton2021,Minato2021}, and it was found to be consistent with the \BKT universality class. The \BKT scenario was confirmed analytically for a corresponding continuum model in Ref.~\cite{Buchhold2021} and for a related model in Ref.~\cite{Bao2021}.}, separating a scale invariant phase from a measurement-induced, pinned phase \cite{Alberton2021,Buchhold2021} (also Refs.~\cite{Chen2020,Bao2021}).

We investigated the question how a residual coupling to a dephasing environment (measurement and bath operators commute) will modify such a transition and identified three qualitatively different phases: a scale invariant, weakly mixed phase \phaseone (robust against weak dephasing), a scale invariant, but more strongly mixed phase \phasetwo (only present for finite dephasing), and a measurement-induced phase \phasethree. Interestingly, \phaseone and \phasetwo cannot be distinguished based on `observables', which only depend on the local densities $\hat{n}_i$, but are separated in terms of a weak or strong mixedness.

The mixedness is also a challenge: the numerical method we used for larger systems is not well-suited to analyze strongly mixed regimes. Therefore, it would be desirable to investigate and classify the phase \phasetwo by means of alternative methods like, e.g., matrix product states \cite{Wolff2019,Bernier2020,Doggen2021arxiv,Wolff2020} and to extract the purity, mutual information or entanglement measures for \emph{mixed} states (e.g., the (fermionic) logarithmic entanglement negativity \cite{Shapourian2017} (see also Refs.~\cite{Sang2021,Alba2021,Minoguchi2021,Sharma2021arxiv,Weinstein2022arxiv})) -- the only requirement being that the expectation values in the \emph{conditional} master equation can be included in the dynamics. The benefit of such an investigation would be to clarify the properties of the phase \phasetwo, and possibly sharpen the character of the transition towards the measurement-induced phase \phasethree. 

On the contrary, the strong suppression of off-diagonal elements in $\rhotraj$ also opens the possibility for simplification (see also Ref.~\cite{Altland2021arxiv}): we introduced a phenomenological perspective, where dephasing and the hopping Hamiltonian lead to effective diffusion on the diagonal entries of the density matrix, counteracted by measurements, favoring localization. An interesting question is whether for related models, this competition could lead to a transition, which is described by an effectively \emph{classical} model.

Finally, we connect the imperfect measurement scenario with practical attempts to calculate `observables' like $\overline{\langle \hat{\mathcal{O}}\rangle^2}$. The main obstruction lies in extracting the expectation value $\langle \hat{\mathcal{O}}\rangle$, which formally requires multiple copies of the state. This poses a tremendous challenge, because it would require multiple measurement trajectories with the \emph{same} measurement outcomes. A follow up question could be whether it were sufficient to have a set of $n$ measurement trajectories, which only have partly agreeing measurement outcomes, say $\vec{m}$, to faithfully detect measurement-induced transitions. Such a set of trajectories corresponds approximately to the density matrix $\rhotraj_{\vec{m}}$, conditioned onto the outcomes $\vec{m}$ (where all other outcomes are unknown):
\begin{align}
    \rhotraj_{\vec{m}} \approx \frac{1}{n} \sum_{i=1}^n |\psi_{\vec{m}}^{(i)}\rangle \langle \psi_{\vec{m}}^{(i)}| .
\end{align}
If we are using $\rhotraj_{\vec{m}}$ to approximate $\langle \hat{\mathcal{O}}\rangle$, we are formally working in a regime of $\eta <1$ (or $\gamma_B \neq 0$). Therefore, depending on the fraction of equivalent outcomes, the observable might indicate the `wrong' phase: as we have seen, tuning $\eta$ can itself lead to a phase transition and therefore a `wrong phase', where, e.g., the phase of weak measurements is detected instead of the phase of strong measurements.

\textit{Acknowledgements} -- We thank A. Altland, T. M\"uller, A. Rosch, J. \AA berg and M. Rudner for useful discussions. We acknowledge support from the Deutsche Forschungsgemeinschaft (DFG) within the CRC network TR 183 (project grant 277101999) as part of project B02. 
S.D. acknowledges support by the Deutsche Forschungsgemeinschaft (DFG, German Research Foundation) under Germany’s Excellence Strategy Cluster of Excellence Matter and Light for Quantum Computing (ML4Q) EXC 2004/1 390534769, and by the European Research Council (ERC) under the Horizon 2020 research and innovation program, Grant Agreement No. 647434 (DOQS). M.B. acknowledges funding via grant DI 1745/2-1 under DFG SPP 1929 GiRyd.

\appendix

\section{Numerical methods \label{App:NumericalMethods}}

In the following, we discuss some details of the numerical implementation of density matrix evolution as well as the ensemble evolution (used for larger system sizes). In both cases, we rely on `Trotterized' time-evolution operators, which are accurate to order $\delta t$. For single trajectories ($\trajindex$), we define three different stochastic time evolution operators ($\hat M_i:=\hat L_i-\langle \hat L_i \rangle$):

\begin{align}
\begin{aligned}
    &\hat U_\trajindex^{(R)} \equiv \\ &\exp\Big[\sum_j \sqrt{\gamma_B}\Delta W_j^{(\trajindex)}(\hat n_j-\langle \hat n_j\rangle_\trajindex) 
    -\gamma_B\delta t (\hat n_j-\langle \hat n_j\rangle_\trajindex)^2\Big],\\
    &\hat U_\trajindex^{(I)}\equiv \exp\Big[i \sum_j \sqrt{\gamma_B}\Delta W_j^{(\trajindex)}\hat n_j \Big], \\
    &\hat V\equiv\\
    &\exp\Big[\sum_j \sqrt{\gamma_M}\Delta W_j(\hat n_j-\langle\langle \hat n_j\rangle\rangle)-\gamma_M\delta t (\hat n_j-\langle\langle \hat n_j\rangle\rangle)^2\Big].
    \label{eq:TrotterizedNoiseOperators}
    \end{aligned}
\end{align}
 \emph{Unravellings:} The operators $\hat{U}_\trajindex$ can be seen as yielding two different unravellings of the master equation $\partial_t \overline{\rhotraj} = -\frac12 \gamma_B \sum_i \left[\hat{n}_i,\left[\hat{n}_i,\overline{\rhotraj} \right]\right]$, such that an average over $n_\text{ens}$ different trajectories gives an approximation to $\overline{\rhotraj}$:
\begin{align}
    \overline{\rhotraj_{t+\delta t}} \approx  \frac{1}{n_\text{ens}} \sum_{\trajindex=1}^{n_\text{ens}} \hat{U}_\trajindex |\psi^{(\trajindex)}_t \rangle \langle \psi^{(\trajindex)}_t | \hat{U}_\trajindex^\dagger ,
\end{align}
which is an approximation for \emph{finite} $n_\text{ens}$. The operators $\hat{U}_\trajindex$ act \emph{independently} on the different states $|\psi^{(\trajindex)}\rangle \langle \psi^{(\trajindex)}|$ and the sum formally corresponds to the \emph{summation} over different, independent measurement trajectories/outcomes. The two different $\hat{U}_\trajindex$'s correspond to different weak measurements, which could be performed on the system (see, e.g., Ref.~\cite{Brun2002}). Here, $\hat{U}_\trajindex^{(R)}$ corresponds to the one discussed in the main text and $\hat{U}_\trajindex^{(I)}$ corresponds to an effectively unitary unravelling. (For some recent work on (complex) unravellings in the setting of fermions, see Ref.~\cite{Jin2021}). The overall sum over trajectories should be independent of the choice of unravelling, but it is important to note that single members in the sum will have very different physical properties, depending on the choice of $\hat{U}_\trajindex^{(R)}$ or $\hat{U}_\trajindex^{(I)}$. \\

\emph{Measurements:} In contrast, $\hat{V}$ describes real measurements and acts the same onto all members of the ensemble; it is controlled by the ensemble expectation value $\langle \langle \hat{n}_j \rangle \rangle= \text{tr}[\rhotraj \hat{n}_j]$.

\subsection{Small-scale simulations}
We numerically solve the conditional master equation \Eq{eq:FullFermionicModel}, using the following scheme (based on a 
Trotterization and re-exponentiation, accurate to first order in $\delta t$):
\begin{align}
    &\hat{U}_H = \exp\left(-i \hat{H} \delta t\right),\\
    & \hat{O}_{ij} := \left(\sum_{l=1}^{L} n^{(i)}_l n^{(j)}_l \right) |\{n\}_i \rangle \langle \{n\}_j |, \\
    & \hat{D} = \exp\left( \left(\hat{O}-\frac{L}{2} \mathbb{1} \right)\gamma_B \delta t \right),\\
    & \rhotraj_{t+\delta t} =  \hat{D}\cdot \left( \hat{V} \hat{U}_H \rhotraj_t \hat{U}_H^\dagger \hat{V}^\dagger \right),
\end{align}
where $\hat{D}\cdot (...)$ indicates the element-wise multiplication, which describes the dephasing of off-diagonal elements in the density matrix. All operators are constructed in the fixed particle-number Hilbert space with $N=L/2$ fermions. The parameters used for the different plots are given in Tab.~\ref{tab:NumericalParametersSmallScale}.

\begin{table}[]
    \centering
    \begin{tabular}{c|c|c|c|c}
         & $L$ & $J \delta t$ & $JT$ &  $n_\text{avg}$ \\ \colrule
         \Fig{fig:OverviewSmallSystemsNew}{ (top)}  & 10 & 0.01 & 40 & 400 \\
       \Fig{fig:OverviewSmallSystemsNew}{ (bottom)} & 10 & 0.02 & 20-80 & 400
    \end{tabular}
    \caption{Numerical parameters used for the small scale simulations (longer times for $\gamma_M/J=0.1,0.2$): $L$ (system size), $J\delta t$ (time step), $JT$ (running time), $n_\text{avg}$ (number of independent runs). Initial state is $\rhotraj_{t=0} = |\psi \rangle \langle \psi |$, $|\psi\rangle=|0101...\rangle$.}
    \label{tab:NumericalParametersSmallScale}
\end{table}

\begin{table}[]
    \centering
    \begin{tabular}{c|c|c|c|c|c}
         & $L$ & $J\delta t$ & $JT$ & $n_\text{ens}$ & $n_\text{avg}$ \\
         \colrule
        \Fig{fig:OverviewLargeSystems}{(a) (top)} & 128 & 0.05 & 200 & 500 & 50 \\ \colrule
       \Fig{fig:OverviewLargeSystems}{(b),(c) (top)}& 192 & 0.05 & 200-300 & 500 & 400 \\
       & 256 & 0.05 & 200 & 500 & 200  \\ \colrule
       \Fig{fig:OverviewLargeSystems}{ (bottom)} & 128, 192 & 0.05 & 200 & 500 & 50 \\ \colrule
        \Fig{fig:OverviewMeasurements}{(a)} & 256-512 & 0.05 & 300 & - & 400 \\
       & 768 & 0.05 & 300 & - & 200
    \end{tabular}
    \caption{Numerical parameters used for the larger scale simulations (with and without a bath): $L$ (system size), $J\delta t$ (time step), $JT$ (running time), $n_\text{ens}$ (ensemble size), $n_\text{avg}$ (number of independent runs). Initial state is $\rhotraj_{t=0} = |\psi \rangle \langle \psi |$, $|\psi\rangle=|0101...\rangle$, only for \Fig{fig:OverviewLargeSystems}{(a) top} an ensemble of random number eigenstates has been chosen.}
    \label{tab:NumericalParametersLargeScale}
\end{table}

\subsection{Ensemble simulations \label{App:EnsembleSimulations}}

The ensemble simulation is based on evolving the density matrix
\begin{align}
    \rhotraj_t \approx \sum_{\trajindex=1}^{n_\text{ens}} p_\trajindex(t) |\psi^{(\trajindex)}_t \rangle \langle \psi^{(\trajindex)}_t |,
\end{align}
where $|\psi^{(\trajindex)}_t\rangle$ are Gaussian states. The entire dynamics is encoded in the time evolution of these states and the weights $p_\trajindex(t)$. 

\emph{General procedure:} The combined dynamics can be split into two steps \cite{Jacobs2010}:
\begin{enumerate}
\item{Bath-step: Evolve each member with $|\psi^{(\trajindex)}_{t+ \delta t}\rangle^{(pre)} = \hat{U}_{\trajindex} |\psi^{(\trajindex)}_t \rangle$ and normalize.}
\item{Measurement-step: Evolve each member with $|\psi^{(\trajindex)}_{t+\delta t}\rangle = \hat{U}_H\hat{V}|\psi^{(\trajindex)}_{t+\delta t} \rangle^{(pre)}$, where $\hat{V}$ is the \emph{same} for all members, update the norm: $p_\trajindex \to p_\trajindex \langle \psi^{(\trajindex)}_{t+\delta t} | \psi^{(\trajindex)}_{t+\delta t} \rangle$, and normalize the states as well as the probabilities.}
\end{enumerate}

The main aspects of the ensemble simulations are: all states $|\psi_t^{(\trajindex)}\rangle$ are subject to the \emph{same} measurement noise and ensemble expectation value (described by $\hat{V}$). The bath is modelled by (independent) noise processes on each state (unravelling with $\hat{U}_\trajindex$). As  said before, the physical properties of single ensemble members will depend on the choice of the unravelling (only the sum should be independent). Since we want to study the stability of the scale invariant, weakly mixed phase against the measurement induced pinning phase, we want to avoid an artificial bias towards the measurement induced phase. Therefore, we model the bath by choosing $\hat{U}_\trajindex^{(I)}$ (also referred to as `fluctuating chemical potential' or `unitary unravelling' \cite{Cao2019}). This model (without additional measurements) is very similar to the one discussed in Ref.~\cite{Bauer2017}. One physical property of individual states for such an unravelling is that they still evolve into a volume-law entangled state \cite{Cao2019} (quite in contrast to the measurements we have considered before). In particular, such a noise process on its own will not `localize' the states into number eigenstates. \\

\emph{Gaussian state evolution}: Based on the approach developed in Ref.~\cite{Cao2019} (see also Refs.~\cite{Turkeshi2021,Kells2021arxiv}), we parametrize a Gaussian state for system size $L$ and fixed particle number $N$ as:
\begin{align}
    &|\psi_t \rangle = \prod_{i=1}^N \left( \sum_{j=1}^L U_{ji}(t) c_j^\dagger \right)|0\rangle \label{eq:GaussianParametrization}, \\
    & U^\dagger(t) U(t) = \mathbb{1},\\
    & \langle \psi' | \psi \rangle = \text{det}((U')^\dagger U).
\end{align}
The time evolution is described by updating $U(t)$ (see Ref.~\cite{Cao2019} for more details), directly encoding the correlation matrix, see \Eq{eq:ObservablesFromGaussianState}. Given an unnormalized state $| \psi \rangle$ ($U$), we can perform a QR-decomposition, such that $U=QR \to \tilde{U}=Q$ and $\langle \psi | \psi \rangle = \prod_j |R_{jj}|^2$, which gives us a normalized state $|\tilde{\psi}\rangle$, corresponding to $\tilde{U}$ and the norm of the old state (needed for the ensemble approach).  \\

\emph{Extracting observables (Gaussian states)}: Physical properties of a state $|\psi^{(\alpha)}\rangle$ are extracted from the correlation matrix $D^{(\alpha)}$ \cite{Cao2019,Alberton2021,Turkeshi2021}:
\begin{align}
    &D^{(\alpha)} = UU^\dagger, && D^{(\alpha)}_{ij}= \langle \psi^{(\alpha)}|c_i^\dagger c_j | \psi^{(\alpha)}\rangle = \langle c_i^\dagger c_j \rangle_\alpha \nonumber, \\
    & C^{(\alpha)}_{ij} = | \langle c_{i}^\dagger c_j \rangle_\trajindex|^2, && \subparity^{(\alpha)} = \det \left(2\cdot \left.D^{(\alpha)}\right|_A-\left.\mathbb{1}\right|_A \right)^2, \label{eq:ObservablesFromGaussianState}
\end{align}
where $\left. D\right|_A$ is the subset of the matrix $D$ with indices in region $A$. \\

\emph{Approximation for the ensemble}: Besides this general formalism, the size of the ensemble has to be limited. The effect of the measurement is to change the weights $p_\trajindex$ in the ensemble, such that after some time most weights become very small. Therefore, we use the simple recycling procedure, discussed in Ref.~\cite{Jacobs2010}: Once some $p_\trajindex$ falls below a threshold $p_{\text{thres}}$, the corresponding state $|\psi^{(\trajindex)}\rangle$ is discarded. To keep the ensemble at a fixed size, the discarded state is replaced by a duplicate of the most likely state $|\psi^{(\beta)}\rangle$ in the ensemble with $p_\beta = p_{\text{max}}$. Since we now have two copies of $|\psi^{(\beta)}\rangle$, we give both copies half the weight: $p_\text{max} \to  p_\text{max}/2$, whereby the overall state $\rhotraj$ is (nearly) unchanged, according to:
\begin{align*}
    &p_\trajindex |\psi^{(\trajindex)}\rangle \langle \psi^{(\trajindex)}| + p_\beta |\psi^{(\beta)}\rangle \langle \psi^{(\beta)}| \\
    &\approx p_\beta/2 |\psi^{(\beta)}\rangle \langle \psi^{(\beta)}| + p_\beta/2 |\psi^{(\beta)}\rangle \langle \psi^{(\beta)}|.
\end{align*}
Afterwards, the set of probabilities is normalized again. If not stated differently, we use $p_\text{thres} = 10^{-4}$ for $n_{\text{ens}}=500$. An overview of the numerical parameters is given in Tab.~\ref{tab:NumericalParametersLargeScale}.\\

\begin{figure}
 \centering
   \includegraphics[width=\textwidth]{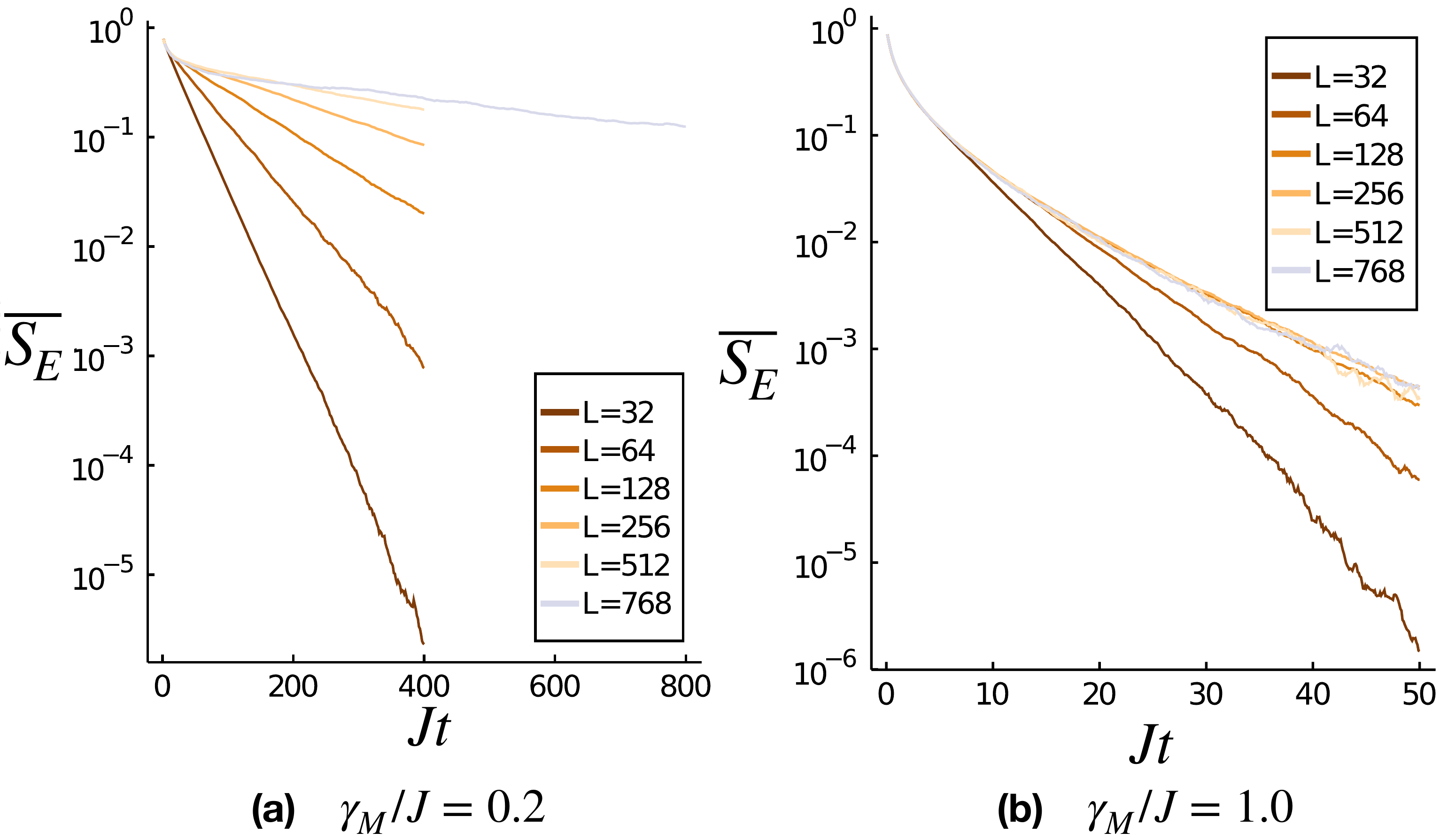}
    \caption{Decay of $\overline{S_E}$ (log-scale) of the ancilla, coupled to the fermionic system for (a) weak measurements $\gamma_M/J=0.2$ ($L=32$: $n_\text{avg}=10^5$; $L=64-256$: $n_\text{avg}=4\cdot 10^3$; $L=512$: $n_\text{avg}=1000$; $L=768$: $n_\text{avg}=250$) and (b) strong measurements $\gamma_M/J=1.0$. For weak measurements, the time scale of relaxation grows linearly in the system size. For strong measurements, the time-scale roughly saturates for large enough system sizes $L\ge 256$ ($L\le 256$, $n_\text{avg}=4\cdot 10^4$; $L=512,768$: $n_\text{avg}=4 \cdot 10^3$).}
    \label{fig:TimeResolvedPurification}
\end{figure}

\emph{Observables (sum of Gaussian states): density-density correlations}: One observable is the density-density correlator: 
\begin{align}
&C_{ij} = \overline{\langle\langle \hat{n}_i \rangle\rangle \cdot  \langle \langle \hat{n}_j \rangle\rangle -\langle\langle \hat{n}_i \hat{n}_j \rangle\rangle}, \\
&\langle \langle \hat{\mathcal{O}} \rangle \rangle = \text{tr}[ \hat{\mathcal{O}}\rhotraj] := \sum_{\trajindex=1}^{n_\text{ens}} p_\trajindex \langle \hat{\mathcal{O}} \rangle_\trajindex.
\end{align}
Here, the overline corresponds to the average over different measurement trajectories. From \Eq{eq:GaussianParametrization} we have direct access to the correlations of the Gaussian ensemble members $D_{ij}^{(\trajindex)}=\langle \psi^{(\trajindex)}| c_i^\dagger c_j | \psi^{(\trajindex)} \rangle= \langle c_i^\dagger c_j \rangle_\trajindex$. To access the second part in the correlator, $\langle\langle \hat{n}_i \hat{n}_j \rangle \rangle$, we can still make use of the Wick theorem, before we average:
\begin{align}
&\langle c_j^\dagger c_j c_k^\dagger c_k \rangle_\trajindex =\\
&\langle c_j^\dagger c_j\rangle_\trajindex \langle c_k^\dagger c_k\rangle_\trajindex - \langle c_j^\dagger c_k\rangle_\trajindex \langle c_k^\dagger c_j\rangle_\trajindex - \langle c_j^\dagger c_k\rangle_\trajindex \delta_{j,k}.
\end{align}
This means, that we can rewrite:
\begin{align}
\overline{\langle\langle \hat{n}_i \hat{n}_j \rangle\rangle} = \overline{\sum_\trajindex p_\trajindex \left( D_{ii}^{(\trajindex)} D_{jj}^{(\trajindex)} - |D_{ij}^{(\trajindex)}|^2 +D_{ii}^{(\trajindex)} \delta_{ij} \right)}
\end{align}
and therefore, knowledge of $D_{ij}^{(\trajindex)}$ and $p_\trajindex$ is enough to calculate the density-density correlator according to:
\begin{align}
\begin{aligned}
&C_{ij} =\overline{\sum_{\trajindex,\trajindex'} p_\trajindex p_{\trajindex'} D_{ii}^{(\trajindex)}D_{jj}^{(\trajindex ')}}\\
&- \overline{ \sum_\trajindex p_\trajindex \left( D_{ii}^{(\trajindex)} D_{jj}^{(\trajindex)} - |D_{ij}^{(\trajindex)}|^2 +D_{ii}^{(\trajindex)} \delta_{ij} \right)}.
\label{eq:DensityCorrelationEnsemble}
\end{aligned}
\end{align}
Remark: For strong noise ($\gamma_B/J$ or $\gamma_M/J$ large) the fluctuations in \Eq{eq:DensityCorrelationEnsemble} become large and can even lead to negative values, which might be avoidable by using larger ensemble-sizes (but which is not very feasible). The strong fluctuations can be seen in, e.g., the lower curves in  \Fig{fig:OverviewLargeSystems}{b} (top, $\gamma_M/J=4.0$).

\emph{Purification dynamics}: As discussed in the main text, the ensemble approach allows us to study the purification dynamics of the system, locally coupled to an ancilla. The corresponding time resolved plots for different measurement strengths are shown in Fig.~\ref{fig:TimeResolvedPurification}.

\section{Details about the Replica Action and RG analysis \label{App:DetailsRGAnalysis}}
In the following, we derive the explicit form of the replica action and describe the details of integrating out the absolute modes (following Ref.~\cite{Buchhold2021}). Afterwards, we discuss the first-order ($\mathcal{O}(\lambda_{ab}^1)$) \RG equations. Finally, we construct the second order ($\mathcal{O}(\lambda_{ab}^2)$) \RG equations and give the full set of flow equations in \Eq{eq:FullSetFlowEquations}. \\

The path integral description of the dynamics of $\rhotworeplica= \overline{\rhotraj \otimes \rhotraj}$ will be the key object to study the large distance properties of our effective, bosonic model. As we already said, $\rhotraj \otimes \rhotraj$ consists of two \emph{identical} copies. The dynamics of each copy is given by \Eq{eq:FullFermionicModel}, where the noise $dW_i$ is \emph{identical} for both copies. The roles of measurements and dephasing are again rather different: (i) measurements will, after averaging, induce a coupling between replicas; (ii) dephasing acts onto each replica individually.

Before averaging, $\rhotraj \otimes \rhotraj$ stays in a product state, because there are no interactions \emph{between} the copies. After averaging though, a coupling is induced and $\rhotworeplica$ will become correlated, the corresponding conditional 2-replica master equation reads:
\begin{align}
\begin{aligned}
&\rhotworeplica_{t+\delta t} = \rhotworeplica_{t} +i\delta t[\rhotworeplica_t,\hat{H}^{(1)}+\hat{H}^{(2)}] \\
 &-\frac12 (\gamma_M+ \gamma_B)\delta t \sum_i \left( [\hat{L}_i^{(1)},[\hat{L}_i^{(1)},\rhotworeplica_t]] \right.\\
 &\left.+[\hat{L}_i^{(2)},[\hat{L}_i^{(2)},\rhotworeplica_t]] \right) \\
 &\underbrace{+\gamma_M \delta t \overline{\sum_i \left\{ \hat{M}_i^{(2)},\{\hat{M}_i^{(1)},\rhotraj_t \otimes \rhotraj_t\} \right\}}}_{\text{measurement-induced interaction between replicas}},
 \label{eq:ReplicaMasterEquation}
\end{aligned}
\end{align}
where $\hat{M}^{(i)} := \hat{L}^{(i)}-\langle \hat{L}^{(i)} \rangle$. The exact expression of the last term will depend on higher replicas (due to the nonlinearities, see Ref.~\cite{Buchhold2021} for further details), but we are seeking for a closed expression for $\rhotworeplica$ only. Introducing 
\begin{align}
    \langle \langle ... \rangle \rangle = \text{tr}[... \rhotworeplica],
\end{align}
an approximate, norm-conserving version can be written as:
\begin{align}
    &\overline{\left\{ \hat{M}_i^{(1)},\{ \hat{M}_i^{(2)},\rhotraj \otimes \rhotraj\} \right\}} \approx -4 \tilde{C}_i \rhotworeplica \\
    &+ \{ \hat{L}_i^{(2)}-\langle \langle \hat{L}_i^{(2)}\rangle \rangle, \{ \hat{L}_i^{(1)}-\langle \langle \hat{L}_i^{(1)} \rangle \rangle, \rhotworeplica \} \}, 
\end{align}
where $\tilde{C}_i$ is defined as
\begin{align}
    \tilde{C}_i = \langle \langle \hat{L}_i^{(1)} \hat{L}_i^{(2)} \rangle \rangle - \langle \langle \hat{L}_i^{(1)} \rangle \rangle \langle \langle \hat{L}_i^{(2)} \rangle \rangle .
\end{align}
The approximation is based on the assumption that the statistical average over expectation values and the two-replica density matrix can be factorized, leading to 
\begin{align}
\begin{aligned}
    \overline{\langle \hat{L}_i\rangle \rhotraj \otimes \rhotraj} &\approx \overline{\langle \hat{L}_i\rangle} \cdot \rhotworeplica \\
    &=\langle \langle \hat{L}_i^{(1)}\rangle \rangle \cdot \rhotworeplica,
    \end{aligned}\\
    \begin{aligned}
     \overline{\langle \hat{L}_i\rangle^2 \rhotraj \otimes \rhotraj} &\approx \overline{\langle \hat{L}_i\rangle^2} \cdot \rhotworeplica\\
    &= \langle \langle \hat{L}_i^{(1)}\hat{L}_i^{(2)}\rangle \rangle \cdot \rhotworeplica.
    \end{aligned}
\end{align}
 It was shown in Ref.~\cite{Buchhold2021} that in the case of linear measurement operators and a quadratic Hamiltonian, this decoupling is exact. In general, the approximation is justified when the averages only contain the center of mass replica mode, and are independent of relative replica fluctuations. Adding decoherence, which is linear in the system operators, does not modify this result and provides exactly solvable relative correlations (replacing $\gamma \to \gamma_M+\gamma_B$ in the linear terms and $\gamma \to \gamma_M$ in the non-linear ones in Ref.~\cite{Buchhold2021} (Eqs.~(B8),(B9)) (see also Ref.~\cite{Minoguchi2021}).

In the following, we switch to the bosonic description, in which case there are two different kinds of measurement operators, but the structure is the same as in \Eq{eq:ReplicaMasterEquation}. Introducing $\hat{\Theta}=\partial_x \hat{\theta} /\pi$, we have a pair of conjugate variables, which we use to construct the path integral based on $\langle \phi_x|\Theta_x\rangle = e^{i\Theta_x \phi_x}$. Based on the conjugate operators $\hat{\phi}_x,\hat{\Theta}_x$ we can express $\text{tr}[\rhotworeplica_t]$ as a (Keldysh) path integral \footnote{See, e.g., Ref.~\cite{Sieberer2016a} for details on such a construction.}:
\begin{align}
&Z= \text{tr}[\rhotworeplica]= \int \mathcal{D}\left[\theta_\pm^{(1,2)},\phi_\pm^{(1,2)}\right] e^{i S}, \\
&S = S_{1}^{(0)}+ S_2^{(0)}+\Delta S_{1,2}, \label{eq:DirectReplicaAction}
\end{align}
where $\Delta S_{1,2}$ incorporates all non-quadratic terms, which couple the replicas. The $\pm$-indices denote the contour-indices, which stem from operators acting from the left or right onto the density matrix.

The formulation in \Eq{eq:DirectReplicaAction} is not ideal, because it contains cross-couplings, even at the quadratic level ($S_{1,2}^{(0)}$). To make this transparent, we rearrange the master equation using only $\hat{\mathcal{O}}^{(1)} \pm \hat{\mathcal{O}}^{(2)}$ instead: 
\begin{align}
&\rhotworeplica_{t+\delta t} = + i \delta t[\rhotworeplica_t,\hat{H}^{(1)}+\hat{H}^{(2)}] \nonumber \\
&+ \frac12 \gamma_B \sum_i \delta t \underbrace{\left( \mathcal{L}_{\hat{M}^{(1)}_i+\hat{M}^{(2)}_i}[\rhotworeplica_t] + \mathcal{L}_{\hat{M}^{(1)}_i-\hat{M}^{(2)}_i}[\rhotworeplica_t] \right)}_{\text{heating Lindblad terms}} \nonumber \\
& + \delta t\sum_i \gamma_M\underbrace{(\hat{M}_i^{(1)}+\hat{M}_i^{(2)})\rhotworeplica_t(\hat{M}_i^{(1)}+\hat{M}_i^{(2)})}_{\text{(heating) contour coupling}} \nonumber \\ &   - \delta t\Big\{ \underbrace{\frac12 \gamma_M \sum_i (\hat{M}_i^{(1)}- \hat{M}_i^{(2)})^2}_{\text{non-Hermitian Hamiltonian}},\rhotworeplica_t\Big\} \nonumber \\
 &- 4\gamma_M \delta t \sum_i\tilde{C}_i \rhotworeplica_t,
\end{align}
where we introduced the Lindblad superoperators 
\begin{align}
    \mathcal{L}_{\hat{\mathcal{O}}}[\hat{\rho}] := -\frac12 \left[ \hat{\mathcal{O}},\left[\hat{\mathcal{O}},\hat{\rho}\right] \right].
\end{align}
If the corresponding operators $\hat{\mathcal{O}}^{(1)} \pm \hat{\mathcal{O}}^{(2)}$ are \emph{linear} in the operators $\hat{\phi}_x, \hat{\theta}_x$, we can rotate the basis into the `relative' and 'absolute' space:
\begin{align}
\hat{\phi}_x^{(a)} := \frac{\hat{\phi}_x^{(1)}+\hat{\phi}_x^{(2)}}{\sqrt{2}}, && \hat{\phi}_x^{(r)} := \frac{\hat{\phi}_x^{(1)} - \hat{\phi}_x^{(2)}}{\sqrt{2}},
\end{align}
which will decouple the dynamics: $\rhotworeplica_t = \hat{\rho}^{(a)}_t \otimes \hat{\rho}^{(r)}_t$ (for factorized initial conditions). 
The action reads accordingly
\begin{align}
&Z := \int \mathcal{D}[\theta_\pm^{(a,r)},\phi_\pm^{(a,r)}] e^{i S}, \\
&S = S_{a}^{(0)}+S_{r}^{(0)}+\Delta S_{r,a},
\end{align}
where $\Delta S_{r,a}$ incorporates all non-quadratic terms, which couple the absolute and relative mode and $S_{l}^{(0)}$ have the form (after integrating out the field $\theta_X$, which only appears quadratically): 
\begin{align}
&S_l^{(0)}= \frac{1}{2} \int \frac{d\omega}{2\pi} \int \frac{dk}{2\pi} (\bosonmodes^{(l)}_{-Q})^{T} G^{-1}_l  \bosonmodes_Q^{(l)}, \\
& \bosonmodes_Q^{(l)}:= (\phi_+^{(l)},\phi_-^{(l)})^T .
\end{align}
Following the strategy outlined in Ref.~\cite{Buchhold2021}, we integrate out the absolute mode, which is 'heating' up. An indicator is already given at the quadratic level (remember: $t \to \nu t$):
\begin{align}
G_a^{-1} =&  \frac{1}{\pi}\left(\omega^2 -k^2\right) \sigma_z + i \frac{\gamma_B}{\pi^2 \nu}k^2 \mathbb{1} \\
&-i \frac{(\gamma_B+2\gamma_M)}{\pi^2\nu} k^2 \sigma_x
\label{eq:AbsoluteModePropagator}
\end{align}

where \emph{real} poles in the frequency plane for the absolute mode emerge (in contrast to the relative mode). In this `contour' description, the roles of the measurement and the bath are somewhat intermixed. A more transparent description is given in terms of the Keldysh coordinates ($c$: classical, $q$: quantum):
\begin{align}
&\phi_c^{(a,r)} = \frac{\phi_+^{(a,r)}+\phi_-^{(a,r)}}{\sqrt{2}}, \,\  \phi_q^{(a,r)} = \frac{\phi_+^{(a,r)} - \phi_-^{(a,r)}}{\sqrt{2}}.
\end{align}
For a Lindblad master equation, this description is favourable, since it makes use of some redundancies in the contour description (see, e.g., Ref.~\cite{Sieberer2016a}). The quadratic part of the action for the relative modes was already given in the main text $G_r^{-1}=G_0^{-1}$, \Eq{eq:RelativeModeGreensFunction}.

In this description, measurements couple symmetrically to the classical and quantum components and the bath couples \emph{only} to the quantum component. 

\begin{widetext}
\subsection{Full replica action}
Using the conjugate variables to construct the path integral, we get the following translation table: If we consider a superoperator:
\begin{align}
\hat{\mathcal{L}}[\hat{\rho}_t] = \int dx \, a L[\hat{\phi}_x] \hat{\rho}_t L[\hat{\phi}_x] +b \left\{ L^2[\hat{\phi}_x],\hat{\rho}_t \right\},
\end{align}
where $L$ are functions of the operators $\hat{\phi}_x$,
it gives rise to a contribution in the action (once re-exponentiated):
\begin{align}
iS_{\mathcal{L}} = \int dt  dx\{ a L[\phi_{x,t}^+]L[\phi_{x,t}^-] + b\left(L^2[\phi_{x,t}^+]+L^2[\phi_{x,t}^-]\right)\}.
\end{align}
With this, we can directly translate the replica master equation into a field theoretic description (keeping the replica indices):
\begin{align}
S = S_{r}^{(0)} + S_{a}^{(0)} + \Delta S_{+,-}.
\end{align}
The quadratic parts are given in \Eq{eq:RelativeModeGreensFunction} and \Eq{eq:AbsoluteModePropagator} (but in the contour-version), the interaction part reads (about dimensions: $[\normalorderingmass]=[x]^{-1}=[t]^{-1}$):
\begin{align}
i\Delta S_{+,-} &= m^2\int dt dx \left\{ \frac{2(\gamma_B+2\gamma_M)}{\nu}\cos(\sqrt{2} \phi_{+,X}^{(a)})\cos(\sqrt{2} \phi_{+,X}^{(r)})\cos(\sqrt{2} \phi_{-,X}^{(a)})\cos(\sqrt{2} \phi_{-,X}^{(r)}) \right.\\
&+ 2\frac{\gamma_B}{\nu}\sin(\sqrt{2} \phi_{+,X}^{(a)})\sin(\sqrt{2} \phi_{+,X}^{(r)})\sin(\sqrt{2} \phi_{-,X}^{(a)})\sin(\sqrt{2} \phi_{-,X}^{(r)}) \\
& -\frac12 \frac{(\gamma_B+\gamma_M)}{\nu} \left[ \cos(2\sqrt{2} \phi_{+,X}^{(a)})\cos(2\sqrt{2}\phi_{+,X}^{(r)}) + \cos(2\sqrt{2} \phi_{-,X}^{(a)})\cos(2\sqrt{2}\phi_{-,X}^{(r)}) \right] \\
&\left.+ \sum_{\sigma = \pm}\frac12 \frac{\gamma_M}{\nu}\left[ \cos(2\sqrt{2}\phi_{\sigma,X}^{(a)})+\cos(2\sqrt{2} \phi_{\sigma,X}^{(r)})\right] \right\}.
\end{align}
We have already left out the expectation values of the measured operators at this level (see Ref.~\cite{Buchhold2021} for more details).

\subsection{Integrating out the absolute mode \label{App:IntegratingOutAbsoluteMode}}
The guiding principle for integrating out the absolute mode are the relations (based on the heating of the absolute mode, see again Ref.~\cite{Buchhold2021}):
\begin{align}
&\left\langle e^{i \phi_{\sigma,X}^{(a)}} \right\rangle_a = e^{-\frac12 \langle (\phi_{\sigma,X}^{(a)})^2\rangle_a} \to 0, \,\ \left\langle e^{i \left(\phi_{\sigma,X}^{(a)}\pm \phi_{\sigma,Y\neq X}^{(a)} \right)} \right\rangle_a = e^{-\frac12 \langle \left(\phi_{\sigma,X}^{(a)} \pm\phi_{\sigma,Y\neq X}^{(a)} \right)^2\rangle_a} \to 0 .
\end{align}
We furthermore make the following \textbf{assumption}:
\begin{align}
\exp\left(-\left\langle \left(\phi_{+,X}^{(a)} \pm \phi_{-,Y}^{(a)} \right)^2 \right\rangle_{a} \right) \to 0 .
\end{align}
Using these inputs, we perturbatively calculate the action for the relative modes, according to 
\begin{align}
S[\phi^{(r)}] \approx S_r^{(0)} +\underbrace{\langle \Delta S_{+,-} \rangle_{(a)} +\frac{i}{2} \left( \langle \Delta S_{+,-}^2\rangle_{(a)} -\langle \Delta S_{+,-}\rangle_{(a)}^2 \right)}_{=:\Delta S_r}. 
\end{align}
The correction $\Delta S_r$ takes the form:
\begin{align}
&\Delta S_r = \normalorderingmass^2\int dt dx\left\{-\frac{i}{2} \frac{\gamma_M}{\nu} \sum_{\sigma=\pm} \cos(2\sqrt{2} \phi_{\sigma,X}^{(r)}) +\frac{i}{2} \frac{\gamma_M (\gamma_B+\gamma_M)}{\nu^2}\frac{1}{8} \sum_{\sigma=\pm} \cos(2\sqrt{2} \phi_{\sigma,X}^{(r)}) \label{eq:BarePerturbativeInteractions}\right.\\
&\left.- \frac{i}{2} \frac{1}{4} \left[\left(\frac{(2\gamma_M+\gamma_B)^2}{\nu^2}-\frac{(\gamma_B)^2}{\nu^2}\right) \sum_{\sigma=\pm} \cos(2\sqrt{2} \phi_{\sigma,X}^{(r)}) + \left(\frac{(2\gamma_M+\gamma_B)^2}{\nu^2} +\frac{(\gamma_B)^2}{\nu^2}\right)\cos(2\sqrt{2}\phi_{+,X}^{(r)})\cos(2\sqrt{2}\phi_{-,X}^{(r)})\right]\right\} \nonumber \\
=:& \int d^2X \left\{ i \left(\tfrac{\lambda_{cq}^{(c)}+i \lambda_{cq}^{(s)}}{2}\right) \cos(2\sqrt{2}\phi_{+,X}^{(r)}) + i \left(\tfrac{\lambda_{cq}^{(c)}-i \lambda_{cq}^{(s)}}{2} \right) \cos(2\sqrt{2} \phi_{-,X}^{(r)}) \right.\\
&\left.+i (\lambda_c+\lambda_q) \cos(2\sqrt{2} \phi_{+,X}^{(r)})\cos(2\sqrt{2} \phi_{-,X}^{(r)}) + i (\lambda_q-\lambda_c)\sin(2\sqrt{2}\phi_{+,X}^{(r)})\sin(2\sqrt{2} \phi_{-,X}^{(r)})\right\}, \nonumber \\
=:& \int d^2 X \left[i \lambda_c \cos(4\phi_{c,X}^{(r)}) + i \lambda_q \cos(4\phi_{q,X}^{(r)}) + i \lambda_{cq}^{(c)} \cos(2\phi_{c,X}^{(r)})\cos(2\phi_{q,X}^{(r)})+  \lambda_{cq}^{(s)} \sin(2\phi_{c,X}^{(r)})\sin(2\phi_{q,X}^{(r)})\right], 
\end{align}
where we have introduced the interactions $\lambda$ with dimensions $[\lambda] = [x]^{-2}$. Some of the couplings are zero initially, but will be generated under the \RG and we have already ignored terms of higher order, e.g., $\cos(4\sqrt{2}\phi_\sigma^{(r)})$. For comparison, we formulate these interaction terms in the contour as well as the Keldysh language and use the couplings introduced in the main text ($X:=(t,x)$), where we leave out the superscript $(r)$ in the following.

\subsection{1st order renormalization \label{App:FirstOrderRG}}
\textbf{Convention} (Fourier-transform): $f(x,t)=f(X) = \int_{-\infty}^{\infty} \frac{d\omega}{2\pi} \frac{dk}{2\pi} e^{i(\omega t +kx)} f(k,\omega) = \int \frac{d^2 Q}{(2\pi)^2} e^{i \vec{Q}\vec{X}}f(Q)$.

Formally, many of the following line of arguments are similar to, e.g., Ref.~\cite{Kogut1979} for the \RG analysis of the sine-Gordon model. To study which of these interaction terms can actually become relevant, we use a perturbative momentum-shell \RG scheme. To this end, we separate the fields into $\phi(X) = \phi^>(X) + \phi^<(X)$: 
\begin{align}
&\phi^>(X) = \int_{|k|<\Lambda/b} \frac{dk}{2\pi} \int_{-\infty}^{\infty} \frac{d\omega}{2\pi} \phi(Q) e^{i\vec{Q}\vec{X}}, \,\ \phi^<(X) = \int_{|k|>\Lambda/b} \frac{dk}{2\pi} \int_{-\infty}^{\infty} \frac{d\omega}{2\pi} \phi(Q) e^{i\vec{Q}\vec{X}}.
\end{align}
The short-wavelength modes $\phi^<(X)$, defined in a shell $\Lambda/b < |k| < \Lambda$, will be integrated out. At the first order, we get \footnote{The averages denote: $\langle \mathcal{O} \rangle_< := \int \mathcal{D}[\phi^<] \mathcal{O} e^{i S_0^<}$, $\langle \phi_a(X) \phi_b(X) \rangle_< := \int_{\Lambda/b < |k|<\Lambda} \frac{dk}{2\pi} \langle \phi_a(t=0,-k) \phi_b(t=0,k) \rangle$.} $S^> \approx S_0^> + \langle \Delta S \rangle_<$. Under the renormalization step, we get for example:
\begin{align}
\langle \cos(4(\phi_c^>(X)+\phi_c^<(X))) \rangle_< = \cos(4\phi_c^>(X)) e^{-8 \langle (\phi_c^<(X))^2 \rangle_<}, \\
\langle \cos(4(\phi_q^>(X)+\phi_q^<(X))) \rangle_< = \cos(4\phi_q^>(X)) e^{-8 \langle (\phi_q^<(X))^2 \rangle_<}.
\end{align}
The other interaction terms give rise to:
\begin{align}
\begin{aligned}
&\langle \cos(2(\phi_c^{>}(X)+\phi_c^<(X)))\cos(2(\phi_q^>(X)+\phi_q^<(X))) \rangle_< = \\
&\frac12 \cos(2(\phi_c^>(X)+\phi_q^>(X)))e^{-2 \langle (\phi_c^<(X)+\phi_q^<(X))^2 \rangle_<} + \frac12 \cos(2(\phi_c^>(X)-\phi_q^>(X))) e^{-2 \langle (\phi_c^<(X) - \phi_q^<(X))^2 \rangle_<} \\
& =\frac12 \left[\cos(2\phi_c^>(X))\cos(2\phi_q^>(X)) - \sin(2\phi_c^>(X))\sin(2\phi_q^>(X))\right]e^{-2 \langle (\phi_c^<(X)+\phi_q^<(X))^2 \rangle_<} \\
&+\frac12 \left[\cos(2\phi_c^>(X))\cos(2\phi_q^>(X)) + \sin(2\phi_c^>(X))\sin(2\phi_q^>(X))\right]e^{-2 \langle (\phi_c^<(X)-\phi_q^<(X))^2 \rangle_<}, 
\end{aligned} \\
\begin{aligned}
&\langle \sin(2(\phi_c^{>}(X)+\phi_c^<(X)))\sin(2(\phi_q^>(X)+\phi_q^<(X))) \rangle_< = \\
&-\frac12 \cos(2(\phi_c^>(X)+\phi_q^>(X)))e^{-2 \langle (\phi_c^<(X)+\phi_q^<(X))^2 \rangle_<} + \frac12 \cos(2(\phi_c^>(X)-\phi_q^>(X))) e^{-2 \langle (\phi_c^<(X) - \phi_q^<(X))^2 \rangle_<} \\
&= \frac12 \cos(2\phi_c^>(X))\cos(2\phi_q^>(X)) \left[-e^{-2 \langle (\phi_c^<(X)+\phi_q^<(X))^2 \rangle_<} + e^{-2 \langle (\phi_c^<(X)-\phi_q^<(X))^2 \rangle_<} \right] \\
&+\frac12 \sin(2\phi_c^>(X))\sin(2\phi_q^>(X)) \left[e^{-2 \langle (\phi_c^<(X)+\phi_q^<(X))^2 \rangle_<} + e^{-2 \langle (\phi_c^<(X)-\phi_q^<(X))^2 \rangle_<} \right]. \\
\end{aligned}
\end{align}
For $b=e^s$ and $s\to 0$, we can formulate the change in the couplings $\lambda(s)$ as differential equations, given in Sec.~\ref{Sec:FirstOrderRG}. An overview is given in  \Fig{fig:RGPhaseDiagramDifferentICs}{(a)}.

\subsection{Second order RG analysis: Details about the derivation}

We calculate the second order (in interaction strengths) correction to the action according to 
\begin{align}
\begin{aligned}
&\Delta S^{(\text{2nd})} = \frac{i}{2}\left( \langle \Delta S_r^2 \rangle_< - \langle \Delta S_r \rangle_<^2\right) \\
&\Delta S_r = \int d^2X \left[i\lambda_c \cos(4\phi_c) +i\lambda_q \cos(4\phi_q) + i\lambda_{cq}^{(c)} \cos(2\phi_c)\cos(2\phi_q) + \lambda_{cq}^{(s)}\sin(2\phi_c)\sin(2\phi_q) \right]\\
& = i \int d^2 X \left[\lambda_{+-}^{(c)} \cos(2\sqrt{2}\phi_+)\cos(2\sqrt{2} \phi_-) + \lambda_{+-}^{(s)} \sin(2\sqrt{2}\phi_+)\sin(2\sqrt{2} \phi_-) + \lambda_+ \cos(2\sqrt{2}\phi_+) + \lambda_- \cos(2\sqrt{2} \phi_-) \right],
\label{eq:InteractionsKeldyshandContour}
\end{aligned}
\end{align}
where we will use the formulation in terms of the Keldysh coordinates and the contour coordinates interchangeably. A qualitative overview of the corrections is given below:
\begin{align*}
&\textbf{2nd order} && \textbf{correction to} && \text{neglected} \\
&\lambda_q^2 && (\nabla \phi_q)^2  && \textcolor{gray}{\cos(8\phi_q)} \\
&\lambda_c^2 && (\nabla \phi_c)^2 && \textcolor{gray}{\cos(8\phi_c)} \\
&\lambda_+^2 &&  (\nabla \phi_q)^2,(\nabla \phi_c)^2, (\nabla \phi_c)(\nabla \phi_q) && \textcolor{gray}{\cos(4\sqrt{2}\phi_+)} \\
&\lambda_-^2 &&  (\nabla \phi_q)^2,(\nabla \phi_c)^2, (\nabla \phi_c)(\nabla \phi_q) && \textcolor{gray}{\cos(4\sqrt{2}\phi_-)} \\
& \lambda_+ \lambda_{+-}^{(c)},\lambda_+\lambda_{+-}^{(s)} && \cos(2\sqrt{2}\phi_-) && \textcolor{gray}{\cos(2\sqrt{2}(2\phi_+ \pm \phi_-))} \\
& \lambda_- \lambda_{+-}^{(c)},\lambda_-\lambda_{+-}^{(s)} && \cos(2\sqrt{2}\phi_+) &&  \textcolor{gray}{\cos(2\sqrt{2}(2\phi_- \pm \phi_+))}\\
& \lambda_+ \lambda_- && \cos(4\phi_c),\cos(4\phi_q)  && \\
&\textcolor{gray}{\lambda_c \lambda_q} && && \textcolor{gray}{\cos(4\sqrt{2}\phi_\pm)}
\end{align*}
[there are also other terms, which are generated, like derivative couplings coupled to an interaction term].

\textbf{Convention:}
In the following, we define the quadratic part of the action as:
\begin{align}
S_0 = \frac12 \int \frac{d\omega}{2\pi} \int \frac{dk}{2\pi} (\phi_c(-Q) \, \phi_q(-Q))  \, \begin{pmatrix} i\left(\eta_{qq}^2k^2 -\epsilon_{qq}^2 \omega^2\right) & \epsilon_{cq}^2 \omega^2 -\eta_{cq}^2 k^2 \\ \epsilon_{cq}^2 \omega^2 -\eta_{cq}^2 k^2 & i\left( \eta_{cc}^2 k^2 -\epsilon_{cc}^2 \omega^2\right) \end{pmatrix} \begin{pmatrix} \phi_c(Q) \\ \phi_q(Q) \end{pmatrix} .
\label{eq:GeneralDefinitionInversePropagator}
\end{align}
This form implies that all (equal-time) correlators can be brought into the form:
 \begin{align}
 \langle \phi_a(0,k) \phi_b(0,-k)\rangle =: \chi_{ab} \int \frac{d\omega}{2\pi} \frac{\eta_{ab}^2 k^2 -\epsilon_{ab}^2 \omega^2}{\Delta \epsilon (\omega-z|k|)(\omega +z|k|)(\omega-z^* |k|)(\omega+z^*|k|)} = \frac{\chi_{ab}}{2} \left[ \frac{\eta_{ab}^2}{\sqrt{\Delta \eta}} -\frac{\epsilon_{ab}^2}{\sqrt{\Delta \epsilon}} \right] \frac{i}{\sqrt{\Delta \epsilon} (z-z^*)},
 \end{align}
where we use (as in Sec.~\ref{Sec:FirstOrderRG})
\begin{align}
    \chi_{ab} = \begin{cases} 1 & a=b \\ -i & a  \neq b \end{cases}.
\end{align}
 The pole structure is determined by $z$, given by:
 \begin{align}
&z^2 = \pm \sqrt{\frac{(\epsilon_{cq}^2\eta_{cq}^2+\frac12 (\epsilon_{qq}^2 \eta_{cc}^2+\epsilon_{cc}^2\eta_{qq}^2))^2-\Delta \eta \Delta \epsilon}{\Delta \epsilon^2}} + \frac{\epsilon_{cq}^2\eta_{cq}^2+\frac12 (\epsilon_{qq}^2\eta_{cc}^2+\epsilon_{cc}^2 \eta_{qq}^2)}{\Delta \epsilon}, \\
&\Delta \epsilon := \epsilon_{cc}^2 \epsilon_{qq}^2+\epsilon_{cq}^4, \,\ \Delta \eta := \eta_{cc}^2 \eta_{qq}^2 + \eta_{cq}^4.
\end{align}
In the following, we present one explicit example to identify the necessary integral expression for the second order renormalization:
\begin{align*}
\lambda_q^2: \Delta S_{qq}^{2nd} = -\frac{i}{4}\lambda_q^2 \int d^2X d^2Y \left[ (e^{-16\langle \phi_q^<(X)\phi_q^<(Y)\rangle}-1)e^{-16 \langle (\phi_q^<)^2\rangle} \cos(4(\phi_q^>(X)+\phi_q^<(Y)) \right. \\
\left.+(e^{+16\langle \phi_q^<(X) \phi_q^<(Y)\rangle}-1)e^{-16 \langle (\phi_q^<)^2\rangle} \cos(4(\phi_q^>(X)-\phi_q^>(Y))\right] .
\end{align*}
The correlators $\langle \phi_a(X)\phi_b(X+\delta X)\rangle$ are quickly decaying, therefore it is sufficient to consider  $X,Y$ being close by (see also, e.g., Ref.~\cite{Kogut1979}). Since a term like $\cos(8\phi_q(X))$ is less relevant (from the first order \RG analysis) compared to $\cos(4\phi_q(X))$, we ignore those terms and only consider the second term:
 \begin{align}
  \cos(4(\phi_q^>(X)-\phi_q^>(X+\delta X))) \approx 1 - 8 \left(\delta X \vec{\nabla} \phi_q^> \right)^2,
 \end{align}
which gives rise to derivative corrections. Finally, we are interested in the flow equations of the couplings and therefore we are only interested in the leading order in $s$ ($b=e^s$) (our small parameter). Expressions like 
\begin{align}
(e^{\pm16\langle \phi_q^<(X)\phi_q^<(Y)\rangle}-1) \approx  \underbrace{(\pm16\langle \phi_q^<(X)\phi_q^<(Y)\rangle)}_{\sim \mathcal{O}(s)} + \mathcal{O}(s^2)
\end{align}
contribute already linearly in $s$ and therefore we can ignore any further corrections (e.g., from $e^{-16 \langle (\phi_q^<)^2\rangle}$ or $b^2$ from the rescaling step). For all expressions at second order, only two basic expressions are required ($\delta X:= (\delta t, \delta x)$):
 \begin{align}
 &\text{\textcolor{gray}{potential corrections:}} \nonumber \\
 &\mathcal{A}_{ab} = \int d^2 (\delta X)  \langle \phi_a(X) \phi_b(X+\delta X) \rangle \approx \underbrace{\chi_{ab}\frac{i}{\sqrt{\Delta \epsilon}(z^* -z)}\left( \frac{\eta_{ab}^2}{\sqrt{\Delta \eta}} - \frac{\epsilon_{ab}^2}{\sqrt{\Delta \epsilon}}\right)}_{=:A_{ab}} (A_1\cdot s)  \\
  &\text{\textcolor{gray}{derivative corrections:}} \nonumber \\
 &\mathcal{B}_{ab}^{(t,x)} = \int d^2 (\delta X) \begin{pmatrix} \delta t^2 \\ \delta x^2 \end{pmatrix}  \langle \phi_a(X) \phi_b(X+\delta X) \rangle \approx \chi_{ab}\begin{pmatrix} \frac{\eta_{ab}^2}{\Delta \eta} \\ -\frac{\epsilon_{ab}^2}{\Delta \epsilon} \end{pmatrix} (A_2 \cdot s) \label{Eq:DerivativeCorrection}
 \end{align}
 where $A_1,A_2$ are real constants, which we discuss in App.~\ref{App:AboutCoefficients}, and we furthermore made the order in $s$ explicit. \\

 \subsubsection{Corrections of the potential terms at second order}
 Depending on the terms under consideration, working either in the contour language or the Keldysh language might be favorable. A novelty compared to the standard \BKT scenario is that we also get corrections of the potential terms at second order. The corrections to the potential terms at leading order are:
\begin{align}
& \Delta S_{+,+-} \approx -2i \lambda_+ b^2 \int d^2X \left[ \lambda_{+-}^{(c)}\mathcal{A}_{++} - \lambda_{+-}^{(s)}\mathcal{A}_{+-}\right] \cos(2\sqrt{2} \phi_-(X)), \\
&\Delta S_{-,+-} \approx-2i \lambda_- b^2 \int d^2X \left[ \lambda_{+-}^{(c)}\mathcal{A}_{--} - \lambda_{+-}^{(s)}\mathcal{A}_{+-}\right] \cos(2\sqrt{2} \phi_+(X)).
\end{align}

 The corrections can be written as
 \begin{align}
 \Delta S _{\pm,+-} =&\int d^2X b^2\left[-4 \left[ i\lambda_c \lambda_{cq}^{(c)} \mathcal{A}_{cc} + i\lambda_q \lambda_{cq}^{(c)}\mathcal{A}_{qq}  - (\lambda_c+\lambda_q)  \lambda_{cq}^{(s)}\mathcal{A}_{cq} \right] \cos(2\phi_c)\cos(2\phi_q) \right.\\
 &\left. -4 \left[ \lambda_c \lambda_{cq}^{(s)} \mathcal{A}_{cc} - \lambda_q \lambda_{cq}^{(s)} \mathcal{A}_{qq} + i(\lambda_c + \lambda_q) \lambda_{cq}^{(c)}\mathcal{A}_{cq} \right] \sin(2\phi_c)\sin(2\phi_q) \right]
 \end{align}
 (there is an additional factor of $2$ due to the cross product in $\Delta S^2$). The correction at lowest order in the $\lambda_+ \lambda_-$-sector reads:
\begin{align}
\Delta S_{+,-} \approx -i 4\lambda_+\lambda_- b^2\int d^2X \left[-(\mathcal{A}_{cc} -\mathcal{A}_{qq})\cos(4\phi_c(X)) +  (\mathcal{A}_{cc} -\mathcal{A}_{qq} )\cos(4\phi_q) \right].
\end{align}

Therefore, the flow of the potential terms up to second order takes the form ($4\pi A_1 =: \bar{A}_1$):
\begin{align}
\partial_s \lambda_c =& \left(2+\frac{4}{\pi} A_{cc}\right) \lambda_c + \frac{1}{4\pi} \left( (\lambda_{cq}^{(c)})^2+(\lambda_{cq}^{(s)})^2\right) \left(A_{cc}-A_{qq}\right) \bar{A}_1, \\
\partial_s \lambda_q =& \left(2+\frac{4}{\pi} A_{qq}\right) \lambda_q - \frac{1}{4\pi} \left( (\lambda_{cq}^{(c)})^2+ (\lambda_{cq}^{(s)})^2\right) \left(A_{cc}-A_{qq}\right) \bar{A}_1, \\
\partial_s \lambda_{cq}^{(c)} =& \left( 2+ \frac{1}{\pi} \left( (1-\lambda_c  \bar{A}_1)A_{cc} +(1-\lambda_q \bar{A}_1)A_{qq} \right)\right)\lambda_{cq}^{(c)}  + i\frac{1}{\pi} \left( \left( 2-(\lambda_c+\lambda_q) \bar{A}_1\right) A_{cq}\right) \lambda_{cq}^{(s)}, \\
\partial_s \lambda_{cq}^{(s)} =& \left( 2 + \frac{1}{\pi}\left( (1+ \lambda_c  \bar{A}_1) A_{cc} +(1+\lambda_q  \bar{A}_1) A_{qq}\right)\right)\lambda_{cq}^{(s)} - i \frac{1}{\pi} \left( (2+(\lambda_c+\lambda_q) \bar{A}_1) A_{cq}\right) \lambda_{cq}^{(c)}.
\end{align}

\subsubsection{Derivative corrections}
Derivative corrections emerge from $\lambda_j^2$-like terms:

\begin{align}
\lambda_q^2: \Delta S_{qq} &= +2i\lambda_q^2 \int d^2X\left[(e^{+16\langle \phi_q^<(X) \phi_q^<(Y)\rangle}-1)(\delta X \nabla \phi_q)^2\right] \approx 32i \lambda_q^2 \int d^2X \left( \mathcal{B}_{qq}^{(t)}(\partial_t \phi_q)^2 +  \mathcal{B}_{qq}^{(x)}(\partial_x \phi_q)^2 \right), \\
\lambda_c^2: \Delta S_{cc} &= +2i\lambda_c^2 \int d^2X \left[(e^{+16\langle \phi_c^<(X) \phi_c^<(Y)\rangle}-1)(\delta X \nabla \phi_c)^2\right]  \approx 32i \lambda_c^2 \int d^2X \left( \mathcal{B}_{cc}^{(t)}(\partial_t \phi_c)^2 +  \mathcal{B}_{cc}^{(x)}(\partial_x \phi_c)^2 \right), \\
\lambda_+^2: \Delta S_{++}&= +i  \lambda_+^2 \int d^2X \left[ (e^{+8\langle \phi_+^<(X)\phi_+^<(Y)\rangle}-1)(\delta X \nabla \phi_+)^2\right]\\
&\approx i 8 \lambda_+^2 \int d^2X \left( \mathcal{B}_{++}^{(t)}\left[(\partial_t \phi_c)^2 + (\partial_t \phi_q)^2 + 2(\partial_t \phi_c)(\partial_t \phi_q) \right] + \mathcal{B}_{++}^{(x)}\left[(\partial_x \phi_c)^2 + (\partial_x\phi_q)^2 + 2(\partial_x \phi_c)(\partial_x \phi_q) \right] \right), \\
\lambda_-^2: \Delta S_{--} &= +i  \lambda_-^2 \int d^2X  \left[ (e^{+8\langle \phi_-^<(X)\phi_-^<(Y)\rangle}-1)(\delta X \nabla \phi_-)^2\right] \\
& \approx i 8 \lambda_-^2 \int d^2X \left( \mathcal{B}_{--}^{(t)}\left[(\partial_t \phi_c)^2 + (\partial_t \phi_q)^2 -2 (\partial_t \phi_c)(\partial_t \phi_q) \right] + \mathcal{B}_{--}^{(x)}\left[(\partial_x \phi_c)^2 + (\partial_x \phi_q)^2 - 2(\partial_x \phi_c)(\partial_x \phi_q) \right] \right) .
\end{align}

The flow equations for the quadratic sector of the action can be written as:
\begin{align}
&\partial_s \eta_{qq}^2 \approx \left[-64 \lambda_c^2 \frac{\epsilon_{cc}^2}{\Delta \epsilon} - 4\left[(\lambda_{cq}^{(c)})^2-(\lambda_{cq}^{(s)})^2\right] \left( \frac{\epsilon_{cc}^2+\epsilon_{qq}^2}{\Delta \epsilon}\right) -  16 \lambda_{cq}^{(c)}\lambda_{cq}^{(s)} \frac{\epsilon_{cq}^2}{\Delta \epsilon} \right]A_2, \\
& \partial_s \epsilon_{qq}^2 \approx \left[ -64 \lambda_c^2 \frac{\eta_{cc}^2}{\Delta \eta}  - 4\left[(\lambda_{cq}^{(c)})^2- (\lambda_{cq}^{(s)})^2\right] \left( \frac{\eta_{cc}^2+\eta_{qq}^2}{\Delta \eta}\right)  - 16 \lambda_{cq}^{(c)}\lambda_{cq}^{(s)} \frac{\eta_{cq}^2}{\Delta \eta}  \right]A_2, \\
& \partial_s \eta_{cq}^2 \approx \left[- 8 \lambda_{cq}^{(c)} \lambda_{cq}^{(s)} \left( \frac{\epsilon_{cc}^2+\epsilon_{qq}^2}{\Delta \epsilon}\right)  + 8 \left[ (\lambda_{cq}^{(c)})^2 - (\lambda_{cq}^{(s)})^2 \right] \frac{\epsilon_{cq}^2}{\Delta \epsilon} \right] A_2 , \\
& \partial_s \epsilon_{cq}^2 \approx \left[- 8 \lambda_{cq}^{(c)} \lambda_{cq}^{(s)} \left( \frac{\eta_{cc}^2+\eta_{qq}^2}{\Delta \eta}\right)  + 8 \left[ (\lambda_{cq}^{(c)})^2 - (\lambda_{cq}^{(s)})^2 \right] \frac{\eta_{cq}^2}{\Delta \eta} \right]A_2.
\end{align}

\textbf{Important property of the flow equations}: We assume that the constants $A_1,A_2$ are real-valued. This can be motivated along two lines: (i) for $A_2$ this will lead to the known result from Ref.~\cite{Buchhold2021} (the sign of $A_2$ can be inferred from the special case of $\gamma_B=0$); (ii) from a symmetry point of view, the bare action is  invariant under 
\begin{align}
\begin{aligned}
    &\phi_+ \to \phi_-,\\
    &\phi_- \to \phi_+, \\
    &\{g_i\} \to \{g_i^*\},
    \label{eq:SymmetryConstants}
    \end{aligned}
\end{align}
where $\{g_i\}$ is the set of all couplings. Choosing $A_1,A_2$ real preserves this structure and terms in the action $S$ like $i\int_X\lambda_+ \cos(2\sqrt{2}\phi_+)$ and $i\int_X \lambda_-\cos(2\sqrt{2}\phi_-)$ (see \Eq{eq:InteractionsKeldyshandContour}) are converted into each other under \Eq{eq:SymmetryConstants} for $\lambda_+=\lambda_-^*$, this relation being preserved during the full \RG-flow, \Eq{eq:FullSetFlowEquations} for $A_1$ real. The initial conditions read (again, for $t\to \nu t$ as given in the main text, \Eq{eq:RelativeModeGreensFunction} together with \Eq{eq:GeneralDefinitionInversePropagator})
\begin{align}
&\eta_{qq}^2 = \frac{2}{\pi^2} \frac{\gamma_M}{\nu}, && \epsilon_{cq}^2 =  \frac{1}{\pi}, \\
&\eta_{cc}^2 = \frac{2}{\pi^2} \frac{(\gamma_M+\gamma_B)}{\nu}, && \eta_{cq}^2 =  \frac{1}{\pi}.
\end{align}
(also $\lambda_c^{(0)},\lambda_q^{(0)},\lambda_{cq}^{(c)}$ are initially real). This has an important consequence: all couplings stay real during the flow. The full set of the corresponding flow equations are given in \Eq{eq:FullSetFlowEquations}.

\subsubsection{About the coefficients \label{App:AboutCoefficients}}
 Our approach to the second order \RG calculation is strongly based on treating space and time on equal footing. We do not specify the regularization scheme here, but assume that such a scheme exists. The idea is that once we introduce the coordinates (reminder: $z$ encodes the pole structure)
  \begin{align}
&\omega^2 = |z| \tilde{\omega}^2, && t^2 = |z|^{-1} \tilde{t}^2, \\
& k^2 = |z|^{-1} \tilde{k}^2, && x^2 = |z| \tilde{x}^2,
\end{align}
 space and time coordinates in the propagators can essentially be exchanged. The main goal of this section is to identify the scaling behaviour of $\mathcal{A}_{ab}$ and $\mathcal{B}_{ab}$. We first discuss $\mathcal{A}_{ab}$: To this end, we introduce the minimal building block:
 \begin{align}
 &\mathcal{A}_1(\alpha) = \int d^2 \tilde{X} \int \frac{d^2\tilde{Q}}{(2\pi)^2} \frac{ \tilde{\omega}^2 e^{-i \tilde{\vec{Q}} \tilde{\vec{X}}}}{ (\alpha \tilde{\omega}^2 -\alpha^{-1} \tilde{k}^2)(\alpha^{-1} \tilde{\omega}^2 - \alpha \tilde{k}^2)}  
 \end{align}
 ($\alpha := \sqrt{\frac{z^*}{z}}$), where the precise integration domain depends on the regularization scheme. There is an analogous expression with a $\tilde{k}^2$-term in the numerator instead of $\tilde{\omega}^2$. Here an important assumption comes in: we assume that both expressions are the same under transformations of the kind: $\tilde{x} \leftrightarrow \tilde{t}, \tilde{k} \leftrightarrow \tilde{\omega}$ (we assume that a proper regularization scheme exists, where space and time can be treated  equally). Using the `symmetrizing' rescalings as introduced above, we can write:
 \begin{align}
 &\mathcal{A}_{ab}= \int d^2X \langle \phi_a(0) \phi_b(X)\rangle_{<} =\chi_{ab} \left(\frac{\eta_{ab}^2}{\Delta \epsilon |z|^3} - \frac{\epsilon_{ab}^2}{\Delta \epsilon |z|} \right) \mathcal{A}_1(\alpha), \\
 & \Delta \epsilon = \epsilon_{cc}^2 \epsilon_{qq}^2+\epsilon_{cq}^4 , \,\ \Delta \eta = \eta_{cc}^2 \eta_{qq}^2+\eta_{cq}^4. 
  \end{align}
  The only inconvenience of this approach is that $\mathcal{A}_1(\alpha)$ has a residual dependence on $\alpha$, which we can get rid of by noting:
 \begin{align}
(\alpha-\alpha^{-1}) \mathcal{A}_1 = \int \frac{d^2 \bar{X} d^2 \bar{Q}}{(2\pi)^2} \frac{1}{\bar{\omega}^2-\bar{k}^2} e^{-i \bar{\vec{Q}}\bar{\vec{X}}}=:A_1(s) \approx A_1 \cdot s,
\end{align}
 where we rescaled $\alpha^{-1}\tilde{\omega}^2= \bar{\omega}^2,\alpha \tilde{k}^2= \bar{k}^2$ etc.. Here, $A_1(s)$ does not depend on the details of the propagator anymore and we assume $A_1$ to be a real number. Therefore, we have $\mathcal{A}_1(\alpha) = \frac{i|z|}{z^*-z} A_1$. The structural form of $\mathcal{A}_1(\alpha)$ is very similar to the structure of the first order calculation, in particular $z-z^*$ appears in the denominator. \\

Regarding the expression relevant for the derivative corrections, $\mathcal{B}_{ab}$, we will first go back to the \emph{symmetric} case $\eta_{cc}=\eta_{qq}$ etc., meaning $\gamma_B=0$ (or $\eta=1$). We use the symmetric case as a starting point to show that the same  $\mathcal{B}_{ab}$ can also be written as a constant times some propagator dependent prefactor. In the symmetric setting, it is much easier to work in the contour language, since the contours essentially decouple. The correlator for the $(+)$-contour reads:
\begin{align}
&\langle \phi_+(0)\phi_+(X)\rangle = \int \frac{d^2 Q}{(2\pi)^2} \frac{- e^{-i \vec{Q} \vec{X}} (\epsilon_-^2 \omega^2 -\eta_-^2 k^2)}{(\epsilon_+^2 \omega^2-\eta_+^2 k^2)(\epsilon_-^2 \omega^2 -\eta_-^2 k^2)} =  \int \frac{d^2Q}{(2\pi)^2} \frac{ - e^{-i \vec{Q}\vec{X}} ((z^*)^{-1} \omega^2 -z^* k^2)}{\epsilon_+ \eta_+ (z^{-1} \omega^2 - zk^2)((z^*)^{-1} \omega^2 -z^* k^2)},\\
&z^2:= \frac{\eta_+^2}{\epsilon_+^2}, \,\ (z^*)^2 = \frac{\eta_-^2}{\epsilon_-^2}.
\end{align}
[See the next section for the connection of $\epsilon_{\pm}/\eta_{\pm}$ with the Keldysh versions.] Symmetrizing the expression as before, we get ($\alpha := \sqrt{\frac{z^*}{z}}$):
\begin{align}
&\langle \phi_+(0)\phi_+(X)\rangle = -\int \frac{d^2 \tilde{Q}}{(2\pi)^2} \frac{  e^{-i \vec{Q}\vec{X}}(\alpha^{-1} \tilde{\omega}^2-\alpha \tilde{k}^2)}{\epsilon_+ \eta_+ (\alpha \tilde{\omega}^2 -\alpha^{-1} \tilde{k}^2)(\alpha^{-1} \tilde{\omega}^2- \alpha \tilde{k}^2)} .
\end{align}
We wish to evaluate the scaling of $\mathcal{B}_{ab}^{(t,x)}$, therefore we consider ($d^2(\delta X) = d (\delta t) d(\delta x))$:
\begin{align}
\begin{pmatrix} \mathcal{B}_{++}^{(t)} \\ \mathcal{B}_{++}^{(x)} \end{pmatrix} = \int d^2 (\delta X) \begin{pmatrix} (\delta t)^2 \\ (\delta x)^2 \end{pmatrix} \langle \phi_+(X)\phi_+(X+\delta X) \rangle = \int d^2 (\delta X) \begin{pmatrix} (\delta t)^2 \\ (\delta x)^2 \end{pmatrix} \langle \phi_+(0)\phi_+(\delta X) \rangle
\end{align}
By explicit evaluation, we get:
\begin{align}
\mathcal{B}_{++}^{(t)} =& -\frac{1}{|z|} \int d^2 \tilde{X} \tilde{t}^2 \int \frac{d^2 \tilde{Q}}{(2\pi)^2} \frac{ e^{-i \tilde{Q}\tilde{X}}}{\epsilon_+ \eta_+ (\alpha \tilde{\omega}^2-\alpha^{-1} \tilde{k}^2)} \\
=& -\underbrace{\frac{\alpha}{|z|} \frac{1}{\epsilon_+ \eta_+}}_{= \frac{1}{\eta_+^2}} \underbrace{\int d^2 \bar{X} \bar{t}^2 \int \frac{d^2 \bar{Q}}{(2\pi)^2} \frac{\pi e^{-i \bar{Q}\bar{X}}}{(\bar{\omega}^2-\bar{k}^2)}}_{=: A_2(s) \approx A_2 \cdot s} = -\frac{1}{\eta_{+}^2} A_2(s) , 
\end{align}
where we used a rescaling $\alpha \tilde{\omega}^2 = \bar{\omega}^2$ etc.. Here, $A_2(s)$ again does not depend on the details of the propagator and $A_2$ is just a constant. In the presence of an additional bath, this approach cannot be used (there will be no direct cancellation between numerator and denominator). Nevertheless, we can also calculate the integral by splitting the numerator:
\begin{align}
&\underbrace{- \int \frac{d^2 \tilde{Q}}{(2\pi)^2} \frac{ e^{-i \tilde{Q} \tilde{X}} \alpha^{-1} \tilde{\omega}^2}{\epsilon_+ \eta_+ (\alpha \tilde{\omega}^2 -\alpha^{-1} \tilde{k}^2)(\alpha^{-1} \tilde{\omega}^2 -\alpha \tilde{k}^2)}}_{:= -\frac{\alpha^{-1}}{\epsilon_+\eta_+} G^{(\omega)}} +\underbrace{\int \frac{d^2 \tilde{Q}}{(2\pi)^2} \frac{ e^{-i \tilde{Q} \tilde{X}} \alpha \tilde{k}^2}{\epsilon_+ \eta_+ (\alpha \tilde{\omega}^2 -\alpha^{-1} \tilde{k}^2)(\alpha^{-1} \tilde{\omega}^2 -\alpha \tilde{k}^2)}}_{:= \frac{\alpha}{\epsilon_+ \eta_+} G^{(k)}}, \\
&G^{(y \in \{k,\omega\})}:= \int \frac{d^2 \tilde{Q}}{(2\pi)^2} \frac{ e^{-i \tilde{Q} \tilde{X}} \tilde{y}^2}{ (\alpha \tilde{\omega}^2 -\alpha^{-1} \tilde{k}^2)(\alpha^{-1} \tilde{\omega}^2 -\alpha \tilde{k}^2)} . 
\end{align}

Our expression of interest now takes the form:
\begin{align}
&\begin{pmatrix} \mathcal{B}_{++}^{(t)} \\ \mathcal{B}_{++}^{(x)} \end{pmatrix} = \int d^2 (\delta X) \begin{pmatrix} (\delta t)^2 \\ (\delta x)^2 \end{pmatrix} \langle \phi_+(X)\phi_+(X+\delta X) \rangle = \int d^2 (\delta \tilde{X}) \begin{pmatrix} |z|^{-1} (\delta\tilde{t})^2 \\ |z| (\delta \tilde{x})^2 \end{pmatrix} \left[ - \frac{\alpha^{-1}}{\epsilon_+ \eta_+} G^{(\omega)} + \frac{\alpha}{\epsilon_+ \eta_+} G^{(k)} \right].  
\end{align}
We treat space and time on equal footing at this level, therefore we make the following `assumptions' (the same as we used for $\mathcal{A}_1$):
\begin{align}
&\underbrace{\int d^2 \tilde{X} \tilde{x}^2 G^{(k)}}_{:=A_x^{(k)}} = \underbrace{\int d^2 \tilde{X} \tilde{t}^2 G^{(\omega)}}_{:= A_t^{(\omega)}} =: C_\parallel, \,\  \underbrace{\int d^2 \tilde{X} \tilde{t}^2 G^{(k)}}_{:=A_t^{(k)}} = \underbrace{\int d^2 \tilde{X} \tilde{x}^2 G^{(\omega)}}_{:= A_x^{(\omega)}} =: C_\perp,
\end{align}
which are the building blocks for the correlators. Using these, we can write
\begin{align}
&\mathcal{B}_{++}^{(t)} = \frac{1}{|z|} \left[ -\frac{\alpha^{-1}}{\epsilon_+ \eta_+} C_\parallel + \frac{\alpha}{\epsilon_+ \eta_+} C_\perp \right] = \left[- \frac{1}{\epsilon_+ \eta_+} \frac{1}{z^*} C_\parallel + \frac{1}{\eta_+^2} C_\perp \right], \\
&\mathcal{B}_{++}^{(x)} = \left[-\frac{1}{\epsilon_+^2} C_\perp + z^* \frac{1}{\epsilon_+ \eta_+} C_\parallel \right]
\end{align}
From before, we have:
\begin{align}
\begin{aligned}
&\mathcal{B}_{++}^{(t)} = -\frac{1}{\eta_+^2} A_2(s) \stackrel{!}{=} \left[-\frac{1}{\epsilon_+ \eta_+} \frac{1}{z^*} C_\parallel + \frac{1}{\eta_+^2} C_\perp \right], \\
&\mathcal{B}_{++}^{(x)} = \frac{1}{\epsilon_+^2} A_2(s) \stackrel{!}{=} \left[ -\frac{1}{\epsilon_+^2} C_\perp + z^* \frac{1}{\epsilon_+ \eta_+} C_\parallel \right].
\label{eq:CoefficientEquationA2}
\end{aligned}
\end{align}
For the symmetric case ($\gamma_B=0$): $\epsilon_+^*=\epsilon_-$, $\eta_+^*=\eta_-$ and $z=\eta_+/\epsilon_+$, see the next section. Therefore, solving \Eq{eq:CoefficientEquationA2}, requires comparing $\frac{\eta_+}{\epsilon_+ }\frac{1}{z^*} = \frac{\eta_+ \epsilon_-}{\eta_- \epsilon_+}$ with  $\frac{\epsilon_+}{\eta_+}z^* = \frac{\eta_- \epsilon_+}{\eta_+ \epsilon_-}$. Since we are dealing with complex couplings, these ratios are not identical and solving the two equations, \Eq{eq:CoefficientEquationA2}, yields $C_\parallel \equiv 0$. Therefore, $\mathcal{B}_{++}^{(t,x)}$ is only related to $C_\perp=-A_2(s)\approx -A_2 \cdot s$. Nevertheless, the relation between the object $A_2(s)$ and $C_\perp(s)$ (which are only functions of $s$, not of any other parameters) holds generally for complex $z$ and we conclude (since $C_\parallel=0$) also for $\gamma_B \neq 0$:
\begin{align}
 \mathcal{B}_{ab}^{(t,x)} = \int d^2 (\delta X) \begin{pmatrix} \delta t^2 \\ \delta x^2 \end{pmatrix}  \langle \phi_a(X) \phi_b(X+\delta X) \rangle \approx 
 \chi_{ab} \begin{pmatrix} \frac{\eta_{ab}^2}{\Delta \eta} C_\perp \\-\frac{\epsilon_{ab}^2}{\Delta \epsilon} C_\perp \end{pmatrix} = -\chi_{ab}\begin{pmatrix} \frac{\eta_{ab}^2}{\Delta \eta} \\ -\frac{\epsilon_{ab}^2}{\Delta \epsilon} \end{pmatrix} (A_2 \cdot s) .
 \end{align}
The only important information we still need is the sign of $A_2$ (which we get by recovering the flow equations for the symmetric case). \\

\subsubsection{Recovering the symmetric case \label{App:RecoveringSymmetricCase}}
The symmetric case, $\eta_{cc}=\eta_{qq}$, $\epsilon_{cc}=\epsilon_{qq}$, leads to many simplifications and allows us to determine the sign of $A_2$. Most importantly, the couplings $\lambda_c$ and $\lambda_q$ do not become relevant (before $\lambda_\pm$). Therefore, we neglect them from the start. For the symmetric case, it is very convenient to work in the contour-description ($\pm$) instead of the Keldysh description ($c,q$):

\begin{align}
    &G_{0,\pm}^{-1} = i \begin{pmatrix} (\eta_{cc}^2+i\eta_{cq}^2)k^2-(\epsilon_{cc}^2+i\epsilon_{cq}^2)\omega^2 & 0 \\ 0 & (\eta_{cc}^2-i\eta_{cq}^2)k^2 - (\epsilon_{cc}^2-i\epsilon_{cq}^2)\omega^2 \end{pmatrix} := i \begin{pmatrix} \eta_+^2 k^2 -\epsilon_+^2 \omega^2 & 0 \\ 0 & \eta_-^2 k^2 - \epsilon_-^2 \omega^2 \end{pmatrix} \\
    & \langle \phi_\sigma (Q) \phi_\sigma (-Q) \rangle  = (\eta_\sigma^2 k^2 -\epsilon_\sigma^2 \omega^2)^{-1}
\end{align}

where we work with the complex couplings $\eta_\pm^2 = \eta_{cc}^2 \pm i \eta_{cq}^2$ and $\epsilon_\pm^2 = \epsilon_{cc}^2 \pm i \epsilon_{cq}^2$, such that $\eta_+^*=\eta_-$, $(\epsilon^2_+)^*=\epsilon_-^2$. Furthermore, we can write $z$ (encoding the pole-structure $\omega_P^2 = z^2 k^2$) as
\begin{align}
&z^2 = \frac{\eta_{cc}^2+i \eta_{cq}^2}{\epsilon_{cc}^2+i \epsilon_{cq}^2} = \frac{\eta_+^2}{\epsilon_+^2}.
\end{align}
This should be seen as a definition (such that $z,-z,z^*,-z^*$ encode the whole pole structure). For the flow equations, we get
\begin{align}
&\partial_s \eta_{cc}^2 = -16 \left[ \lambda_+^2 \frac{1}{\epsilon_+^2} + \lambda_-^2 \frac{1}{\epsilon_-^2}\right] A_2, \\
&\partial_s \eta_{cq}^2 = 16 \left[ \lambda_+^2 \frac{1}{\epsilon_+^2} - \lambda_-^2 \frac{1}{\epsilon_-^2} \right] A_2, \\
& \partial_s \eta_+^2 = -32 \lambda_+^2 \frac{1}{\epsilon_+^2} A_2.
\end{align}
Analogously, we obtain:
\begin{align}
\partial_s \epsilon_+^2 = -32 \lambda_+^2 \frac{1}{\eta_+^2} A_2.
\end{align}
For the interaction couplings, we get:
\begin{align}
&\partial_s \lambda_+ = \partial_s \left(\frac{\lambda_{cq}^{(c)}+i\lambda_{cq}^{(s)}}{2}\right) = \left( 2 - \frac{2}{\pi} i \frac{\alpha_z}{\sqrt{\Delta \epsilon} (z-z^*)} \left[ \frac{\eta_{cc}^2-i\eta_{cq}^2}{\sqrt{\Delta \eta}} - \frac{\epsilon_{cc}^2-i \epsilon_{cq}^2}{\sqrt{\Delta \epsilon}}\right]\right) \lambda_+ = \left( 2 + \frac{2}{\pi}\frac{i \alpha_z}{\eta_+ \epsilon_+}\right) \lambda_+, \\
& \Delta \eta = \eta_+^2 \eta_-^2,\quad  \Delta \epsilon = \epsilon_+^2 \epsilon_-^2.
\end{align}
We define $K_+ := \frac{i \epsilon_+ \eta_+}{\alpha_z}$ as the effective derivative coupling. Therefore, we have the flow equations
\begin{align}
&\partial_s \lambda_+ = \left( 2- \frac{2}{\pi} \frac{1}{K_+} \right) \lambda_+, \\
& \partial_s K_+ ^2 = 64 \lambda_+^2 A_2,
\label{eq:SymmetricFlowEquationDerivative}
\end{align}
where the sign of $A_2$ has to be \emph{negative} to recover the symmetric flow equations. The initial conditions read $K_+(0) = \frac{1}{\pi} \sqrt{1- i \frac{2\gamma_M}{\pi \nu}}$ (with $\alpha_z = -1$). The main difference to Ref.~\cite{Buchhold2021} is the $\partial_s K_+^2$ instead of $\partial_s K_+$, which we get by explicitly treating the $K_+$-dependence in the propagator. One might wonder if these equations are really closed, since $K_+$ still depends on $\alpha_z$. Nevertheless, considering the full set of flow equations again, we realize
\begin{align}
 \partial_s z^2 =\partial_s \left( \frac{\eta_+^2}{\epsilon_+^2}\right) =  0.
 \end{align}
 Therefore, the sign $\alpha_z$ is soley determined by the initial conditions (assuming that there can be no sudden jump in the otherwise trivial dynamics of $\partial_s z$). \\ 
 
 \subsubsection{First order RG in this framework}
 In the last sections, we assumed a regularization scheme, treating space and time on equal footing. Under the same assumption, we can re-derive the first order flow equations (but with an unknown constant). As an example consider again:
 \begin{align}
&\lambda_q \int d^2X \cos(4\phi_q)  \to b^2 e^{-8\langle (\phi_q)^2\rangle_<} \lambda_q \int d^2X \cos(4\phi_q(X)), \\
&\partial_s \lambda_q \approx (2- 8 s^{-1}\underbrace{\langle \phi_q^2\rangle_<}_{\mathcal{O}(s)}) \lambda_q,
\end{align}
where we can write for the equal-time, equal-space correlator (in the spirit of `symmetrizing'):
\begin{align}
&\langle \phi_a \phi_b \rangle_< = \chi_{ab}\left( \frac{\eta_{ab}^2}{\Delta \epsilon |z|^3} - \frac{\epsilon_{ab}^2}{\Delta \epsilon |z|} \right) \mathcal{A}_0, \\
&\mathcal{A}_0 := \int \frac{d^2 \tilde{Q}}{(2\pi)^2} \frac{\tilde{k}^2}{(\alpha \omega^2 -\alpha^{-1} k^2)(\alpha^{-1}\omega^2 -\alpha k^2)}, \\
&(\alpha - \alpha^{-1})\mathcal{A}_0 = i A_0(s), 
\end{align}
where $A_0(s) \approx A_0 \cdot s$ and $A_0$ is a real number. This gives rise to the flow equation
\begin{align}
\partial_s \lambda_q \approx \left( 2 - 8 \frac{i}{z-z^*} \left( \frac{\eta_{qq}^2}{|z|^2} -\frac{\epsilon_{qq}^2}{\Delta \epsilon}\right) |A_0|\right) \lambda_q,
\end{align}
where we put $A_0 = -|A_0|$ to recover the case with the sharp cutoff. \\ \\

\subsubsection{Collection of the full set of flow-equations in explicit form} 

\begin{align}
\begin{aligned}
&\partial_s \eta_{cc}^2 = \left[ -64 \lambda_q^2\frac{\epsilon_{qq}^2}{\Delta \epsilon} - 4\left( \frac{\epsilon_{cc}^2+\epsilon_{qq}^2}{\Delta \epsilon}\right) \left( (\lambda_{cq}^{(c)})^2 - (\lambda_{cq}^{(s)})^2\right) - 16 \lambda_{cq}^{(c)}\lambda_{cq}^{(s)} \frac{\epsilon_{cq}^2}{\Delta \epsilon} \right] A_2, \\
& \partial_s \eta_{qq}^2 = \left[ -64 \lambda_c^2\frac{\epsilon_{cc}^2}{\Delta \epsilon} - 4\left( \frac{\epsilon_{cc}^2+\epsilon_{qq}^2}{\Delta \epsilon}\right) \left( (\lambda_{cq}^{(c)})^2 - (\lambda_{cq}^{(s)})^2\right) - 16 \lambda_{cq}^{(c)}\lambda_{cq}^{(s)} \frac{\epsilon_{cq}^2}{\Delta \epsilon} \right] A_2, \\
& \partial_s \eta_{cq}^2 = \left[ -8\lambda_{cq}^{(c)}\lambda_{cq}^{(s)} \left( \frac{\epsilon_{cc}^2+\epsilon_{qq}^2}{\Delta \epsilon}\right) + 8 \left( (\lambda_{cq}^{(c)})^2 -(\lambda_{cq}^{(s)})^2\right) \frac{\epsilon_{cq}^2}{\Delta \epsilon}\right]A_2, \\
&\partial_s \epsilon_{cc}^2 = \left[ -64 \lambda_q^2\frac{\eta_{qq}^2}{\Delta \epsilon z_1^2z_2^2} - 4\left( \frac{\eta_{cc}^2+\eta_{qq}^2}{\Delta \epsilon z_1^2z_2^2}\right) \left( (\lambda_{cq}^{(c)})^2 - (\lambda_{cq}^{(s)})^2\right) - 16 \lambda_{cq}^{(c)}\lambda_{cq}^{(s)} \frac{\eta_{cq}^2}{\Delta \epsilon z_1^2 z_2^2} \right] A_2, \\
& \partial_s \epsilon_{qq}^2 = \left[ -64 \lambda_c^2\frac{\eta_{cc}^2}{\Delta \epsilon z_1^2 z_2^2} - 4\left( \frac{\eta_{cc}^2+\eta_{qq}^2}{\Delta \eta z_1^2 z_2^2}\right) \left( (\lambda_{cq}^{(c)})^2 - (\lambda_{cq}^{(s)})^2\right) - 16 \lambda_{cq}^{(c)}\lambda_{cq}^{(s)} \frac{\eta_{cq}^2}{\Delta \epsilon z_1^2 z_2^2} \right] A_2, \\
& \partial_s \epsilon_{cq}^2 = \left[ -8\lambda_{cq}^{(c)}\lambda_{cq}^{(s)} \left( \frac{\eta_{cc}^2+\eta_{qq}^2}{\Delta \epsilon z_1^2 z_2^2}\right) + 8 \left( (\lambda_{cq}^{(c)})^2 -(\lambda_{cq}^{(s)})^2\right) \frac{\eta_{cq}^2}{\Delta \epsilon z_1^2 z_2^2}\right]A_2 \\
&\partial_s \lambda_c = \left( 2+ \frac{4}{\pi} \frac{i}{\Delta \epsilon (z_1+z_2)} \left( \frac{\eta_{cc}^2}{z_1z_2} +\epsilon_{cc}^2\right) \right) \lambda_c \\
& \quad + \frac{1}{4\pi} \left( (\lambda_{cq}^{(c)})^2 + (\lambda_{cq}^{(s)})^2\right) \frac{i}{\Delta \epsilon (z_1+z_2)} \left( \frac{\eta_{cc}^2-\eta_{qq}^2}{z_1 z_2} +  \left(\epsilon_{cc}^2-\epsilon_{qq}^2\right) \right)A_1 ,\\
&\partial_s \lambda_q = \left( 2+ \frac{4}{\pi} \frac{i}{\Delta \epsilon (z_1+z_2)} \left( \frac{\eta_{qq}^2}{z_1z_2} +\epsilon_{qq}^2\right) \right) \lambda_q \\
& \quad - \frac{1}{4\pi} \left( (\lambda_{cq}^{(c)})^2 + (\lambda_{cq}^{(s)})^2\right) \frac{i}{\Delta \epsilon (z_1+z_2)} \left( \frac{\eta_{cc}^2-\eta_{qq}^2}{z_1 z_2} +  \left(\epsilon_{cc}^2-\epsilon_{qq}^2\right) \right)A_1 ,\\
& \partial_s \lambda_{cq}^{(c)} =  \left( 2+ \frac{1}{\pi} \frac{i}{\Delta \epsilon (z_1+z_2)} \left( (1-\lambda_c A_1) \left( \frac{\eta_{cc}^2}{z_1 z_2} +\epsilon_{cc}^2\right) + (1-\lambda_q A_1) \left( \frac{\eta_{qq}^2}{z_1 z_2} + \epsilon_{qq}^2 \right) \right) \right) \lambda_{cq}^{(c)} \\ 
& \quad + \frac{1}{\pi} \frac{i}{\Delta \epsilon (z_1+z_2)} (2- (\lambda_c+\lambda_q)A_1) \left( \frac{\eta_{cq}^2}{z_1 z_2} + \epsilon_{cq}^2 \right) \lambda_{cq}^{(s)}, \\
& \partial_s \lambda_{cq}^{(s)} =  \left( 2+ \frac{1}{\pi} \frac{i}{\Delta \epsilon (z_1+z_2)} \left( (1+\lambda_c A_1) \left( \frac{\eta_{cc}^2}{z_1 z_2} +\epsilon_{cc}^2\right) + (1+\lambda_q A_1) \left( \frac{\eta_{qq}^2}{z_1 z_2} + \epsilon_{qq}^2 \right) \right) \right) \lambda_{cq}^{(s)} \\ 
& \quad - \frac{1}{\pi} \frac{i}{\Delta \epsilon (z_1+z_2)} (2 + (\lambda_c+\lambda_q)A_1) \left( \frac{\eta_{cq}^2}{z_1 z_2} + \epsilon_{cq}^2 \right) \lambda_{cq}^{(c)}.
\label{eq:FullSetFlowEquations}
\end{aligned}
\end{align}
The pole structure is encoded in:
\begin{align}
\begin{aligned}
&z^2_\pm = \pm \sqrt{\frac{(\epsilon_{cq}^2\eta_{cq}^2+\frac12 (\epsilon_{qq}^2 \eta_{cc}^2+\epsilon_{cc}^2\eta_{qq}^2))^2-\Delta \eta \Delta \epsilon}{\Delta \epsilon^2}} + \frac{\epsilon_{cq}^2\eta_{cq}^2+\frac12 (\epsilon_{qq}^2\eta_{cc}^2+\epsilon_{cc}^2 \eta_{qq}^2)}{\Delta \epsilon}, \\
&\Delta \epsilon = \epsilon_{cc}^2 \epsilon_{qq}^2+\epsilon_{cq}^4, \\
&\Delta \eta = \eta_{cc}^2 \eta_{qq}^2 + \eta_{cq}^4.
\end{aligned}
\end{align}

In all these expressions, $z_1$ and $z_2$ are the roots of $z^2_\pm$ with poles in the upper half plane, respectively. In the simplest case, we have $z_2 = -z_1^*$. We always assume that there are no real poles. Initially, the `microscopic' initial interaction couplings result from integrating out the absolute modes, as given in \Eq{eq:BarePerturbativeInteractions} (which depend on $\gamma_M/\nu$ and $\gamma_B/\nu$):
\begin{align}
\begin{aligned}
&\eta_{cc}^2(0) = \frac{2}{\pi^2}\frac{(\gamma_M+\gamma_B)}{\nu},\quad \eta_{qq}^2(0) = \frac{2}{\pi^2} \frac{\gamma_M}{\nu}, \quad \eta_{cq}^2(0) =  \frac{1}{\pi},\quad \epsilon_{cc}^2(0)=\epsilon_{qq}^2(0)=0,\quad \epsilon_{cq}^2(0) =  \frac{1}{\pi}, \\
&\lambda_c(0)/\normalorderingmass^2=\lambda_q(0)/\normalorderingmass^2=-\frac{1}{16} \left(\frac{(2\gamma_M+\gamma_B)^2}{\nu^2}+\frac{\gamma_B^2}{\nu^2}\right),\\
&\lambda_{cq}^{(c)}(0)/\normalorderingmass^2=-\frac{\gamma_M}{\nu} + \frac18 \frac{\gamma_M(\gamma_M+\gamma_B)}{\nu^2}-\frac14 \left(\frac{(2\gamma_M+\gamma_B)^2}{\nu^2}-\frac{\gamma_B^2}{\nu^2}\right).
\label{eq:MicroscopicInitialConditions}
\end{aligned}
\end{align}

 For the two remaining constants $A_1,A_2$, we choose $A_2 \cdot \normalorderingmass^4=-1/16$ (which just compensates some prefactors in \eq{eq:FullSetFlowEquations}) and $A_1 \cdot \normalorderingmass^2=-0.1$ with a similar quantitative value.

\subsubsection{Solving the flow equations \label{App:DiscussionFlowEquations}}
{
The general feature of this class of flow equations, here \Eq{eq:FullSetFlowEquations}, is that they lead to a run away flow beyond the Gaussian regime. In our case, an additional complication arises because poles of the propagator can turn real during the flow, leading to a breakdown of our framework at some finite $s_f \sim \mathcal{O}(1)$. We solve the flow equations, \Eq{eq:FullSetFlowEquations}, with the aforementioned initial conditions, \Eq{eq:MicroscopicInitialConditions} up to $s_\text{max}=100$ using the scheme discussed below. In Fig.~\ref{fig:RGPhaseDiagramDifferentICs}, the coupling with the strongest growth at $s=s_f$ is indicated, which we use as a hint for the qualitative features of the phase. In  Fig.~\ref{fig:RGFlowResolved}, examples of the resolved flows are shown. The main features in Fig.~\ref{fig:RGPhaseDiagramDifferentICs} are: 
\begin{enumerate}
    \item{Stable regime \phaseone, where no interactions are relevant (see \Fig{fig:RGFlowResolved}{(left)}).}
    \item{Blurry region directly above \phaseone, with no unique interaction coupling being most relevant (we observe oscillations in this regime), see \Fig{fig:RGFlowResolved}{(right)}. For longer running times the regime increases.}
    \item{Small orange region, surrounding \phaseone, where $\lambda_c$ grows the strongest and (above) red region, where $\lambda_q$ grows the strongest. In both cases, we physically interpret the result as only a mass term of the form $(\mathbb{1}\pm \sigma_z)m_D$ being induced, which in both cases leads to the same phenomenology of \phasetwo.}
    \item{Towards $\gamma_M/\nu \approx 0.4$, $\lambda_{cq}$'s are most strongly growing (dark colors) [actually multiple couplings will be strongly growing, see \Fig{fig:RGFlowResolved}{(mid right)}], which we physically interpret as generating a mass term $m_M$, inducing short ranged correlations.}
    \item{Red spot, for $\gamma_M/\nu \approx 0.55$: since we expect multiple couplings to grow in this regime, we do not expect this region to have a different physical interpretation than the surrounding one.}
\end{enumerate}

Overall, the regime \phasetwo (red) shrinks by increasing $\gamma_B/\nu$, whereas the behaviour changes once $\gamma_B/\nu \approx 1.5$ (though a perturbative treatment becomes more questionable in this regime). We emphasize again that this analysis beyond the Gaussian, scale invariant and weakly mixed regime is not controlled. Additionally, we show the flow of the different couplings for four different points in the phase diagram, corresponding to $\phaseone$, $\phasetwo$, $\phasethree$ and the oscillatory regime. \\

The scheme we use to determine this phase diagram is based on:
\begin{itemize}
    \item{We terminate the flow once an interaction coupling $\lambda$ grows larger than $|\lambda|> 10^{2}$, and if the magnitude of the imaginary parts of any pole gets smaller than $|\text{Im}[z_{1,2}]| < 10^{-10}$. If the flow is not terminated before reaching $s_\text{max}$, we check if any $|\lambda|>10^{-2}$. If so, we conclude that not all couplings are vanishing, and we plot the coupling with the largest derivative $|\partial_s \lambda|$. Another possibility is that higher-order poles emerge in the propagator, so $z_1 \to z_2$, which as well is not covered by our flow equations.}
    \item{We solve the flow equations in Julia, using the `DifferentialEquations.jl'-package and the algorithm `AutoTsit5(Rosenbrock23())' with `reltol=1e-7,abstol=1e-7' up to $s_\text{max}=100$ (we have also checked different $A_1=-0.5,\pm 0.1$, $A_2= -1/2,-1/5,-1/10$ and $s_{\text{max}}=50$, yielding a qualitatively similar picture (for `reltol=1e-6,abstol=1e-6')). Increasing $|A_2|$ leads to a decreased region \phaseone. Using different algorithms yields a different fine structure in the region \phasethree (where, again, multiple couplings grow) with, e.g., larger `red' islands.}
\end{itemize}

\begin{figure}
    \centering
    \includegraphics[width=0.5\textwidth]{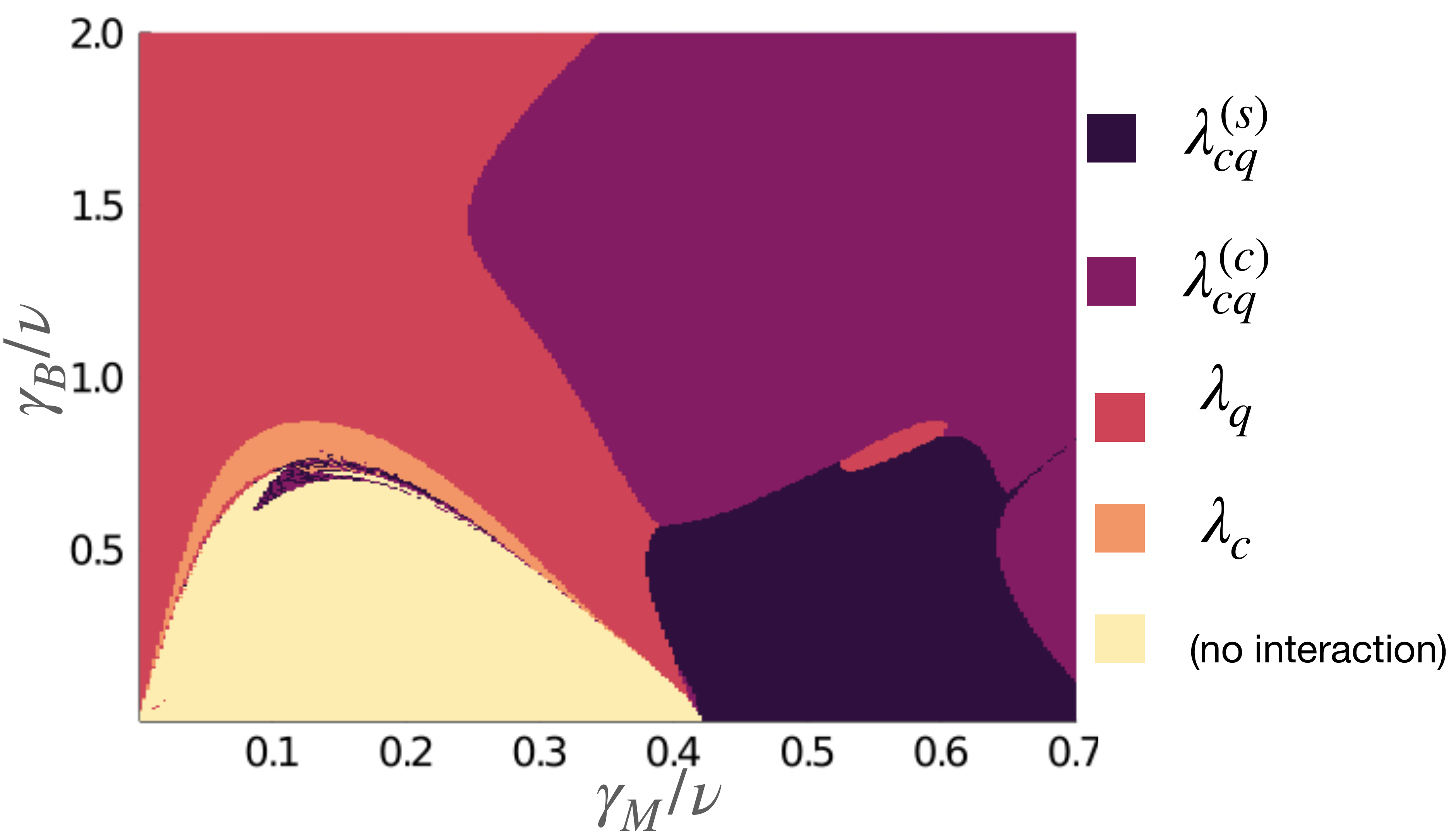}
    \caption{Phase diagram obtained from solving the second order flow equations, \Eq{eq:FullSetFlowEquations}. At each point, the color indicates, whether no coupling is relevant (light orange) or which coupling is most relevant (orange to black), as specified in the text. Beyond the regime of no relevant interaction (light orange), the \RG flow breaks down at a finite flow-parameter $s=s_f$ due to real poles emerging in the propagator. In these cases, we estimate the most relevant interaction by tracking the derivatives of the couplings. The one with the largest (absolute) change is indicated in the plot. The initial conditions are given in \Eq{eq:MicroscopicInitialConditions}.}
    \label{fig:RGPhaseDiagramDifferentICs}
\end{figure}

\begin{figure}
    \centering
    \includegraphics[width=0.95\textwidth]{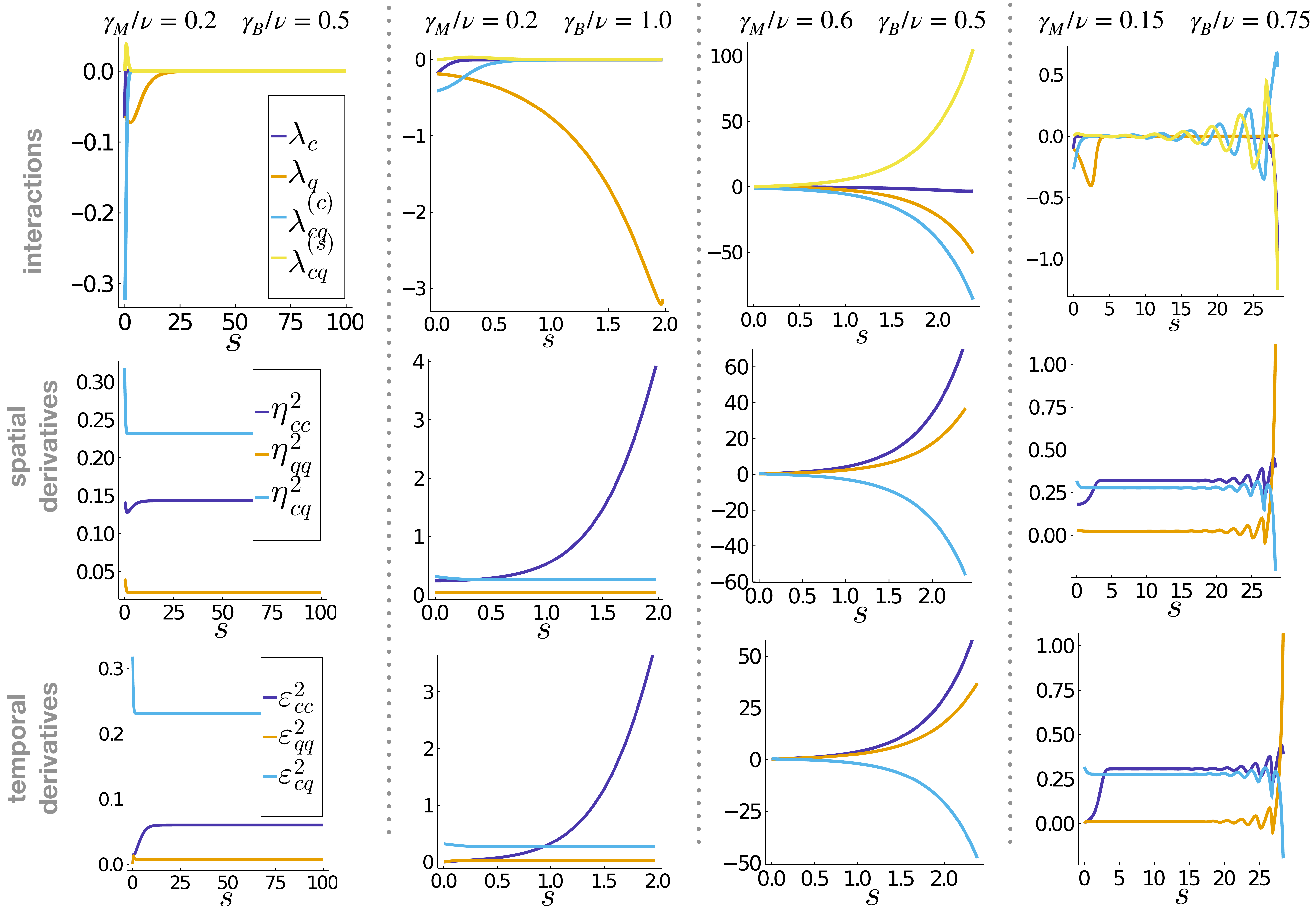}
    \caption{Resolved flow of the different couplings, interactions and derivative couplings, for different parameters, using initial conditions \Eq{eq:MicroscopicInitialConditions}. The first column shows the results inside the scale invariant, weakly mixed regime \phaseone: all interactions vanish and the flow of all couplings saturates. The second column shows the results in the regime of $\lambda_q$ being most relevant (see plot above), which we identify with \phasetwo. Note that these flows terminate at a finite $s_f$, where the flow breaks down. It is also important to note that none of the derivative couplings are vanishing. The third column shows the flow for the phase \phasethree: multiple couplings grow in magnitude. Column four shows the couplings in the `blurry', oscillatory region on top of \phaseone.}
    \label{fig:RGFlowResolved}
\end{figure}
}
\end{widetext}



%

\end{document}